\documentclass[11pt,english,eqsecnum,prd,aps,nofootinbib,superscriptaddress,longbibliography,tightenlines]{revtex4-2}
\usepackage[T1]{fontenc}
\usepackage[latin9]{inputenc}
\usepackage{xcolor}
\usepackage{babel}
\usepackage{amsmath}
\usepackage{amssymb}
\usepackage{mathtools}
\usepackage{graphicx}
\usepackage{float}
\PassOptionsToPackage{normalem}{ulem}
\usepackage{ulem}
\usepackage{subfig}
\usepackage[unicode=true,pdfusetitle,
bookmarks=true,bookmarksnumbered=false,bookmarksopen=false,
breaklinks=false,pdfborder={0 0 1},backref=false,colorlinks=true]
{hyperref}
\hypersetup{allcolors=blue}

\makeatother

\begin{document}
	\title{Probing the interior of the Schwarzschild black hole using congruences:\\ LQG vs. GUP}
	\author{Saeed Rastgoo}
	\email{srastgoo@yorku.ca}
	\affiliation{Department of Physics and Astronomy, York University, 4700 Keele Street, Toronto, Ontario M3J 1P3 Canada}

	\author{Saurya Das}
	\email{saurya.das@uleth.ca}
	\affiliation{Theoretical Physics Group and Quantum Alberta, Department of Physics and Astronomy, University of Lethbridge, 4401 University Drive, Lethhbridge, Alberta T1K 3M4, Canada}

	\selectlanguage{english}%

	\begin{abstract}
		We review, as well as provide some new results regarding the study of
		the structure of spacetime and the singularity in the interior of
		the Schwarzschild black hole in both loop quantum gravity and generalized
		uncertainty principle approaches, using congruences and their associated
		expansion scalar and the Raychaudhuri equation. 
		We reaffirm previous results
		that in loop quantum gravity, in all three major schemes of polymer quantization, the expansion scalar, Raychaudhuri equation and the
		Kretschmann scalar remain finite everywhere in the interior. In the
		context of the eneralized uncertainty principle, we show that only two of the four models we study
		lead to similar results. These two models have the property that their
		algebra is modified by configuration variables rather than the momenta.
	\end{abstract}
	\maketitle
	
\section{Introduction}

Black holes are one the most important objects in the Universe with regards to quantum gravity. 
The singularity in their interior is a
prediction of general relativity (GR), which in turn
is a prediction of its eventual breakdown. 
Furthermore, it is believed that 
this singularity resides in a small spatial region where quantum effects cannot be neglected. 
Thus, one has the natural expectation that 
a final theory of quantum gravity should be able to resolve this singularity.
Various theories of quantum gravity or effective gravity have been
utilized to study such objects. Among these are loop quantum gravity
(LQG) \citep{Thiemann:2007pyv}, a nonperturbative canonical theory
of quantization of the gravitational field, and the generalized uncertainty
principle (GUP), which is a rather phenomenological approach resulting
from the assumption of noncommutativity of spacetime or existence
of a minimum length. 

In LQG, there have been numerous works studying both the interior
and the full spacetime of the Schwarzschild black hole \citep{Ashtekar:2005qt,Bojowald:2004af,Bojowald:2005cb,Bohmer:2007wi,Corichi:2015xia,Ashtekar:2018cay,BenAchour:2018khr,Bodendorfer:2019cyv,Bodendorfer:2019nvy,Bojowald:2018xxu,Campiglia:2007pb,Chiou:2008nm,Corichi:2015vsa,Cortez:2017alh,Gambini:2008dy,Gambini:2009ie,Gambini:2011mw,Gambini:2013ooa,Gambini:2020nsf,Husain:2004yz,Husain:2006cx,Kelly:2020uwj,Modesto:2005zm,Modesto:2009ve,Olmedo:2017lvt,Thiemann:1992jj,Zhang:2020qxw,Ziprick:2016ogy,Campiglia:2007pr,Gambini:2009vp,Rastgoo:2013isa,Corichi:2016nkp,Morales-Tecotl:2018ugi,BenAchour:2020bdt,Blanchette:2020kkk,Husain:2022gwp,Addazi:2021xuf,Assanioussi:2020ezr,Husain:2021ojz,Assanioussi:2019twp}.
In particular, the interior of such a black hole has been studied in various ways. 
One of the most common approaches uses the so called
polymer quantization \citep{Ashtekar:2002sn,Corichi:2007tf,Morales-Tecotl:2016ijb,Tecotl:2015cya,Flores-Gonzalez:2013zuk}, which was originally inspired by loop quantum cosmology (LQC), dealing with a certain quantization of the isotropic Friedmann-Lemaitre-Robertson-Walker
(FLRW) model \citep{Ashtekar:2006rx,Ashtekar:2006uz}. Since the interior
of the Schwarzschild black hole is isometric to the Kantowski-Sachs
cosmological model, the idea in this polymer approach is to apply
the same techniques of the polymer quantization of the Kantowski-Sachs
model to the Schwarzschild interior \citep{Joe:2014tca,Saini:2016vgo}.
Polymer quantization introduces a parameter in the theory called the polymer scale, that sets the minimal scale of the model close to
which the quantum gravity effects become important. The approach in
which such a parameter is taken to be constant is called the $\mu_{0}$
scheme (which in this paper we refer to as the $\mathring{\mu}$ scheme),
while approaches where it depends on the phase space variables are
denoted by $\bar{\mu}$ schemes. 
The various approaches were introduced
to deal with some important issues resulting from quantization, namely,
to have the correct classical limit (particularly in LQC), to avoid large quantum corrections near the horizon, and to have final physical
results that are independent of auxiliary or fiducial parameters.
Other approaches to this model in LQG such as Refs. \citep{Alesci:2019pbs,Alesci:2020zfi}
provide a derivation of a Schwarzschild black hole modified dynamics, not relying on minisuperspace
models. Starting from the full LQG theory, this model performs the symmetry reduction at the quantum level. This has led to several differences
in the effective dynamics with respect to previous polymer quantization-inspired
models, one of which is the absence of the formation of a white hole
in the extended spacetime region replacing the classical singularity.
All of these past studies in LQG and some other approaches point to
the resolution of the singularity at the effective level.

Another approach to quantum gravity uses the so-called 
Generalized Uncertainty Principle (GUP). GUP extends the standard canonical commutation relation to 
include additional (small) momentum dependent terms, such that Heisenberg's uncertainty principle gets modified as well. It can be shown that as a result, there must exist a minimum measurable length, which can be for example, a multiple of the Planck length. 
Furthermore, such a modification affects practically all quantum Hamiltonians, even at low energies, giving rise to 
potentially measurable predictions of various quantum gravity theories 
\citep{Das:2008kaa,Ali:2011fa,Blanchette:2021vid,Bosso:2020ztk,Villalpando:2019usm,Das:2021nbq}. 
%
It may be noted that in the infrared limit, there is also an Extended Uncertainty Principle (EUP) which may apply to the black hole spacetimes under consideration 
\cite{Mignemi:2009ji,CostaFilho:2016wvf,Schurmann:2019jxe,Dabrowski:2019wjk,Hamil:2019pum,Gine:2020izd,Wagner:2021thc}. 


A particularly powerful approach to study the singularities in classical
and semiclassical/effective gravity is the use of congruences and
the associated expansion scalar and the Raychaudhuri equation to probe
the structure of spacetime. This approach which is the backbone of
the Hawking--Penrose singularity theorem, was particularly used,
among other works, in several of our recent studies \citep{Blanchette:2021vid,Blanchette:2020kkk,Blanchette:2021jcw}.
In this approach, a particular choice of congruence is made by choosing
the velocity vector field of the associated geodesics. In previous
works we have mainly used timelike congruences, while here we systematically
use both timelike and null ones.

This paper serves as both a review of our recent works in studying
the congruences in the interior of the Schwarzschild black hole in
LQG and GUP approaches, and also includes new results, particularly
with regard to GUP and the nonperturbative behavior of the Kretschmann
scalar in both approaches. The structure of the paper is as follows:
In Sec. \ref{sec:Raychaudhuri-equation}, we brief review the geodesic
deviation, expansion scalar and the Raychaudhuri equation and their
significance in studying the structure of spacetime. In Sec. \ref{sec:General-Schwarzschild-Congruence}
we use these results to choose certain congruences to study the interior
of the Kantowski-Sachs metric which is isometric to the Schwarzschild
black hole interior. In Sec. \ref{sec:Classical-Schwarzschild-interior}
we review the classical formulation of the interior of the Schwarzschild
balck hole based on the Ashtekar-Barbero connection and derive general
expressions for the expansion scalar and the Raychaudhuri equation
for both timelike and null cases. In Sec. \ref{subsec:Loop-quantum-gravity}
we apply these results to the effective black hole interior in LQG
and show that in all the three common schemes, and using either timelike
or null congruences, not only expansion scalar and the Raychaudhuri
equation always remain finite in the interior, but also the Kretschmann
scalar does so. In Sec. \ref{subsec:Generalized-uncertainty-principle},
we do the same for four most common model in GUP and show that only
two of them have the property that their expansion scalar and Raychaudhuri
equation together with he Kretschmann scalar always remain finite.
Finally, in Sec. \ref{sec:Discussion}, we conclude and present an
outlook for future work.

\section{General Schwarzschild interior and Congruences\label{sec:General-Schwarzschild-Congruence}}

Given that the radial spacelike and timelike coordinates switch their
causal nature one we cross the horizon in the Schwarzschild black
hole, we can simply switch $t\leftrightarrow r$ in the Schwarzschild
metric to obtain the metric of the interior as
\begin{equation}
	ds^{2}=-\left(\frac{2GM}{t}-1\right)^{-1}dt^{2}+\left(\frac{2GM}{t}-1\right)dr^{2}+t^{2}d\Omega^{2},
\end{equation}
where $t,\,r,\,\theta,\,\phi$ are the standard Schwarzschild coordinates
and $d\Omega^{2}=d\theta^{2}+\sin^{2}(\theta)d\phi^{2}$. As it is
seen, $t^{2}$ now plays the role of the radius of the infalling 2-spheres.
Notice that this model is not a field theory anymore since the metric
components (and hence the degrees of freedom) are independent of $r$.
So we are dealing with a system with finite degrees of freedom, i.e.,
a minisuperspace model. The above metric is a special case of the
Kantowski-Sachs cosmological model
\begin{equation}
	ds_{KS}^{2}=-N(t)^{2}dt^{2}+g_{xx}(t)dx^{2}+g_{\theta\theta}(t)d\theta^{2}+g_{\phi\phi}(t)d\phi^{2},
\end{equation}
which describes a homogeneous but anisotropic spacetime. 

In order to obtain a general result for such models, we consider a
metric of the form
\begin{equation}
	ds^{2}=-N(t)^{2}dt^{2}+X^{2}(t)dr^{2}+Y^{2}(t)d\Omega^{2}.\label{eq:gen-metric}
\end{equation}
We will study the null and timelike congruences propagating on this spacetime. To be self-contained, a brief review of the geodesic deviation, expansion scalar, and Raychaudhuri equation is given in Appendix \ref{sec:Raychaudhuri-equation}.

\subsection{Timelike case}

Let us consider a radial timelike congruence of geodesics where their
velocity vector in the coordinates given in \eqref{eq:gen-metric}
is
\begin{equation}
	U^{\mu}=\left(U^{0},U^{1},0,0\right).
\end{equation}
Given that $U^{a}$ is a unit timelike vector field, the above vector
can be written as 
\begin{equation}
	U^{\mu}=\left(U^{0},\frac{\sqrt{-1+N^{2}\left(U^{0}\right)^{2}}}{X},0,0\right).
\end{equation}
Hence, to simplify our analysis we choose the free component $U^{0}$
as
\begin{equation}
	U^{0}=\frac{1}{N},
\end{equation}
to obtain
\begin{equation}
	U^{\mu}=\left(\frac{1}{N},0,0,0\right).\label{eq:U-timelike}
\end{equation}
Using this form of the velocity vectors, we can easily obtain the
transverse metric from \eqref{eq:g-h-uu} as 
\begin{equation}
	h_{\mu\nu}=\begin{pmatrix}0 & 0 & 0 & 0\\
		0 & X^{2} & 0 & 0\\
		0 & 0 & Y^{2} & 0\\
		0 & 0 & 0 & Y^{2}\sin^{2}\left(\theta\right)
	\end{pmatrix}.\label{eq:transverse-h-timelike}
\end{equation}
The expansion tensor corresponding to \eqref{eq:U-timelike} also
becomes
\begin{equation}
	B_{\mu\nu}=\nabla_{\nu}U_{\mu}=\begin{pmatrix}0 & 0 & 0 & 0\\
		0 & \frac{X\dot{X}}{N} & 0 & 0\\
		0 & 0 & \frac{Y\dot{Y}}{N} & 0\\
		0 & 0 & 0 & \frac{Y\dot{Y}}{N}\sin^{2}\left(\theta\right)
	\end{pmatrix}.
\end{equation}
Using this tensor, the metric and the transverse metric \eqref{eq:transverse-h-timelike},
it is straightforward to obtain
\begin{align}
	\theta= & \frac{\dot{X}}{NX}+2\frac{\dot{Y}}{NY},\label{eq:theta-timlike-gen}\\
	\sigma^{2}= & \frac{2}{3N^{2}}\left(\frac{\dot{X}}{X}-\frac{\dot{Y}}{Y}\right)^{2},\\
	\omega_{ab}= & 0.
\end{align}
It is clear from here that in order to be able to find these quantities,
we need to obtain the equations of motions, i.e., the Einstein's equations.
Here is where the difference between the classical and the effective
cases show up. As we will see later, either the Hamiltonian or the
canonical algebra of the interior is changed and this leads to modified
equations of motion, which consequently results in modified expansion
scalar and its rate of change. 

We can now compute the Raychaudhuri equation \eqref{eq:RE-timelike}
either by finding the Ricci tensor components and replacing them in
the last term of \eqref{eq:RE-timelike}, or simply by using the chain
rule $\frac{d\theta}{d\tau}=\frac{d\theta}{dt}\frac{dt}{d\tau}=\frac{1}{N}\frac{d\theta}{dt}$.
The result is
\begin{equation}
	\frac{d\theta}{d\tau}=-\frac{\dot{N}}{N^{3}}\frac{\dot{X}}{X}+\frac{1}{N^{2}}\frac{\ddot{X}}{X}-\frac{1}{N^{2}}\left(\frac{\dot{X}}{X}\right)^{2}-2\frac{\dot{N}}{N^{3}}\frac{\dot{Y}}{Y}+\frac{2}{N^{2}}\frac{\ddot{Y}}{Y}-\frac{2}{N^{2}}\left(\frac{\dot{Y}}{Y}\right)^{2}\label{eq:RE-timlike-gen}
\end{equation}

\subsection{Null case}

In this case we choose a congruence of radial null geodesics and due
to the null property of their tangent vector $k^{a}$, we obtain
\begin{equation}
	k^{\mu}=\left(k^{0},-\frac{Nk^{0}}{X},0,0\right).
\end{equation}
A simplifying choice for $k^{0}$ is thus $k^{0}=\frac{1}{N}$ which
results in
\begin{equation}
	k^{\mu}=\left(\frac{1}{N},-\frac{1}{X},0,0\right).
\end{equation}
The auxiliary radial null vector field $l^{a}$ has two nonvanishing
components that can be fixed by using the null property of $l^{a}$
and the condition \eqref{eq:k-l-inner}. This way we obtain
\begin{equation}
	l^{\mu}=\left(\frac{1}{2N},\frac{1}{2X},0,0\right)
\end{equation}
Using these vectors and the spacetime metric, we can find the transverse
metric \eqref{eq:transverse-h-null} as
\begin{equation}
	h_{\mu\nu}=\begin{pmatrix}0 & 0 & 0 & 0\\
		0 & 0 & 0 & 0\\
		0 & 0 & Y^{2} & 0\\
		0 & 0 & 0 & Y^{2}\sin^{2}\left(\theta\right)
	\end{pmatrix}\label{eq:transverse-h-null-2}
\end{equation}
which is a two dimensional metric as it should be. Next, we can compute
$B_{ab}$ as in \eqref{eq:B-k} and then find $\tilde{B}_{ab}$ using
$B_{ab}$ and $k^{a},\,l^{a}$ above as
\begin{equation}
	\tilde{B}_{\mu\nu}=\begin{pmatrix}0 & 0 & 0 & 0\\
		0 & 0 & 0 & 0\\
		0 & 0 & \frac{Y\dot{Y}}{N} & 0\\
		0 & 0 & 0 & \frac{Y\dot{Y}}{N}\sin^{2}\left(\theta\right)
	\end{pmatrix}.
\end{equation}
As mentioned before, we can use these data to compute the expansion
scalar and shear and vorticity parameters as
\begin{align}
	\tilde{\theta}= & 2\frac{\dot{Y}}{NY},\label{eq:theta-null-gen}\\
	\tilde{\sigma}^{2}= & 0,\\
	\tilde{\omega}_{ab}= & 0.\label{eq:omega2tilde-null}
\end{align}
While quantities are simpler compared to the timelike case, we still
need the equations of motion in order to be able to compute the expansion.
The Raychaudhuri equation can be computed as before by using the Ricci
tensor, as
\begin{equation}
	\frac{d\theta}{d\lambda}=-\frac{2\dot{N}}{N^{3}}\frac{\dot{Y}}{Y}-\frac{2}{N^{2}}\frac{\dot{X}}{X}\frac{\dot{Y}}{Y}+\frac{2}{N^{2}}\frac{\ddot{Y}}{Y}-\frac{2}{N^{2}}\left(\frac{\dot{Y}}{Y}\right)^{2}.\label{eq:RE-null-gen}
\end{equation}

\section{Classical Schwarzschild interior\label{sec:Classical-Schwarzschild-interior}}

\subsection{Metric and classical Hamiltonian}

Before considering the quantum effects, let us first analyze the classical
interior in the light of the expansion scalar and the Raychaudhuri
equation. For this we need the metric of the interior and the classical
Hamiltonian. Since crossing the event horizon of the Schwarzschild
black hole results in change of causal nature (spacelike/timelike)
of $r,\,t$, the metric of the interior can be obtained by switching
$t\leftrightarrow r$ of the usual Schwarzschild metric as 
\begin{equation}
	ds^{2}=-\left(\frac{2GM}{t}-1\right)^{-1}dt^{2}+\left(\frac{2GM}{t}-1\right)dr^{2}+t^{2}\left(d\theta^{2}+\sin^{2}\theta d\phi^{2}\right).\label{eq:sch-inter}
\end{equation}
Here and throughout the paper, $t$ is the Schwarzschild time coordinate
(in the exterior) which has a range $t\in(0,2GM)$ in the interior.
Such a metric is a special case of a Kantowski-Sachs cosmological
spacetime that is given by the metric \citep{Collins:1977fg} 
\begin{align}
	ds_{KS}^{2}= & -N(T)^{2}dT^{2}+g_{xx}(T)dx^{2}+g_{\theta\theta}(T)d\theta^{2}+g_{\phi\phi}(T)d\phi^{2}\nonumber \\
	= & -d\tau^{2}+g_{xx}(\tau)dx^{2}+g_{\Omega\Omega}(\tau)d\Omega^{2}.\label{eq:K-S-gener}
\end{align}
Note that $x$ here is not necessarily the radius $r$ of the 2-spheres
with area $A=4\pi r^{2}$, but it can be chosen to be. Here $N(T)$
is the lapse function corresponding to a generic time, and $\tau$
is the proper time. The metric \eqref{eq:K-S-gener} represents a
cosmology with spatial homogeneous but anisotropic foliations.

To canonically analyze the model, one decomposes the spacetime into
space and time by foliating spacetime into spatial hypersurfaces with
constant coordinate time using the ADM method. This induces a spatial
metric $q_{ab}$ on the hypersurfaces. The classical Hamiltonian we
will be working with is the one written in terms of Ashtekar--Barbero
connection $A_{a}^{i}$, and its conjugate the densitized triad $\tilde{E}_{i}^{a}$.
The Ashtekar--Barbero connection
\begin{equation}
	A_{a}^{i}=\Gamma_{a}^{i}+\gamma K_{a}^{i}
\end{equation}
is an $su(2)$ connection with $i$ being an $su(2)$ index and $a$
an spatial index. It is the sum of two terms. The hodge dual of the
spin connection $\omega_{a}{}^{ij}$ denoted by $\Gamma_{a}^{i}=\frac{1}{2}\epsilon^{i}{}_{ij}\omega_{a}{}^{ij}$
where $\omega_{a}{}^{ij}$ associated to the symmetry under the Lorentz
transformations, and the extrinsic curvature $K_{a}^{i}\coloneqq\omega_{a}{}^{0i}$.
The parameter $\gamma$ is called the Barbero--Immirzi parameter
which is a free parameter of the theory, and $\epsilon_{ijk}$ is
the totally antisymmetric Levi-Civita symbol. The densitized triad
is related to the spatial metric $q_{ab}$ via 
\begin{equation}
	qq^{ab}=\delta^{ij}\tilde{E}_{i}^{a}\tilde{E}_{j}^{b}\label{eq:qq-EE}
\end{equation}
with $q=\det\left(q_{ab}\right)$. The full gravitational Hamiltonian
constrain in terms of Ashtekar--Barbero connection and densitized
triad is
\begin{equation}
	H_{\textrm{full}}=\frac{1}{8\pi G}\int d^{3}x\frac{N}{\sqrt{\det|\tilde{E}|}}\left\{ \epsilon_{i}^{jk}F_{ab}^{i}\tilde{E}_{j}^{a}\tilde{E}_{k}^{b}-2\left(1+\gamma^{2}\right)K_{[a}{}^{i}K_{b]}^{j}\tilde{E}_{i}^{a}\tilde{E}_{j}^{b}\right\} ,\label{eq:Full-H-gr-class}
\end{equation}
Here, $F=dA+A\wedge A$ is the curvature of the Ashtekar--Barbero
connection and $N$ is the lapse function.

To obtain the classical Hamiltonian of the model, we take the above
Hamiltonian constraint and reduce it by replacing the canonical variables
with the ones adapted to the model,
\begin{align}
	A_{a}^{i}\tau_{i}dx^{a}= & \frac{c}{L_{0}}\tau_{3}dx+b\tau_{2}d\theta-b\tau_{1}\sin\theta d\phi+\tau_{3}\cos\theta d\phi,\label{eq:A-AB}\\
	\tilde{E}_{i}^{a}\tau_{i}\partial_{a}= & p_{c}\tau_{3}\sin\theta\partial_{x}+\frac{p_{b}}{L_{0}}\tau_{2}\sin\theta\partial_{\theta}-\frac{p_{b}}{L_{0}}\tau_{1}\partial_{\phi}.\label{eq:E-AB}
\end{align}
Here $b$, $c$, $p_{b}$ and $p_{c}$ are functions that only depend
on time, and $\tau_{i}=-i\sigma_{i}/2$ are a $su(2)$ basis satisfying
$\left[\tau_{i},\tau_{j}\right]=\epsilon_{ij}{}^{k}\tau_{k}$, with
$\sigma_{i}$ being the Pauli matrices. Substituting these into the
full Hamiltonian of gravity written in Ashtekar connection variables,
one obtains the symmetry reduced Hamiltonian constraint adapted to
this model as \citep{Ashtekar:2005qt} 
\begin{equation}
	H=-\frac{N\mathrm{sgn}(p_{c})}{2G\gamma^{2}}\left[2bc\sqrt{|p_{c}|}+\left(b^{2}+\gamma^{2}\right)\frac{p_{b}}{\sqrt{|p_{c}|}}\right],\label{eq:H-class-N}
\end{equation}
while the diffeomorphism constraint vanishes identically due to homogenous
nature of the model. This classical Hamiltonian is not different from
other classical Hamiltonian since we have only changed the variables
from metric to connection ones. The real difference comes about once
we write \eqref{eq:Full-H-gr-class} in terms of holonomies instead
of connection components.

Since the spatial hypersurfaces have a topology $\mathbb{R}\times\mathbb{S}^{2}$,
the symplectic 2-form is 
\begin{equation}
	\Omega=\frac{1}{8\pi G\gamma}\int_{\mathbb{R}\times\mathbb{S}^{2}}d^{3}x\,dA_{a}^{i}(\mathbf{x})\wedge d\tilde{E}_{i}^{a}(\mathbf{y}).
\end{equation}
However, the part of the integral over $\mathbb{R}$ diverges and
we will not be able to obtain a kinematical structure, i.e., a Poisson
bracket. To remedy this and since the model is homogeneous, one can
restrict the range of integration in $\mathbb{R}$ to $\mathcal{I}=[0,L_{0}]$
and later take the limit $L_{0}\to\infty$. This we symplectic 2-form
becomes
\begin{align}
	\Omega= & \frac{1}{8\pi G\gamma}\int_{\mathcal{I}\times\mathbb{S}^{2}}d^{3}x\,dA_{a}^{i}(\mathbf{x})\wedge d\tilde{E}_{i}^{a}(\mathbf{y})\nonumber \\
	= & \frac{1}{2G\gamma}\left(dc\wedge dp_{c}+2db\wedge dp_{b}\right),
\end{align}
and consequently the fundamental Poisson brackets are
\begin{align}
	\{c,p_{c}\}= & 2G\gamma, & \{b,p_{b}\}= & G\gamma.\label{eq:classic-PBs-bc}
\end{align}
Using \eqref{eq:K-S-gener}, \eqref{eq:qq-EE}, and \eqref{eq:E-AB},
one obtains
\begin{align}
	g_{xx}\left(T\right)= & \frac{p_{b}\left(T\right)^{2}}{L_{0}^{2}p_{c}\left(T\right)},\label{eq:grrT}\\
	g_{\theta\theta}\left(T\right)= & \frac{g_{\phi\phi}\left(T\right)}{\sin^{2}\left(\theta\right)}=g_{\Omega\Omega}\left(T\right)=p_{c}\left(T\right).\label{eq:gththT}
\end{align}
These results correspond to a generic lapse function associated to
a generic time coordinate $T$. If in the above we choose the time
and the lapse function to be the Schwarzschild time $t$ and its lapse
respectively, and then compare the results with \eqref{eq:sch-inter},
we obtain
\begin{align}
	N\left(t\right)= & \left(\frac{2GM}{t}-1\right)^{-\frac{1}{2}},\label{eq:Sch-corresp-1}\\
	g_{xx}\left(t\right)= & \frac{p_{b}\left(t\right)^{2}}{L_{0}^{2}p_{c}\left(t\right)}=\left(\frac{2GM}{t}-1\right),\label{eq:Sch-corresp-2}\\
	g_{\theta\theta}\left(T\right)= & \frac{g_{\phi\phi}\left(T\right)}{\sin^{2}\left(\theta\right)}=g_{\Omega\Omega}\left(T\right)=p_{c}\left(t\right)=t^{2}.\label{eq:Sch-corresp-3}
\end{align}
This shows that 
\begin{align}
	p_{b}= & 0, & p_{c}= & 4G^{2}M^{2}, &  & \textrm{on the horizon\,}t=2GM,\label{eq:t-horiz}\\
	p_{b}\to & 0, & p_{c}\to & 0, &  & \textrm{at the singularity\,}t\to0.\label{eq:t-singular}
\end{align}

\subsection{Dynamics, expansion scalar and Raychaudhuri equation}

\subsubsection{Generic $\theta$ and $\frac{d\theta}{d\tau}$}

Comparing the metric \eqref{eq:gen-metric} with \eqref{eq:grrT}
and \eqref{eq:gththT}, and also using \eqref{eq:lapsNT}, we notice
\begin{align}
	X^{2}= & \frac{p_{b}\left(T\right)^{2}}{L_{0}^{2}p_{c}\left(T\right)},\label{eq:X-in-pbpc}\\
	Y^{2}= & p_{c}\left(T\right).\label{eq:Y-in-pbpc}
\end{align}
Replacing \eqref{eq:X-in-pbpc}--\eqref{eq:Y-in-pbpc} in the timelike
expansion \eqref{eq:theta-timlike-gen} yields
\begin{equation}
	\theta=\pm\left(\frac{\dot{p}_{b}}{Np_{b}}+\frac{\dot{p}_{c}}{2Np_{c}}\right).\label{eq:expansion-pbpc-gen-timelike}
\end{equation}
Notice that the above results are generic for any lapse and its associated
time and also valid in both classical and effective regimes. The difference
between these two regimes comes later due to the different equations
of motion which we will replace in the above expansion formula. For
the null case, we again replace \eqref{eq:X-in-pbpc} and \eqref{eq:Y-in-pbpc}
into \eqref{eq:theta-null-gen} to obtain
\begin{equation}
	\theta=\pm\frac{\dot{p}_{c}}{Np_{c}}.\label{eq:expansion-pbpc-gen-null}
\end{equation}
Using the above expressions for $X,\,Y,\,N$ in the timelike Raychaudhuri
equation \eqref{eq:RE-timlike-gen}, we obtain
\begin{equation}
	\frac{d\theta}{d\tau}=\frac{1}{N^{2}}\left(-\frac{\dot{N}\dot{p}_{b}}{Np_{b}}-\frac{\dot{N}\dot{p}_{c}}{2Np_{c}}+\frac{\ddot{p}_{b}}{p_{b}}-\frac{\dot{p}_{b}^{2}}{p_{b}^{2}}+\frac{\ddot{p}_{c}}{2p_{c}}-\frac{\dot{p}_{c}^{2}}{2p_{c}^{2}}\right).\label{eq:RE-pbpc-gen-timelike}
\end{equation}
The same method for the null Raychaudhuri equation \eqref{eq:RE-null-gen}
yields
\begin{equation}
	\frac{d\theta}{d\lambda}=\frac{1}{N^{2}}\left(-\frac{\dot{N}\dot{p}_{c}}{Np_{c}}-\frac{\dot{p}_{b}\dot{p}_{c}}{p_{b}p_{c}}+\frac{\ddot{p}_{c}}{p_{c}}-\frac{\dot{p}_{c}^{2}}{2p_{c}^{2}}\right).\label{eq:RE-pbpc-gen-null}
\end{equation}
These last two expressions are also generic results and are valid
for any time, lapse, and in both classical and effective regimes.

Notice that the fiducial parameter $L_{0}$ is not explicitly present
neither in $\theta$ nor in $\frac{d\theta}{d\tau}$ above. Of course,
it is hidden in the classical solutions of $p_{b}$ and $c$ (see
below), but wherever we have a term such as $\frac{\dot{p}_{b}}{p_{b}}$
or $\frac{\ddot{p}_{b}}{p_{b}}$, etc., $L_{0}$ will be canceled
out. Hence the above physical expressions are independent of $L_{0}$
as they should be.

\subsubsection{Classical dynamics}

In order to obtain the explicit expressions for $\theta$ and $\frac{d\theta}{d\tau}$
from above relations, we need the equations of motion and their solutions.
To this end we choose a lapse function
\begin{equation}
	N\left(T\right)=\frac{\gamma\,\mathrm{sgn}(p_{c})\sqrt{|p_{c}\left(T\right)|}}{b\left(T\right)}.\label{eq:lapsNT}
\end{equation}
The advantage of this lapse function is that the equations of motion
of $c,\,p_{c}$ decouple from those of $b,\,p_{b}$ and it makes it
possible to solve them. Replacing this lapse function into \eqref{eq:H-class-N}
yields
\begin{equation}
	H=-\frac{1}{2G\gamma}\left[\left(b^{2}+\gamma^{2}\right)\frac{p_{b}}{b}+2cp_{c}\right].\label{eq:H-class-1}
\end{equation}
Using this Hamiltonian together with the Poisson brackets \eqref{eq:classic-PBs-bc},
we can obtain the classical equations of motion 
\begin{align}
	\frac{db}{dT}= & \left\{ b,H\right\} =-\frac{1}{2}\left(b+\frac{\gamma^{2}}{b}\right),\label{eq:EoM-diff-b}\\
	\frac{dp_{b}}{dT}= & \left\{ p_{b},H\right\} =\frac{p_{b}}{2}\left(1-\frac{\gamma^{2}}{b^{2}}\right).\label{eq:EoM-diff-pb}\\
	\frac{dc}{dT}= & \left\{ c,H\right\} =-2c,\label{eq:EoM-diff-c}\\
	\frac{dp_{c}}{dT}= & \left\{ p_{c},H\right\} =2p_{c}.\label{eq:EoM-diff-pc}
\end{align}
These equations should be supplemented by the weakly vanishing ($\approx0$)
of the Hamiltonian constraint \eqref{eq:H-class-1}, 
\begin{equation}
	\left(b^{2}+\gamma^{2}\right)\frac{p_{b}}{b}+2cp_{c}\approx0.\label{eq:weak-van}
\end{equation}
This system can be solved to yield the solutions in generic time $T$.
In order to write the solutions in Schwarzschild time $t$, one compares
the form of $p_{c}(T)$ with its Schwarzschild couterpart $p_{c}=t^{2}$
and this reveals that to go from $T$ to $t$, we should make a transformation
of the form $T=\ln\left(t\right)$. Doing that we obtain
\begin{align}
	b\left(t\right)= & \pm\gamma\sqrt{\frac{2GM}{t}-1},\label{eq:sol-cls-b}\\
	p_{b}\left(t\right)= & L_{0}t\sqrt{\frac{2GM}{t}-1},\label{eq:sol-cls-pb}\\
	c\left(t\right)= & \mp\frac{\gamma GML_{0}}{t^{2}},\label{eq:sol-cls-c}\\
	p_{c}\left(t\right)= & t^{2},\label{eq:sol-cls-pc}
\end{align}
where the constants of integration in these solutions are fixed using
\eqref{eq:Sch-corresp-2}, \eqref{eq:Sch-corresp-3}, and \eqref{eq:t-horiz}. 

\subsubsection{Classical $\theta$ and $\frac{d\theta}{d\tau}$: timelike congruence}

To obtain the expressions for expansion and Raychaudhuri equation
for a timelike congruence, we replace \eqref{eq:EoM-diff-b}--\eqref{eq:EoM-diff-pc}
and \eqref{eq:lapsNT} in \eqref{eq:expansion-pbpc-gen-timelike}
to
\begin{equation}
	\theta=\pm\frac{1}{2\sqrt{p_{c}}}\left(\frac{3b}{\gamma}-\frac{\gamma}{b}\right)=\pm\frac{-2t+3GM}{t^{2}\sqrt{\frac{2GM}{t}-1}}.\label{eq:theta-class}
\end{equation}
where in the last step to get an explicit expression in terms of the
Schwarzschild time $t$ we have made use of \eqref{eq:sol-cls-b}--\eqref{eq:sol-cls-pc}.
The $\pm$ corresponds to ingoing vs outgoing geodesics. Since in
the interior $t\leq2GM$, from $\frac{3}{2}GM\leq t\leq2GM$, the
ingoing (negative branch) of the expansion is positive while for $t<\frac{3}{2}GM$,
this branch becomes negative and continues to become more negative
until at $t\to0$ it goes to $\theta\to-\infty$. 

\begin{figure}
	\begin{centering}
		\includegraphics[scale=0.8]{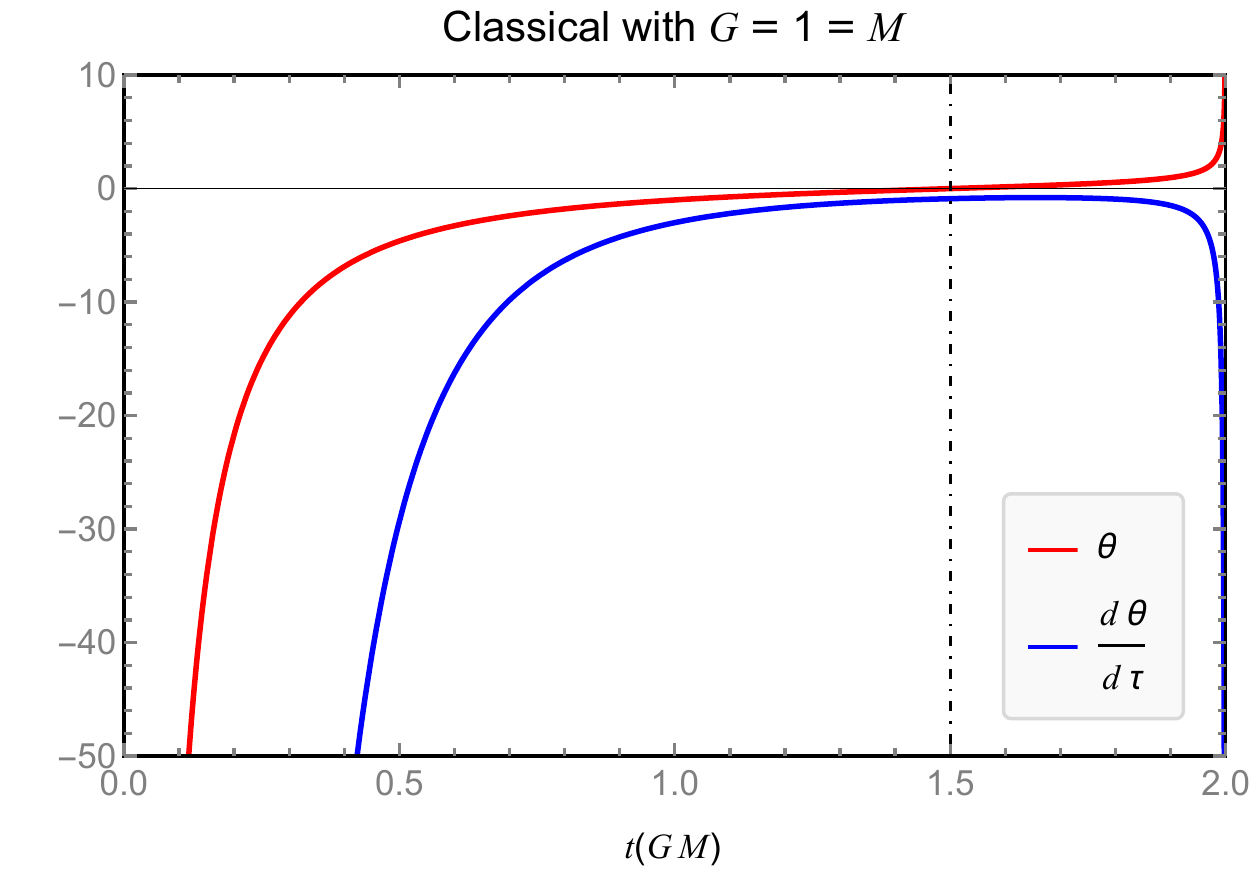}
		\par\end{centering}
	\caption{Classical timelike $\theta$ and $\frac{d\theta}{d\tau}$ diverge
		as we approach $t\to0$. The divergence at the horizon is due to the
		choice of Schwarzschild coordinate system. The dashed line is where
		$\theta$ changes sign. \label{fig:theta-RE-t-class}}
\end{figure}

We can use \eqref{eq:lapsNT}, \eqref{eq:EoM-diff-b}--\eqref{eq:EoM-diff-pc},
and \eqref{eq:sol-cls-b}--\eqref{eq:sol-cls-pc} in the same way
in \eqref{eq:RE-pbpc-gen-timelike} to obtain
\begin{equation}
	\frac{d\theta}{d\tau}=-\frac{1}{2p_{c}}\left(1+\frac{9b^{2}}{2\gamma^{2}}+\frac{\gamma^{2}}{2b^{2}}\right)=\frac{-2t^{2}+8GMt-9G^{2}M^{2}}{\left(2GM-t\right)t^{3}}.\label{eq:Class-RE}
\end{equation}
Notice that the above expression in terms of $b,\,p_{c}$ contains
three terms that are all negative (since $p_{c}$ is always positive
as is seen from \eqref{eq:sol-cls-pc}). This guarantees that there
will be a caustic point at the region where classically we identify
as the singularity. The plots of expansion and Raychaudhuri equation
are in Schwrazschild time are presented in Fig. \ref{fig:theta-RE-t-class}.
It is clear from this plot that both of them diverge at the singularity
at $t\to0$.

\subsubsection{Classical $\theta$ and $\frac{d\theta}{d\tau}$: null congruence}

The expression \eqref{eq:theta-null-gen} for the null expansion is
actually simpler than its timelike counterpart. As in the previous
section, once we replace \eqref{eq:EoM-diff-b}--\eqref{eq:EoM-diff-pc}
and \eqref{eq:lapsNT} into \eqref{eq:theta-null-gen} we get
\begin{equation}
	\theta=\pm\frac{2b}{\gamma\sqrt{p_{c}}}=\pm\frac{2}{t}\sqrt{\frac{2GM}{t}-1}.\label{eq:theta-class-null}
\end{equation}
where once again in the last step we have used \eqref{eq:sol-cls-b}--\eqref{eq:sol-cls-pc}.
Here, as opposed to the timelike case \eqref{eq:theta-class}, $\theta$
remain negative everywhere in the interior where $t<2GM$ and there
are no roots to the expansion scalar. This makes sense since usually
the existence of roots of the expansion scalar points to the existence
of a horizon. Also the ingoing branch of the expansion scalar goes
to $\theta\to-\infty$ as $t\to0$.

\begin{figure}
	\begin{centering}
		\includegraphics[scale=0.8]{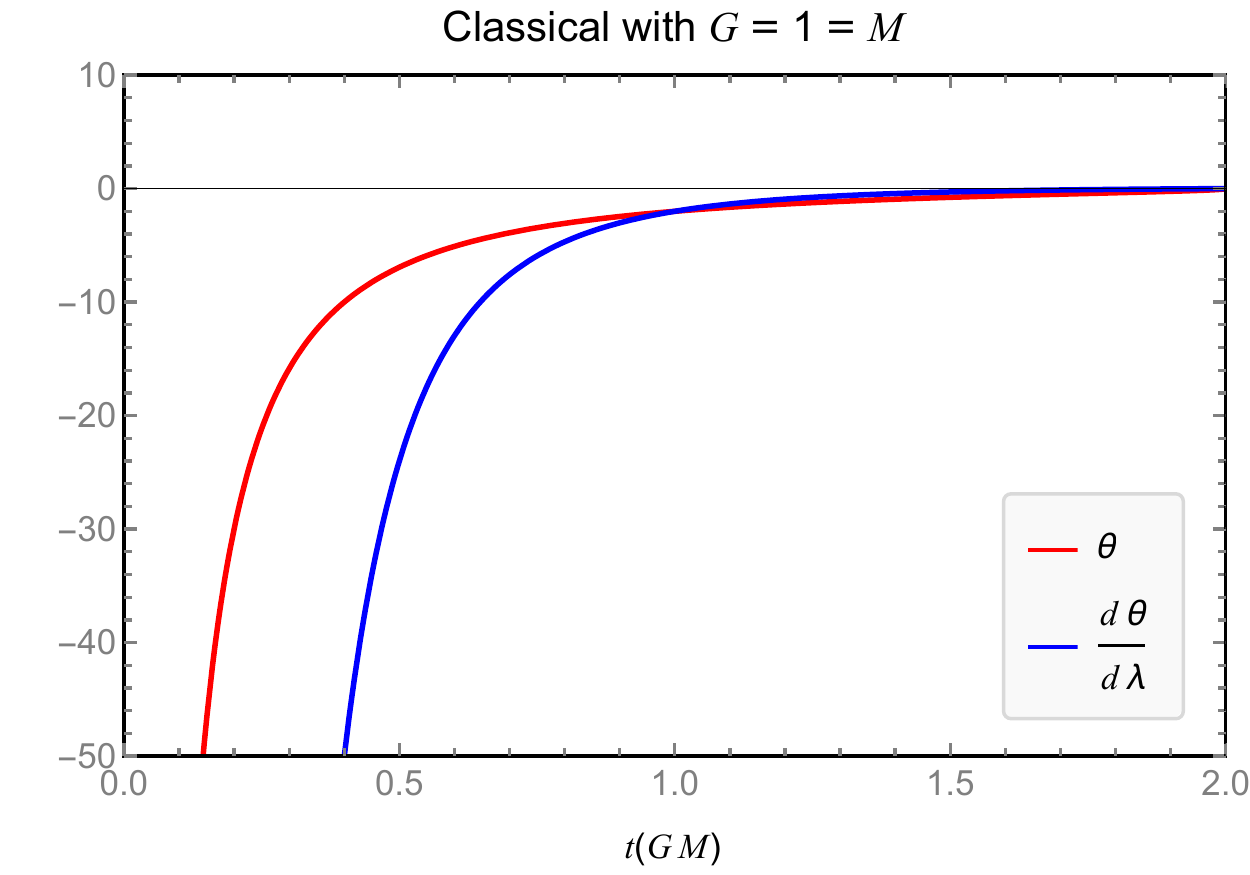}
		\par\end{centering}
	\caption{Classical null $\theta$ and $\frac{d\theta}{d\lambda}$ diverge as
		we approach $t\to0$. Notice that $\theta$ always remains negative
		and has no roots. \label{fig:theta-RE-t-null}}
\end{figure}

To obtain the form of the Raychaudhuri equation, we use \eqref{eq:lapsNT},
\eqref{eq:EoM-diff-b}--\eqref{eq:EoM-diff-pc}, and \eqref{eq:sol-cls-b}--\eqref{eq:sol-cls-pc}
in \eqref{eq:RE-pbpc-gen-null} to obtain
\begin{equation}
	\frac{d\theta}{d\lambda}=-\frac{2b^{2}}{\gamma^{2}p_{c}}=-\frac{2}{t^{2}}\left(\frac{2GM}{t}-1\right),\label{eq:Class-RE-null}
\end{equation}
which clearly is always negative in the interior. Since $\theta$
is negative at least at one point in the interior and $\tilde{\omega}_{ab}=0$,
the theorem we mentioned in Sec. \ref{sec:Raychaudhuri-equation}
guarantees the existence of caustic point(s) in the interior, which
from the above is seen to be at $t\to0$. This can also simply be
deduced by noting that both $\theta$ and $\frac{d\theta}{d\tau}$
are always negative in the interior and tend to $-\infty$ as $t\to0$.
This behavior can be seen in Fig. \ref{fig:theta-RE-t-null}.

\subsubsection{Classical Kretschmann scalar}

In our variables the Kretchammn scalar becomes
\begin{equation}
	K=\frac{12\left(b^{2}+\gamma^{2}\right)^{2}}{\gamma^{4}p_{c}^{2}}\label{eq:Kretsch-Classic}
\end{equation}
which in terms of the Schwarzschild time turns out to be
\begin{equation}
	K=\frac{48G^{2}M^{2}}{t^{6}}
\end{equation}
and unsurprisingly it diverges at $t\to0$ or equivalently at $p_{c}\to0$.
Notice that $p_{c}$ is the radius of the infalling 2-spheres as can
be seen from \eqref{eq:sol-cls-pc}.

\section{Effective Schwarzschild interior\label{sec:Effective-Schwarzschild-interior} }

The main idea in this section is to find the modified equations of
motion of the interior, and use them to compute the effective expansion
and Raychaudhuri equation for null and timelike cases. We consider
two models, one coming from loop quantum gravity (LQG) and the other
one from generalized uncertainty principle (GUP). 

\subsection{Loop quantum gravity\label{subsec:Loop-quantum-gravity}}

In LQG, the configuration variable is not the connection, but the
holonomy of the connection $h_{\xi}[A]$, .i.e., path-ordered exponential
of the connection $A_{a}^{i}$ along some curves $\xi$ in space.
The canonically conjugate momenta to this variable is the smeared
flux of the densitized triad over a two dimensional spatial surface.
As a consequence, to derive a quantum Hamiltonian, one goes back to
\eqref{eq:Full-H-gr-class} and writes the curvatures $F_{ab}^{i}$
in terms of holonomies instead of the connection. Once this expression
is derived classically, then one quantizes the Hamiltonian on a suitable
Hilbert space. On this Hilbert space, only the operators $\hat{h}_{\xi}[A]$
exists. There is no operator corresponding to $A$. As a consequence
the Hilbert space of LQG is unitarily inequivalent to the usual Schrodinger
representation. Another type of quantization which mimics LQG quantization
which is usually used is called polymer quantization. This quantiztion
introduces parameters into the theory called polymer scales that set
the minimal scale of the model. Close to this scale quantum effects
become important. In case of the present model such a polymer quantization
leads to polymer scales $\mu_{b},\,\mu_{c}$ associated with the radial
and angular minimum scales \citep{Ashtekar:2005qt,Corichi:2015xia,Chiou:2008nm,Chiou:2008eg}.

After applying the polymer quantization to the model and obtaining
the quantum Hamiltonian as mentioned above, one finds an effective
Hamiltonian by either using a path integral approach, or by acting
the quantum Hamiltonian on suitable states \citep{Ashtekar:2002sn,Corichi:2007tf,Morales-Tecotl:2016ijb,Tecotl:2015cya,Flores-Gonzalez:2013zuk,Morales-Tecotl:2016dma,Morales-Tecotl:2018ugi}.
These methods will lead to an effective Hamiltonian that can also
be heuristically obtained by replacing 
\begin{align}
	b\to & \frac{\sin\left(\mu_{b}b\right)}{\mu_{b}},\label{eq:b-to-sinb}\\
	c\to & \frac{\sin\left(\mu_{c}c\right)}{\mu_{c}}\label{eq:c-to-sinc}
\end{align}
in the classical Hamiltonian which yields an effective Hamiltonian
constraint, 
\begin{equation}
	H_{\textrm{eff}}^{(N)}=-\frac{N}{2G\gamma^{2}}\left[\left(\frac{\sin^{2}\left(\mu_{b}b\right)}{\mu_{b}^{2}}+\gamma^{2}\right)\frac{p_{b}}{\sqrt{p_{c}}}+2\frac{\sin\left(\mu_{b}b\right)}{\mu_{b}}\frac{\sin\left(\mu_{c}c\right)}{\mu_{c}}\sqrt{p_{c}}\right].\label{eq:H-eff-gen}
\end{equation}
In LQG, there exist two general schemes regarding these $\mu$ parameters.
In one, called the $\mu_{0}$ scheme, $\mu$ parameters are considered
to be constants \citep{Ashtekar:2005qt,Modesto:2005zm,Modesto:2008im,Campiglia:2007pr}.
Applying such a scheme to isotropic and Bianchi-I cosmological models,
however, has shown to lead to incorrect semiclassical limit. To remedy
this and other issues regarding the appearance of large quantum effects
at the horizon or dependence of physical quantities on fiducial variables,
new schemes referred to as the $\bar{\mu}$ scheme or ``improved
dynamics'' have been proposed in which $\mu$ parameters depend on
canonical variables \citep{Bohmer:2007wi,Chiou:2008nm,Chiou:2008eg,Joe:2014tca}.
This scheme is itself divided into various different ways of expressing
the dependence of $\mu$ parameters on canonical variables. In addition,
new $\mu_{0}$ schemes have also been put forward (e.g., Refs. \citep{Corichi:2015xia,Olmedo:2017lvt})
with the intent of resolving the aforementioned issues. In case of
the Schwarzschild interior due to lack of matter content, it is not
clear which scheme does not lead to the correct semiclassical limit.
Hence for completeness, in this paper, we will study the modifications
to the Raychaudhuri equation in the constant $\mu$ scheme, which
here we call the $\mathring{\mu}$ scheme, as well as in two of the
most common improved schemes, which we denote by $\bar{\mu}$ and
$\bar{\mu}^{\prime}$ schemes. These schemes were originally introduced in \citep{Ashtekar:2006wn,Ashtekar:2009vc}. In the $\mathring{\mu}$ scheme, the polymer parameter is taken to be a constant, while in the $\bar{\mu}$ and
$\bar{\mu}^{\prime}$ schemes, this parameter depends on the canonical momenta, but this dependence is difference for each of the last two schemes, as we will see in the following sections.

In order to be able to find the deviations from the classical behavior,
we need to use the same lapse as we did in the classical part. Using
\eqref{eq:b-to-sinb}, the lapse \eqref{eq:lapsNT} becomes (assuming
$p_{c}\geq0$)
\begin{equation}
	N=\frac{\gamma\mu_{b}\sqrt{p_{c}}}{\sin\left(\mu_{b}b\right)}.\label{eq:laps-eff}
\end{equation}
Using this in \eqref{eq:H-eff-gen} yields 
\begin{equation}
	H_{\textrm{eff}}=-\frac{1}{2\gamma G}\left[p_{b}\left[\frac{\sin\left(\mu_{b}b\right)}{\mu_{b}}+\gamma^{2}\frac{\mu_{b}}{\sin\left(\mu_{b}b\right)}\right]+2p_{c}\frac{\sin\left(\mu_{c}c\right)}{\mu_{c}}\right].\label{eq:H-eff-spec}
\end{equation}
The parameters $\mu_{b},\,\mu_{c}$ here are written in a generic
form meaning that they can be either $\mathring{\mu}$, $\bar{\mu}$
or $\bar{\mu}^{\prime}$ depending on the scheme we are considering.
Also note that both \eqref{eq:H-eff-gen} and \eqref{eq:H-eff-spec}
reduce to their classical counterparts \eqref{eq:H-class-N} and \eqref{eq:H-class-1}
respectively for $\mu_{b},\,\mu_{c}\to0$, as expected.

\subsubsection{$\mathring{\mu}$ scheme\label{subsec:mu0-scheme}}

\begin{figure}
	\subfloat[Solutions to the equations of motion in $\mathring{\mu}$ case as
	a function of the Schwarzschild time $t$.]{\begin{centering}
			\includegraphics[scale=0.51]{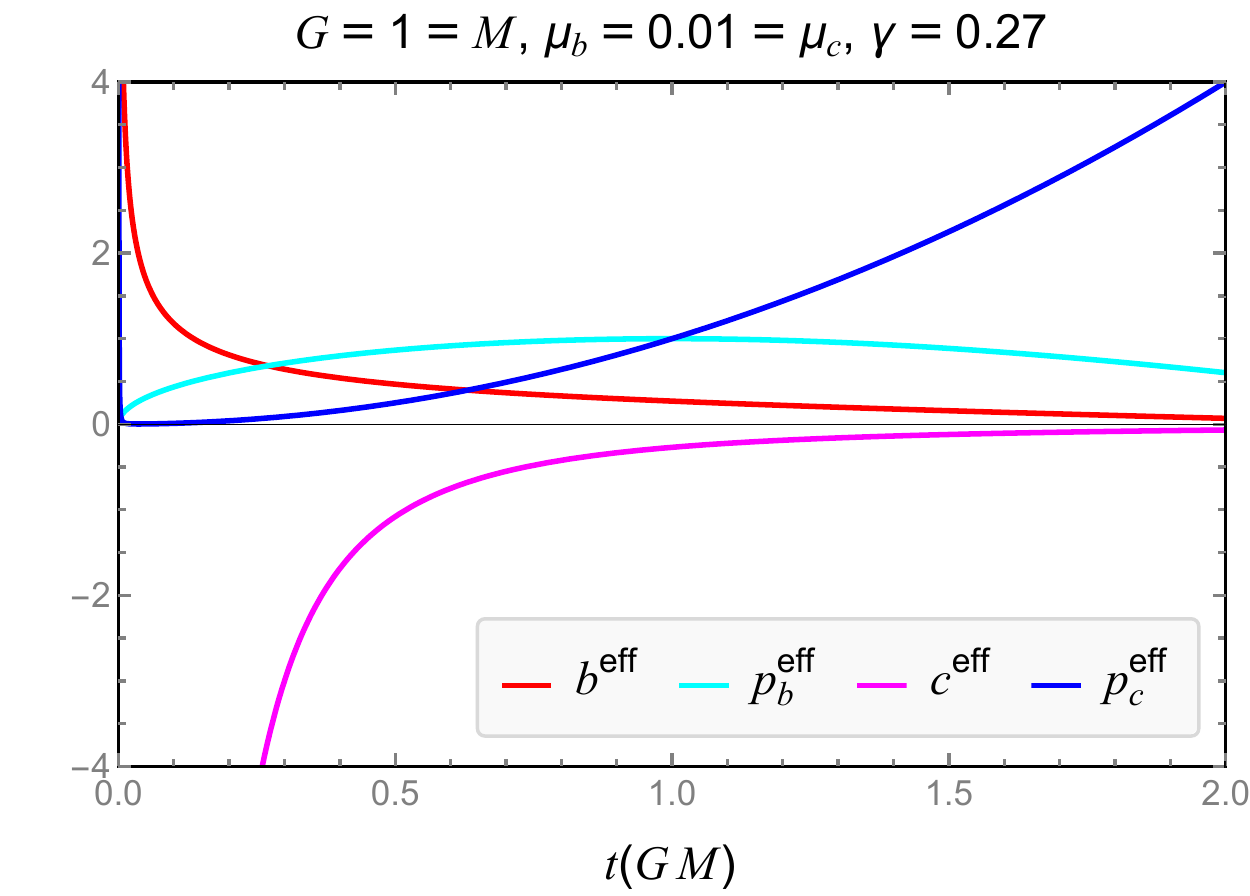}
			\par\end{centering}
	}\hfill{}\subfloat[Close up of the EoM of $p_c$ close to $t=0$. We can see that $p_{c}$ never
	vanishes.]{\begin{centering}
			\includegraphics[scale=0.51]{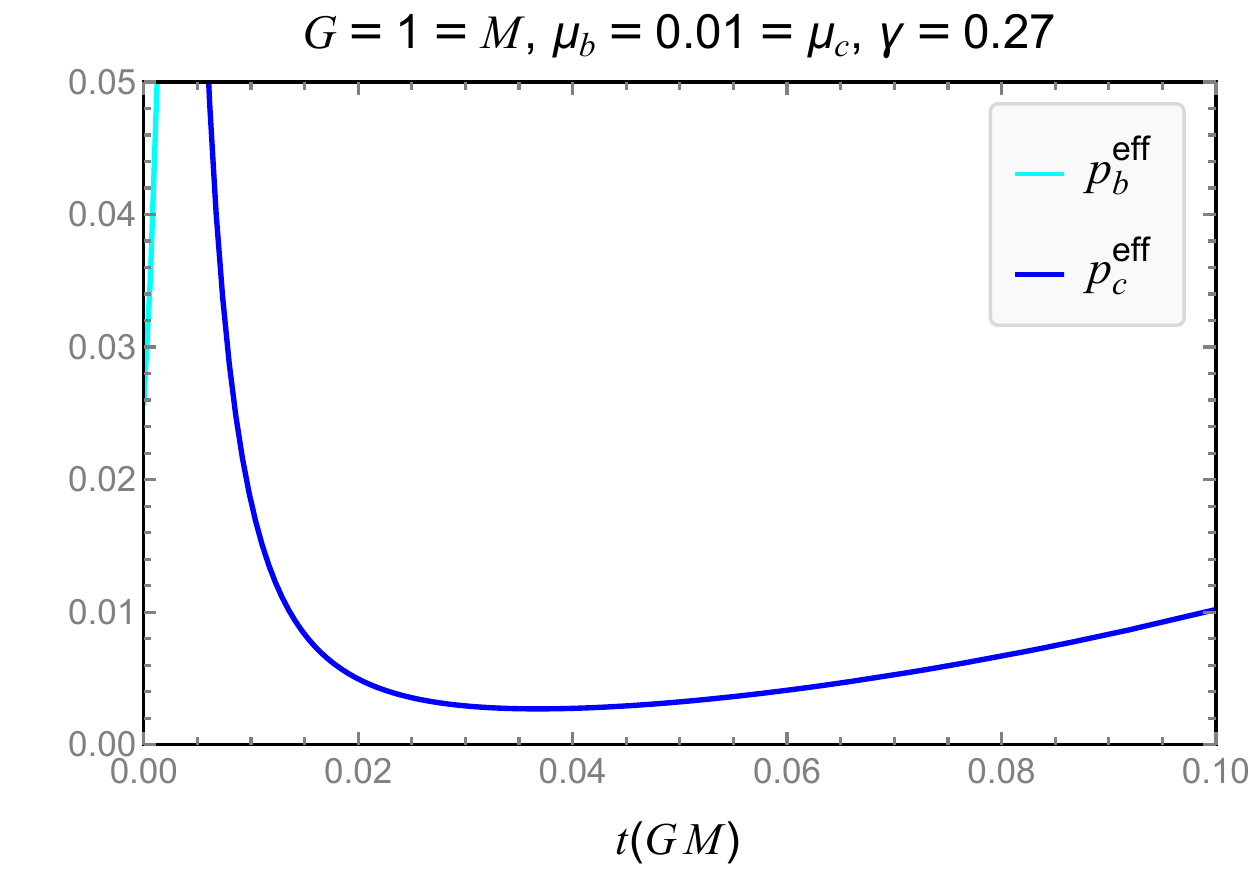}
			\par\end{centering}
	}
	
	\caption{Solutions of the EoM of the $\mathring{\mu}$ case \label{fig:Sol-LQG-mu0}}
\end{figure}

As mentioned before, in this scheme, one assumes that the polymer
or minimal scales $\mathring{\mu}_{b},\,\mathring{\mu}_{c}$ are constants.
The equations of motion corresponding to \eqref{eq:H-eff-spec} become
\begin{align}
	\frac{db}{dT}= & \left\{ b,H_{\textrm{eff}}\right\} =-\frac{1}{2}\left[\frac{\sin\left(\mathring{\mu}_{b}b\right)}{\mathring{\mu}_{b}}+\gamma^{2}\frac{\mathring{\mu}_{b}}{\sin\left(\mathring{\mu}_{b}b\right)}\right],\label{eq:b-diff-mu0}\\
	\frac{dp_{b}}{dT}= & \left\{ p_{b},H_{\textrm{eff}}\right\} =\frac{1}{2}p_{b}\cos\left(\mathring{\mu}_{b}b\right)\left[1-\gamma^{2}\frac{\mathring{\mu}_{b}^{2}}{\sin^{2}\left(\mathring{\mu}_{b}b\right)}\right],\label{eq:pb-diff-mu0}\\
	\frac{dc}{dT}= & \left\{ c,H_{\textrm{eff}}\right\} =-2\frac{\sin\left(\mathring{\mu}_{c}c\right)}{\mathring{\mu}_{c}},\label{eq:c-diff-mu0}\\
	\frac{dp_{c}}{dT}= & \left\{ p_{c},H_{\textrm{eff}}\right\} =2p_{c}\cos\left(\mathring{\mu}_{c}c\right).\label{eq:pc-diff-mu0}
\end{align}
Notice that the $\mathring{\mu}_{b}\to0$ and $\mathring{\mu}_{c}\to0$
limit of these equations corresponds to the classical equations of
motion. The solutions to these equations in terms of the Schwarzschild
time $t$ (after a transformation $T=\ln(t)$) and finding the integration
constants by matching the limit $\mathring{\mu}_{b},\,\mathring{\mu}_{c}\to0$
to classical solutions, are given by 
\begin{align}
	b(t)= & \frac{\cos^{-1}\left[\sqrt{1+\gamma^{2}\mathring{\mu}_{b}^{2}}\tanh\left(\sqrt{1+\gamma^{2}\mathring{\mu}_{b}^{2}}\ln\left[\frac{2\sqrt{\frac{t}{2GM}}}{\gamma\mathring{\mu}_{b}}\right]\right)\right]}{\mathring{\mu}_{b}},\label{eq:bt-eff-mu0}\\
	p_{b}(t)= & \frac{\gamma\mathring{\mu}_{b}L_{0}GM\left(\frac{\gamma^{2}\mathring{\mu}_{c}^{2}L_{0}^{2}G^{2}M^{2}}{4t^{2}}+t^{2}\right)\sqrt{1-\left(1+\gamma^{2}\mathring{\mu}_{b}^{2}\right)\tanh^{2}\left(\sqrt{\gamma^{2}\mathring{\mu}_{b}^{2}+1}\ln\left[\frac{2\sqrt{\frac{t}{2GM}}}{\gamma\mathring{\mu}_{b}}\right]\right)}}{t^{2}\sqrt{\frac{\gamma^{2}\mathring{\mu}_{c}^{2}L_{0}^{2}G^{2}M^{2}}{4t^{4}}+1}\left(\gamma^{2}\mathring{\mu}_{b}^{2}-\left(1+\gamma^{2}\mathring{\mu}_{b}^{2}\right)\tanh^{2}\left(\sqrt{1+\gamma^{2}\mathring{\mu}_{b}^{2}}\ln\left[\frac{2\sqrt{\frac{t}{2GM}}}{\gamma\mathring{\mu}_{b}}\right]\right)+1\right)},\label{eq:pbt-eff-mu0}\\
	c(t)= & -\frac{\tan^{-1}\left(\frac{\gamma\mathring{\mu}_{c}L_{0}GM}{2t^{2}}\right)}{\mathring{\mu}_{c}},\label{eq:ct-eff-mu0}\\
	p_{c}(t)= & \frac{\gamma^{2}\mathring{\mu}_{c}^{2}L_{0}^{2}G^{2}M^{2}}{4t^{2}}+t^{2}.\label{eq:pct-eff-mu0}
\end{align}
Since $p_{c}$ represents the radius of two spheres, it is interesting
to see that it never reaches zero. This is the first sign that the
singularity is resolved as we will see in the following. The plot
of the behavior of these solutions can be seen in Fig. \ref{fig:Sol-LQG-mu0}.

\begin{figure}
	\subfloat[Left: Classical vs timlike $\theta$ in the $\mathring{\mu}$ scheme
	as a function of the Schwarzschild time $t$. The effective expansion $\theta_{(\mathring{\mu})}^{\mathrm{TL}}$
	goes to zero as $t\to0$. Right: Close up of the left figure close
	to $t=0$.]{%
		\begin{minipage}[t]{0.45\textwidth}%
			\begin{center}
				\includegraphics[scale=0.50]{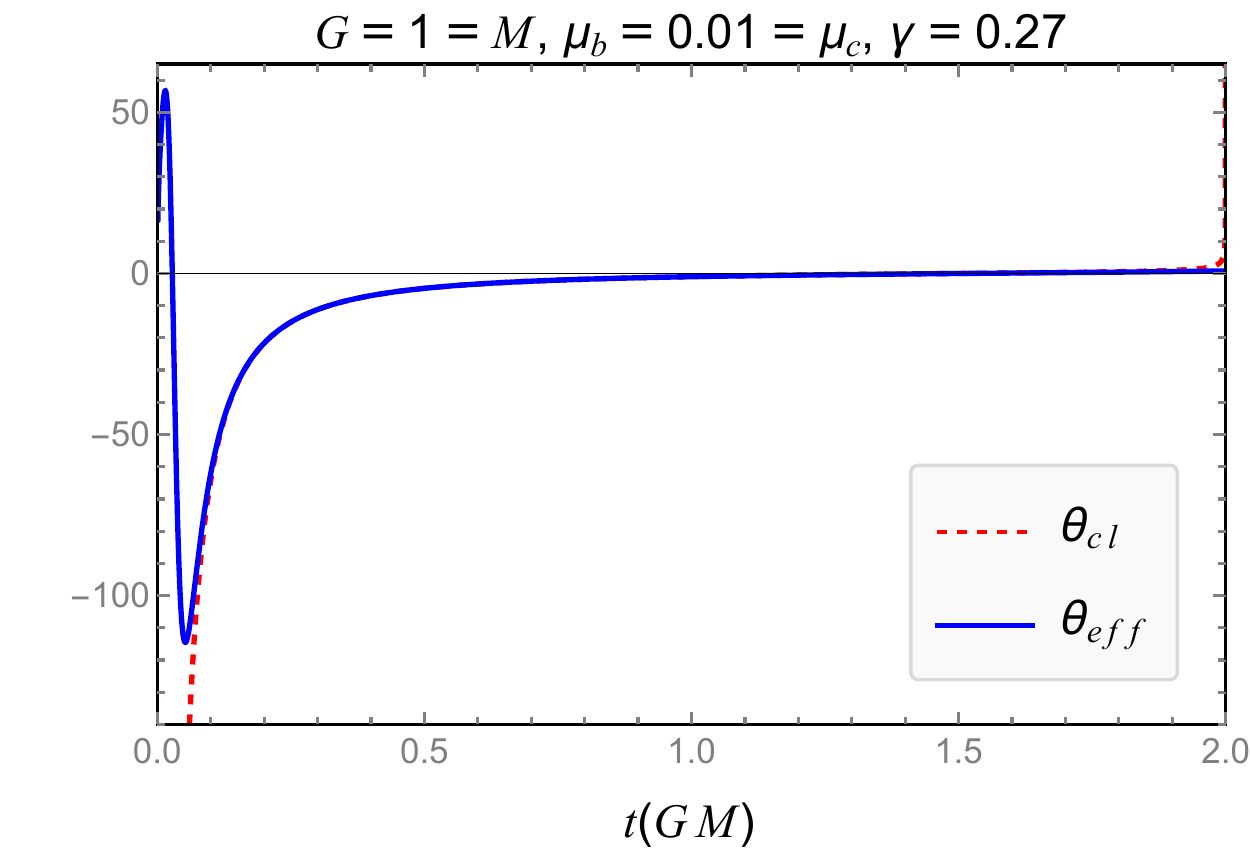}
				\par\end{center}%
		\end{minipage}\hfill{}%
		\begin{minipage}[t]{0.45\textwidth}%
			\begin{center}
				\includegraphics[scale=0.50]{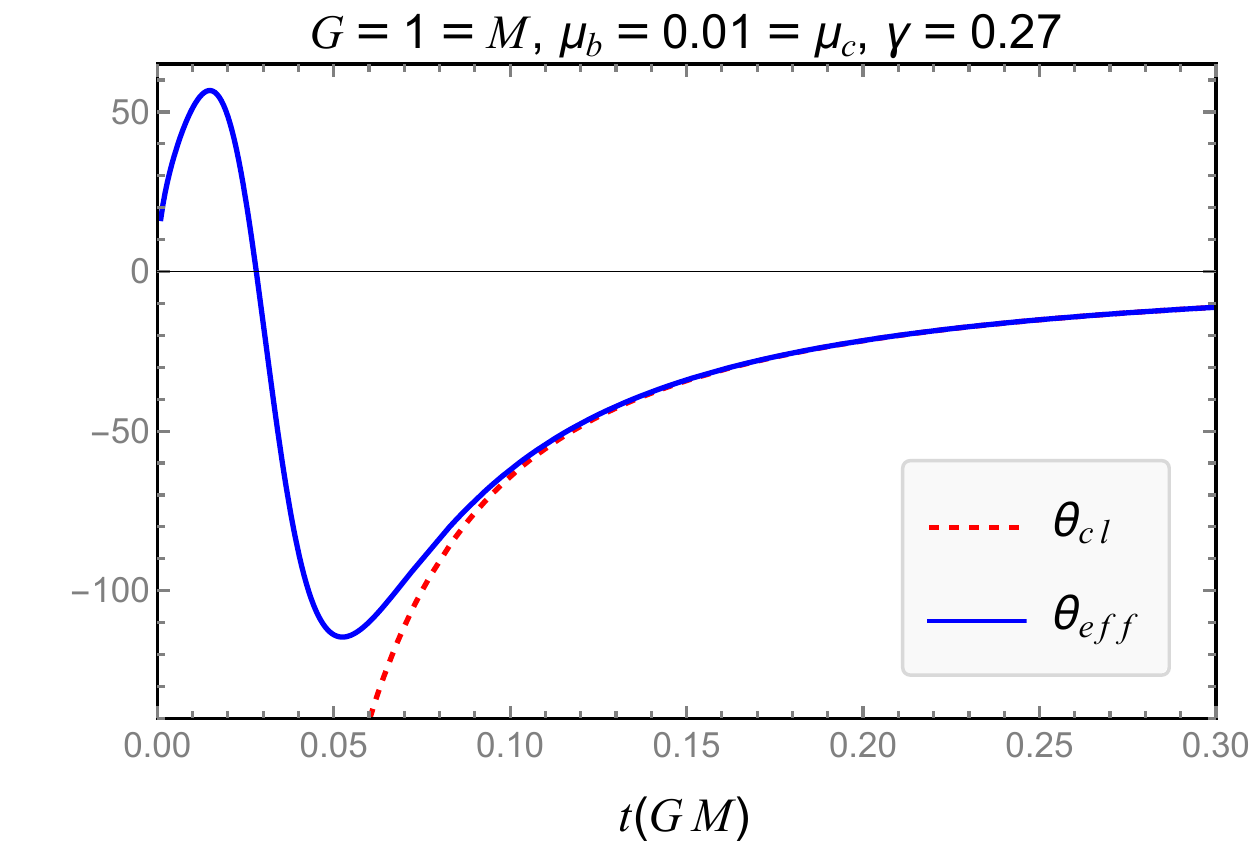}
				\par\end{center}%
		\end{minipage}
		
	}\hfill{}\subfloat[Left: Classical vs timlike $\frac{d\theta}{d\tau}$ in the $\mathring{\mu}$
	scheme as a function of the Schwarzschild time $t$. The effective
	$\frac{d\theta_{(\mathring{\mu})}^{\mathrm{TL}}}{d\tau}$ goes to
	zero as $t\to0$. Right: Close up of the left figure close
	to $t=0$.]{%
		\begin{minipage}[t]{0.45\textwidth}%
			\begin{center}
				\includegraphics[scale=0.50]{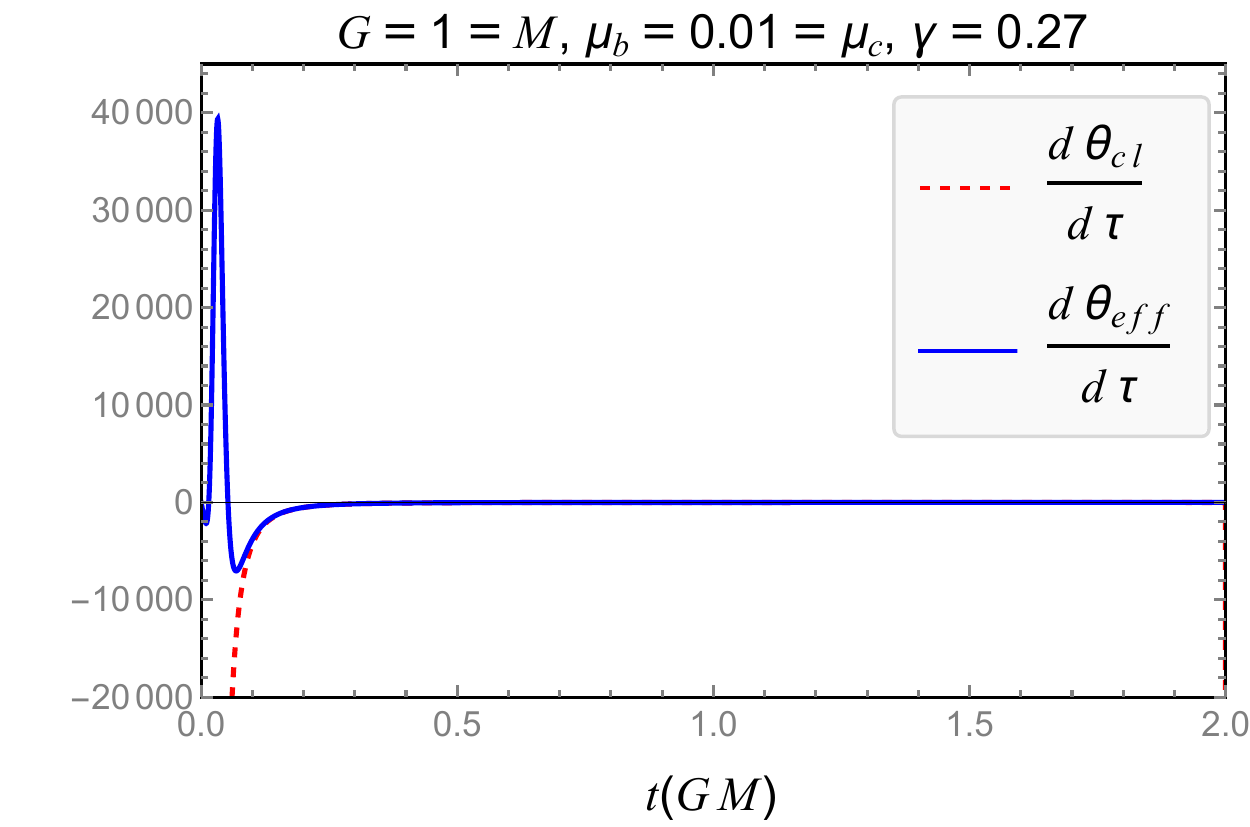}
				\par\end{center}%
		\end{minipage}\hfill{}%
		\begin{minipage}[t]{0.45\textwidth}%
			\begin{center}
				\includegraphics[scale=0.50]{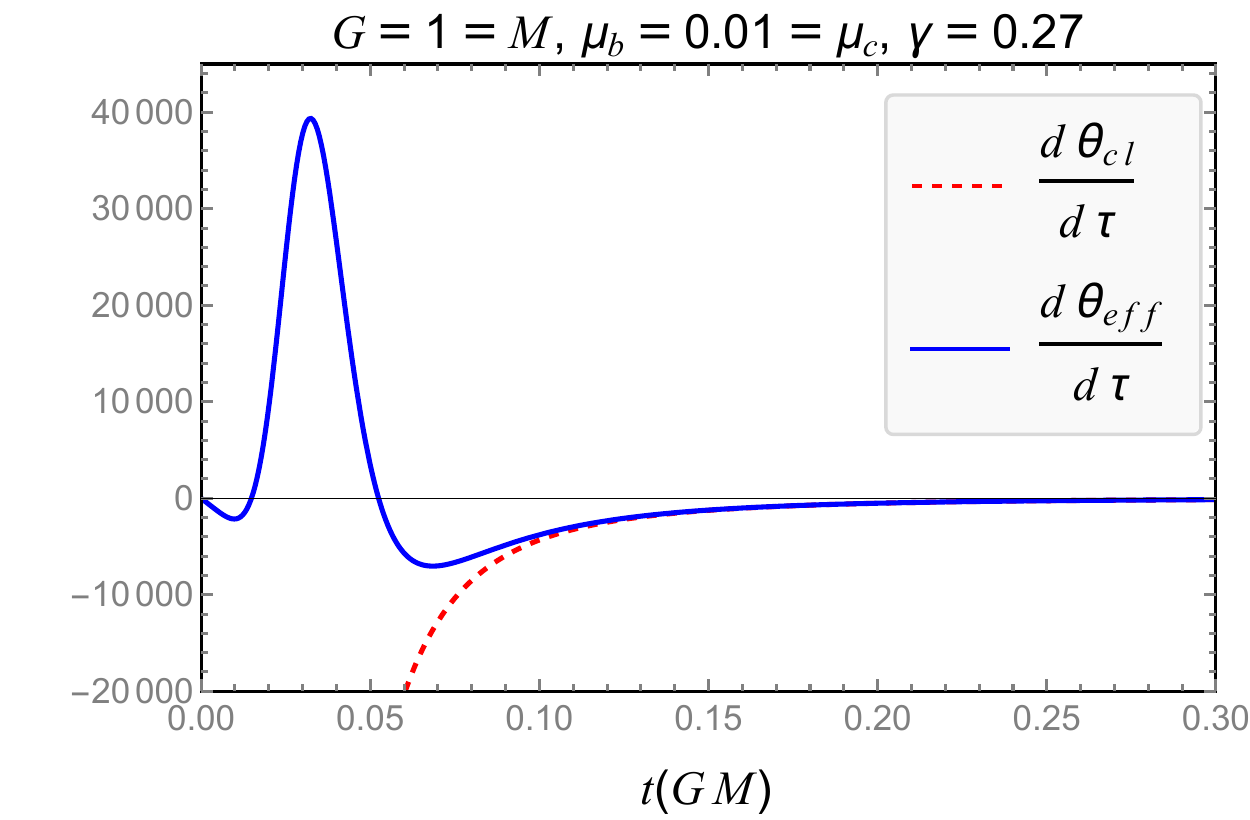}
				\par\end{center}%
		\end{minipage}
		
	}
	
	\caption{$\theta_{(\mathring{\mu})}^{\mathrm{TL}}$ and $\frac{d\theta_{(\mathring{\mu})}^{\mathrm{TL}}}{d\tau}$\label{fig:theta-RE-TL-mu0-vs-classic}}
\end{figure}

The timelike expansion \eqref{eq:expansion-pbpc-gen-timelike} in
this case becomes
\begin{equation}
	\theta_{(\mathring{\mu})}^{\mathrm{TL}}=\pm\frac{1}{\gamma\sqrt{p_{c}}}\left[\frac{\sin\left(\mathring{\mu}_{b}b\right)}{\mathring{\mu}_{b}}\cos\left(\mathring{\mu}_{c}c\right)-\frac{\gamma^{2}}{2}\frac{\mathring{\mu}_{b}}{\sin\left(\mathring{\mu}_{b}b\right)}\cos\left(\mathring{\mu}_{b}b\right)+\frac{\sin\left(2\mathring{\mu}_{b}b\right)}{2\mathring{\mu}_{b}}\right].\label{eq:theta-TL-pbpc-mu0}
\end{equation}
Up to the second order in $\mathring{\mu}_{b},\,\mathring{\mu}_{c}$,
the negative branch of this expression can be written as
\begin{equation}
	\theta_{(\mathring{\mu})}^{\mathrm{TL}}=-\frac{1}{2\sqrt{p_{c}}}\left(\frac{3b}{\gamma}-\frac{\gamma}{b}+b\gamma\left(\frac{1}{3}-\frac{b^{2}}{\gamma^{2}}\right)\mathring{\mu}_{b}^{2}-\frac{b}{\gamma}c^{2}\mathring{\mu}_{c}^{2}\right)+\mathcal{O}\left(\mathring{\mu}^{4}\right).
\end{equation}
The first two terms are the classical ones that contribute to a negative
expansion or focusing. The last term, which is an effective term,
is always positive and given the behavior of $b,\,c$ in this scheme
seen from \eqref{eq:bt-eff-mu0} and \eqref{eq:ct-eff-mu0}, it becomes
very large as $t\to0$. The third term is also an effective term and
becomes positive for $b^{2}>\frac{\gamma^{2}}{3}$, which is indeed
the case from the solution \eqref{eq:bt-eff-mu0}. These two effective
terms take over close to where the classical singularity used to be
and stop the congruence from infinitely focusing. In fact the full
nonperturbative plot \ref{fig:theta-RE-TL-mu0-vs-classic}, obtained
by replacing the solutions \eqref{eq:bt-eff-mu0}--\eqref{eq:pct-eff-mu0}
into \eqref{eq:theta-TL-pbpc-mu0}, reveals that the effective terms
perfectly cancel the classical focusing terms such that $\theta_{(\mathring{\mu})}^{\mathrm{TL}}$
becomes zero at $t=0$. 

The Raychaudhuri equation \eqref{eq:RE-pbpc-gen-timelike} in this
case also turns out to be
\begin{align}
	\frac{d\theta_{(\mathring{\mu})}^{\mathrm{TL}}}{d\tau}= & \frac{1}{4\gamma^{2}p_{c}}\left\{ -\gamma^{2}+\frac{5}{4\mathring{\mu}_{b}^{2}}+\frac{6\sin^{2}\left(\mathring{\mu}_{b}b\right)}{\mathring{\mu}_{b}^{2}}\left[\sin^{2}\left(\mathring{\mu}_{c}c\right)-\cos^{2}\left(\mathring{\mu}_{c}c\right)\right]-\frac{3\cos^{2}\left(\mathring{\mu}_{b}b\right)}{2\mathring{\mu}_{b}^{2}}\right.\nonumber \\
	& +\frac{7\sin^{4}\left(\mathring{\mu}_{b}b\right)}{4\mathring{\mu}_{b}^{2}}+\frac{\cos^{4}\left(\mathring{\mu}_{b}b\right)}{4\mathring{\mu}_{b}^{2}}-\frac{4\sin^{2}\left(\mathring{\mu}_{b}b\right)}{\mathring{\mu}_{b}^{2}}\cos\left(\mathring{\mu}_{b}b\right)\cos\left(\mathring{\mu}_{c}c\right)\nonumber \\
	& \left.-\gamma^{4}\frac{\mathring{\mu}_{b}^{2}}{\sin^{2}\left(\mathring{\mu}_{b}b\right)}+\gamma^{2}\left[\sin^{2}\left(\mathring{\mu}_{b}b\right)-\cos^{2}\left(\mathring{\mu}_{b}b\right)\right]\right\} .\label{eq:RE-TL-pbpc-mu0}
\end{align}
Once again we see that perturbatively
\begin{equation}
	\frac{d\theta_{(\mathring{\mu})}^{\mathrm{TL}}}{d\tau}\approx-\frac{1}{2p_{c}}\left(1+\frac{9b^{2}}{2\gamma^{2}}+\frac{\gamma^{2}}{2b^{2}}-\left(b^{2}+\frac{7b^{4}}{2\gamma^{2}}-\frac{\gamma^{2}}{6}\right)\mathring{\mu}_{b}^{2}-\frac{7b^{2}c^{2}}{\gamma^{2}}\mathring{\mu}_{c}^{2}\right)+\mathcal{O}\left(\mathring{\mu}^{4}\right).
\end{equation}
We see that the first three terms are the classical ones leading to
focusing. The terms proportional to $\mathring{\mu}_{b}^{2}$ are
all positive contributing to defocusing except the term $\frac{\gamma^{2}}{6}$
which is small and close to $t\to0$ is much smaller that the other
two terms. The term proportional to $\mathring{\mu}_{c}^{2}$ is always
positive. These effective terms take over close to $t\to0$ and stop
focusing of the congruence. The full nonperturbative behavior of $\frac{d\theta_{(\mathring{\mu})}^{\mathrm{TL}}}{d\tau}$
can be derived numerically by replacing \eqref{eq:bt-eff-mu0}--\eqref{eq:pct-eff-mu0}
into \eqref{eq:RE-TL-pbpc-mu0}. It is plotted in Fig. \ref{fig:theta-RE-TL-mu0-vs-classic}.
It is seen that $\frac{d\theta_{(\mathring{\mu})}^{\mathrm{TL}}}{d\tau}$
becomes quite large and positive close to $t\to0$, then dips a little
bit below zero and the goes to zero at $t=0$.

Hence, a common theme in Fig. \ref{fig:theta-RE-TL-mu0-vs-classic}
is that both $\theta_{(\mathring{\mu})}^{\mathrm{TL}}$ and $\frac{d\theta_{(\mathring{\mu})}^{\mathrm{TL}}}{d\tau}$
become zero as $t\to0$, and neither of them ever blows up anywhere
inside the black hole. This is a clear proof that the singularity
is resolved. Also it can be seen that the effective and classical
expansion and Raychaudhuri equation match very well far from the region
used to be the singularity at $t=0$. But close to this region, quantum
effects starts to take over and turn the curves around, stopping them
from blowing up. 

\begin{figure}
	\subfloat[Left: Classical vs null $\theta$ in the $\mathring{\mu}$ scheme
	as a function of the Schwarzschild time $t$. The effective expansion
	$\theta_{(\mathring{\mu})}^{\mathrm{NL}}$ goes to zero as $t\to0$.
	Right: Close up of the left figure close to $t=0$.]{%
		\begin{minipage}[t]{0.45\textwidth}%
			\begin{center}
				\includegraphics[scale=0.5]{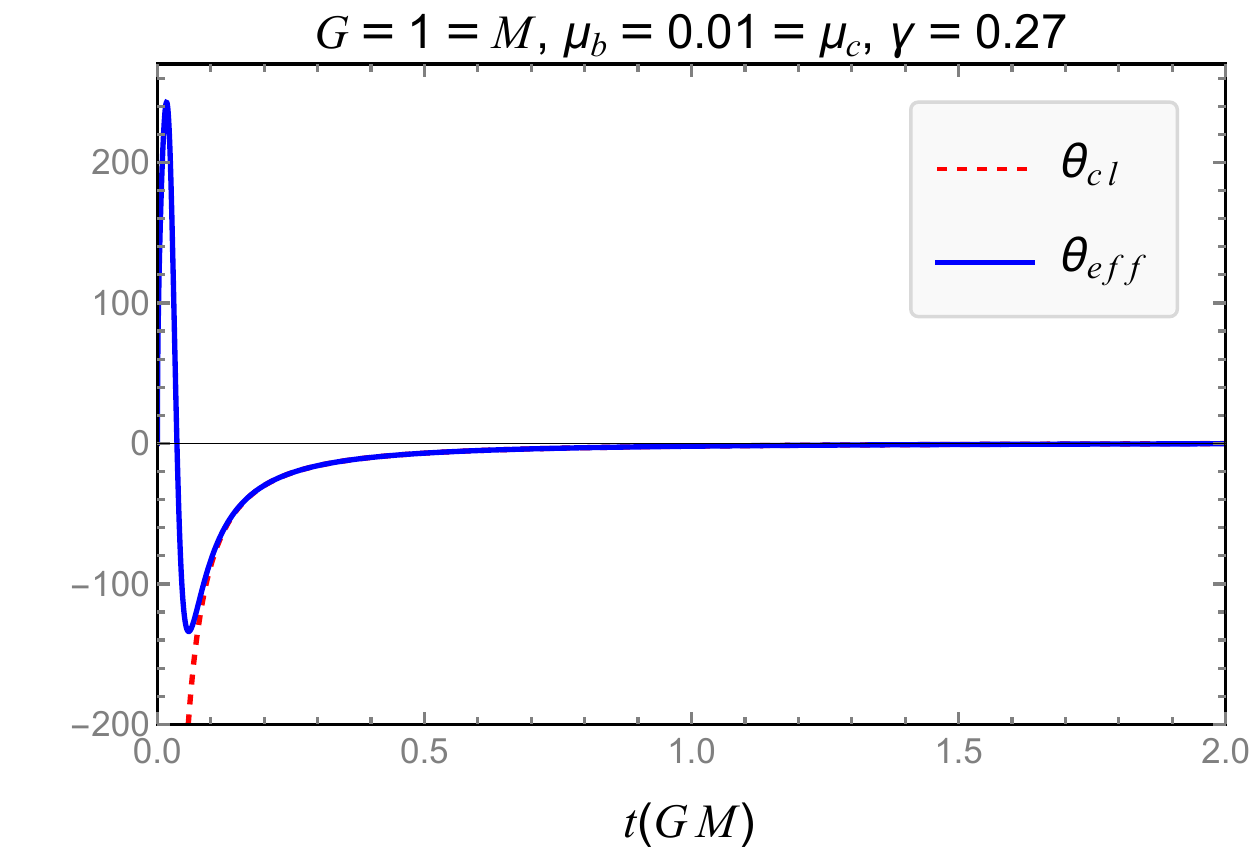}
				\par\end{center}%
		\end{minipage}\hfill{}%
		\begin{minipage}[t]{0.45\textwidth}%
			\begin{center}
				\includegraphics[scale=0.5]{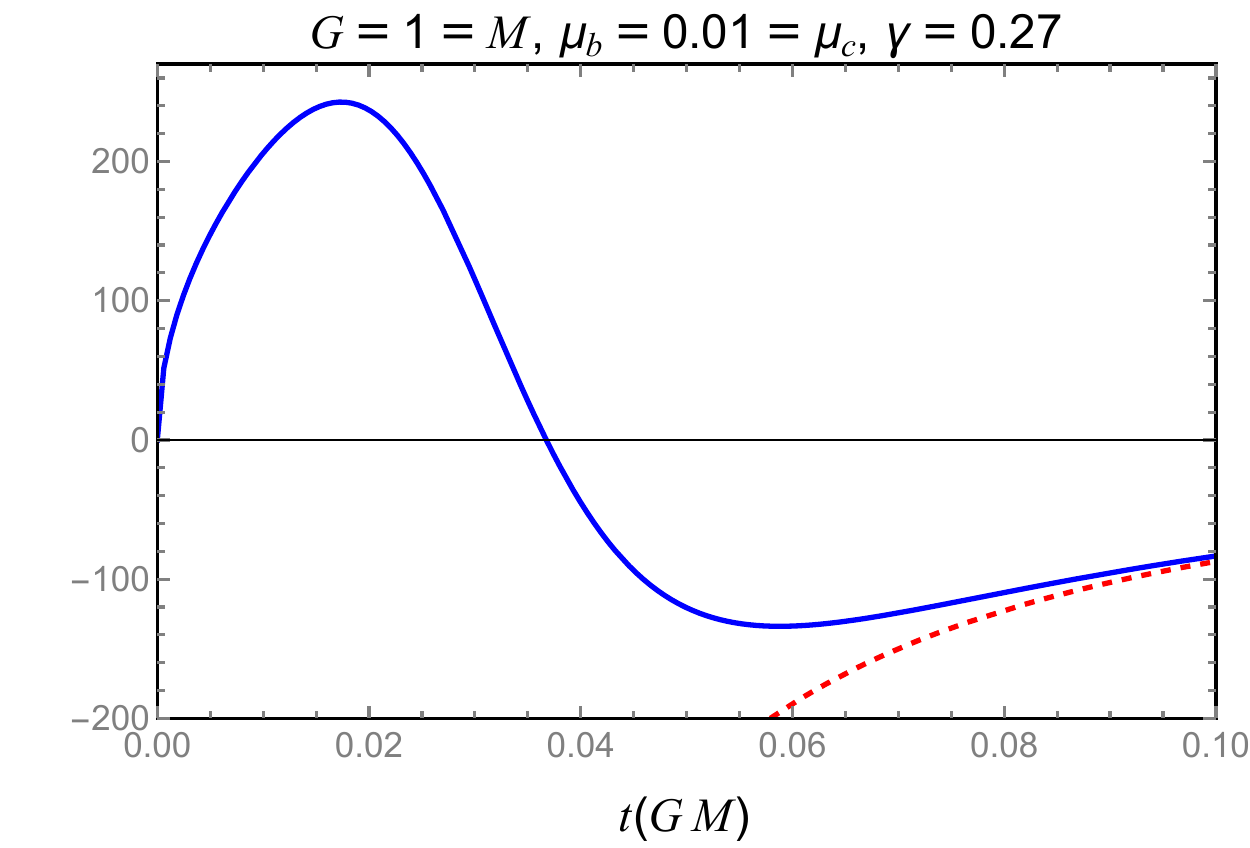}
				\par\end{center}%
		\end{minipage}
		
	}\hfill{}\subfloat[Left: Classical vs null $\frac{d\theta}{d\lambda}$ in the $\mathring{\mu}$
	scheme as a function of the Schwarzschild time $t$. The effective
	$\frac{d\theta_{(\mathring{\mu})}^{\mathrm{NL}}}{d\lambda}$ goes
	to zero as $t\to0$. Right: Close up of the left figure close
	to $t=0$.]{%
		\begin{minipage}[t]{0.45\textwidth}%
			\begin{center}
				\includegraphics[scale=0.5]{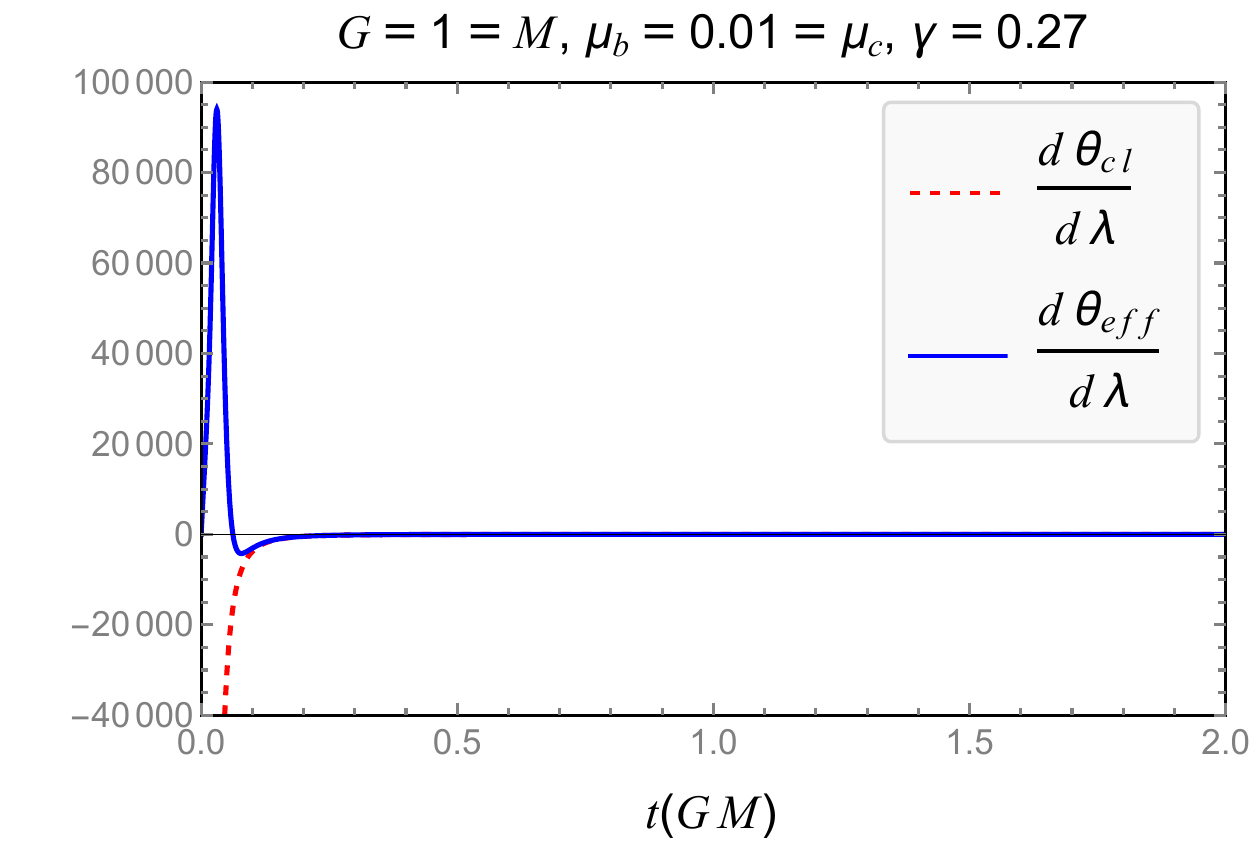}
				\par\end{center}%
		\end{minipage}\hfill{}%
		\begin{minipage}[t]{0.45\textwidth}%
			\begin{center}
				\includegraphics[scale=0.5]{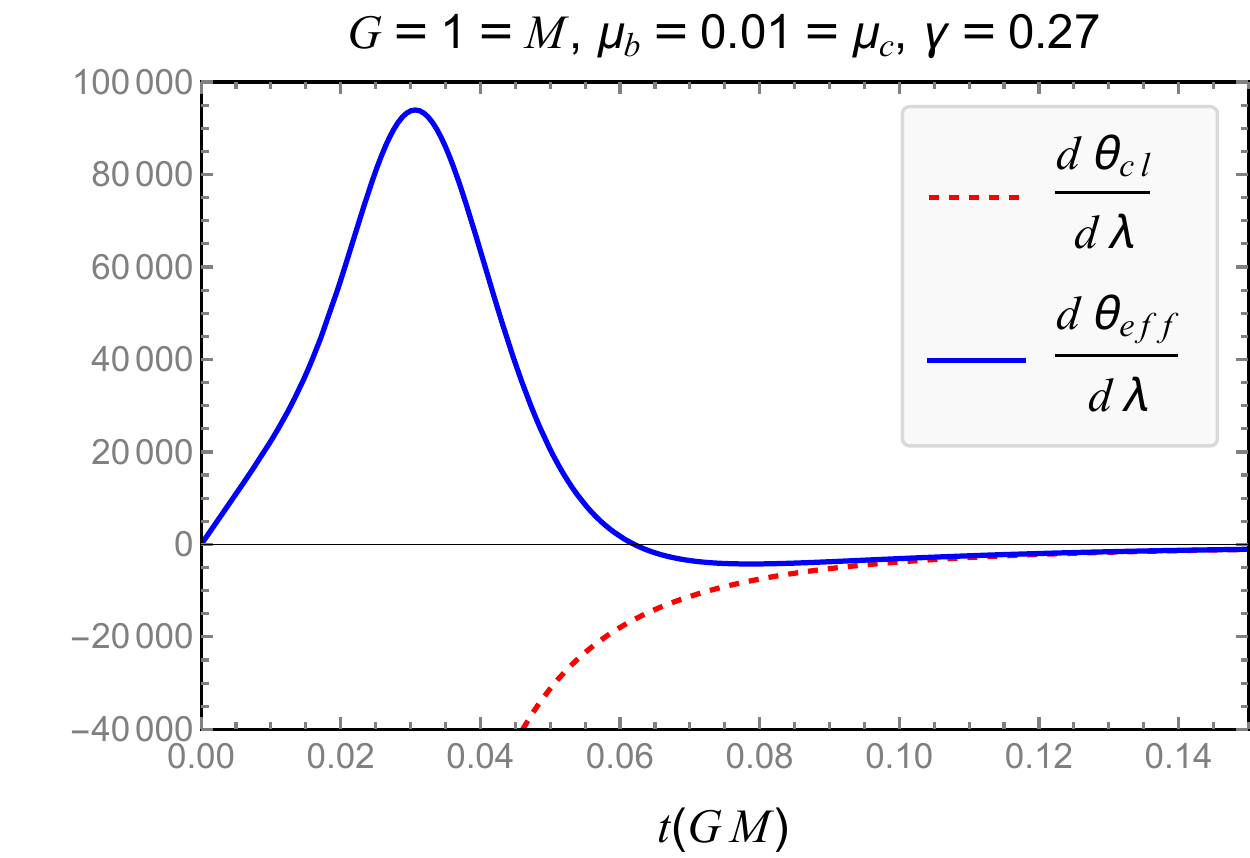}
				\par\end{center}%
		\end{minipage}
		
	}
	
	\caption{$\theta_{(\mathring{\mu})}^{\mathrm{NL}}$ and $\frac{d\theta_{(\mathring{\mu})}^{\mathrm{NL}}}{d\lambda}$
		\label{fig:theta-RE-NL-mu0-vs-classic}}
\end{figure}

Null expansion and Raychaudhuri equations have nicer expressions.
For the null expansion from \eqref{eq:expansion-pbpc-gen-null}, \eqref{eq:laps-eff},
and the equations \eqref{eq:b-diff-mu0}--\eqref{eq:pc-diff-mu0}
we obtain
\begin{equation}
	\theta_{(\mathring{\mu})}^{\mathrm{NL}}=\pm\frac{2}{\gamma\sqrt{p_{c}}}\frac{\sin\left(\mathring{\mu}_{b}b\right)}{\mu_{b}}\cos\left(\mathring{\mu}_{c}c\right),
\end{equation}
whose negative branch for small $\mathring{\mu}_{b},\,\mathring{\mu}_{c}$
is
\begin{equation}
	\theta_{(\mathring{\mu})}^{\mathrm{NL}}=\frac{1}{\gamma\sqrt{p_{c}}}\left(-2b+\frac{b^{3}}{3}\mathring{\mu}_{b}^{2}+bc^{2}\mathring{\mu}_{c}^{2}\right)+\mathcal{O}\left(\mathring{\mu}^{4}\right).
\end{equation}
The first term is the classical term and while it is negative, contributing
to the focusing, the other terms that are effective are always positive
given the solutions \eqref{eq:bt-eff-mu0}--\eqref{eq:pct-eff-mu0}.
They actually take over and stop the focusing of the congruence. The
corresponding nonperturbative behavior is depicted in Fig. \ref{fig:theta-RE-NL-mu0-vs-classic},
where it is seen that the expansion stops, turns around and reaches
zero at $t=0$.

We can also derive an expression for the Raychaudhuri equation in
this null case. Using \eqref{eq:laps-eff} and \eqref{eq:b-diff-mu0}--\eqref{eq:pc-diff-mu0}
in \eqref{eq:RE-pbpc-gen-null} yields
\begin{equation}
	\frac{d\theta_{(\mathring{\mu})}^{\mathrm{NL}}}{d\lambda}=\frac{2}{\gamma^{2}p_{c}}\frac{\sin^{2}\left(\mathring{\mu}_{b}b\right)}{\mathring{\mu}_{b}^{2}}\left[2\sin^{2}\left(\mathring{\mu}_{c}c\right)-\cos\left(\mathring{\mu}_{b}b\right)\cos\left(\mathring{\mu}_{b}c\right)\right],
\end{equation}
which, if Taylor expanded for small $\mathring{\mu}_{b},\,\mathring{\mu}_{c}$
yields
\begin{equation}
	\frac{d\theta_{(\mathring{\mu})}^{\mathrm{NL}}}{d\lambda}\approx\frac{1}{\gamma^{2}p_{c}}\left(-2b^{2}+\frac{5b^{4}}{3}\mathring{\mu}_{b}^{2}+5b^{2}c^{2}\mathring{\mu}_{c}^{2}\right)+\mathcal{O}\left(\mathring{\mu}^{4}\right).
\end{equation}
The effective terms are both positive and based on our previous discussion,
become very large close to $t=0$. In this case too, the full nonperturbative
solution depicted in Fig. \ref{fig:theta-RE-NL-mu0-vs-classic} shows
that $\frac{d\theta_{(\mathring{\mu})}^{\mathrm{NL}}}{d\lambda}$
becomes positive close to $t=0$ before vanishing at $t=0$.

\begin{figure}
	\subfloat[The Kretschmann scalar $K$ for the $\mathring{\mu}$ case as a function
	of the Schwarzschild time $t$.]{\begin{centering}
			\includegraphics[scale=0.51]{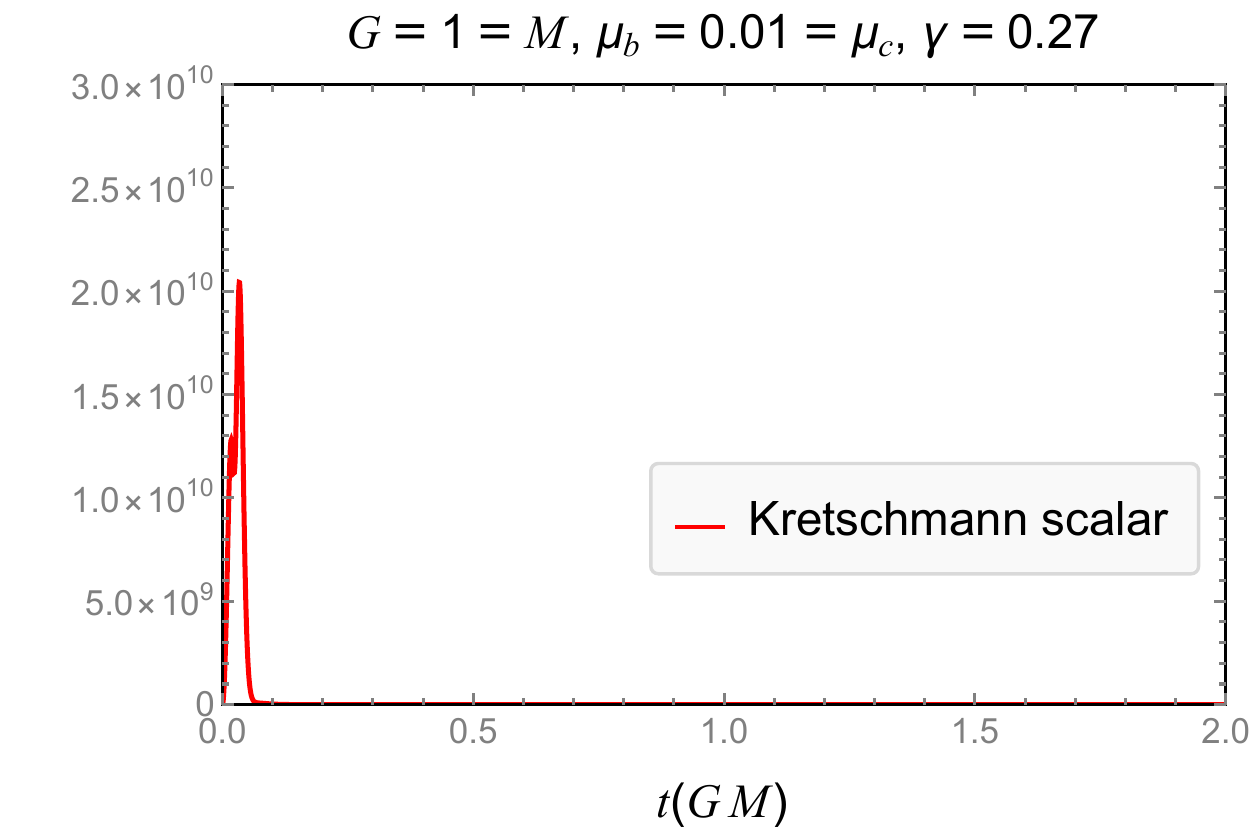}
			\par\end{centering}
	}\hfill{}\subfloat[Close up of $K$ close to $t=0$. It is seen that $K$ remains finite
	everywhere in the interior and vanishes for $t=0$.]{\begin{centering}
			\includegraphics[scale=0.51]{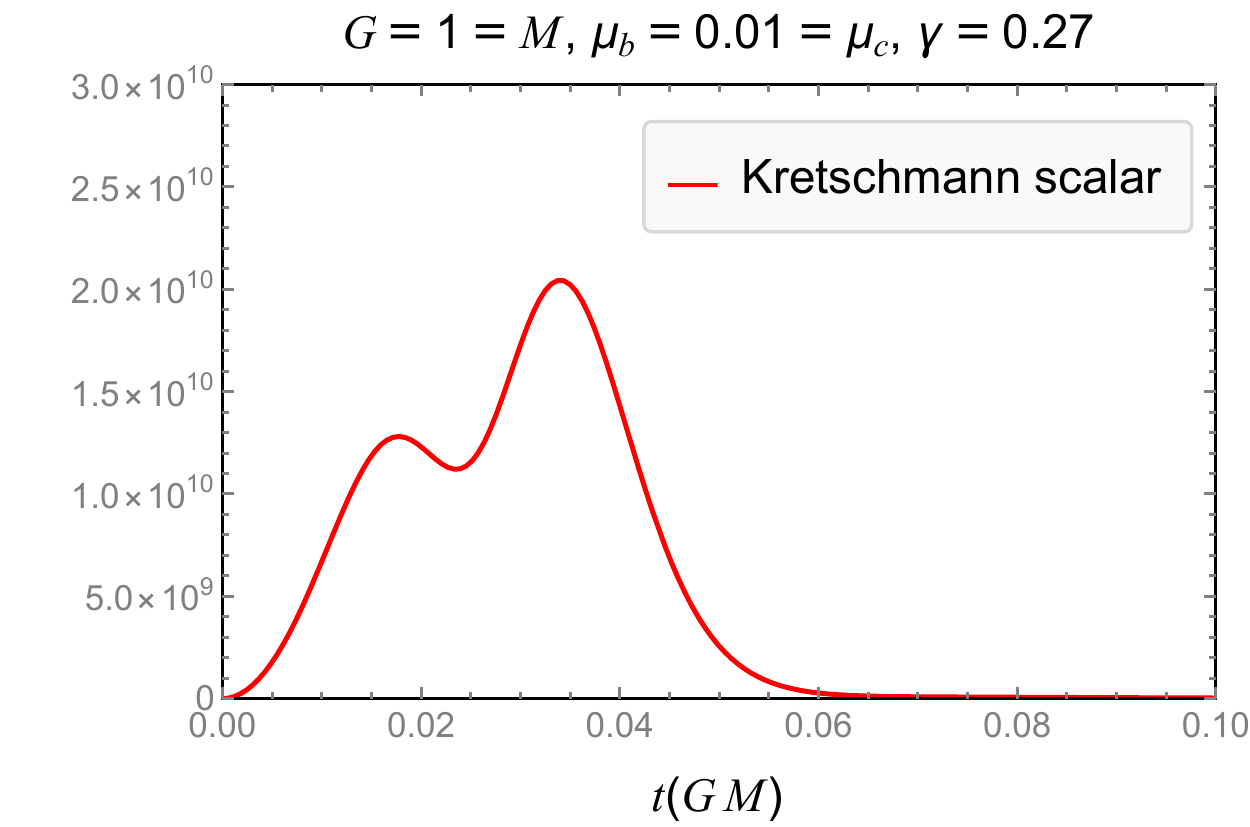}
			\par\end{centering}
	}
	
	\caption{$K$ in the $\mathring{\mu}$ case \label{fig:K-LQG-mu0}}
\end{figure}

We can also look at the Kretschmann scalar to confirm the resolution
of the singularity in the effective regime. The expression for the
Kretschmann scalar $K$ in terms of the canonical variables for $\mathring{\mu}$
case is quite large and we do not write it down here. However, we
can first look at its terms up to the second order in $\mathring{\mu}_{b},\,\mathring{\mu}_{c}$,
\begin{equation}
	K=\frac{2}{\gamma^{2}p_{c}^{2}}\left[12b^{2}+\frac{6b^{4}}{\gamma^{2}}+6\gamma^{2}-\mathring{\mu}_{b}^{2}\left(\frac{b^{6}}{\gamma^{2}}+3b^{4}+3\gamma^{2}b^{2}+\gamma^{4}\right)\right]+\mathcal{O}\left(\mathring{\mu}^{4}\right).
\end{equation}
Interestingly, while the classical terms (the first three terms) are
always positive, the effective terms proportional to $\mathring{\mu}_{b}^{2}$
are all negative and counter the classical terms. These terms become
large close to the region used to be the singularity and stop $K$
from diverging. In fact $K$ becomes zero at $t=0$. This can be checked
from the full numerical nonperturbative behavior of $K$ as a function
of the Schwarzschild time $t$ in Fig. \ref{fig:K-LQG-mu0}. As we
can see, $K$ never diverges in the interior, and although it has
a large increase close to $t=0$, the quantum effects become so large
in that region that turn the curve back towards zero, and we obtain
$K\to0$ as $t\to0$. This together with the above results regarding
the expansion scalar and the Raychaudhuri equation definitely prove
that the singularity is removed in this model.

\subsubsection{$\bar{\mu}$ scheme\label{subsec:mu-bar-scheme}}

\begin{figure}
	\begin{centering}
		\includegraphics[scale=0.7]{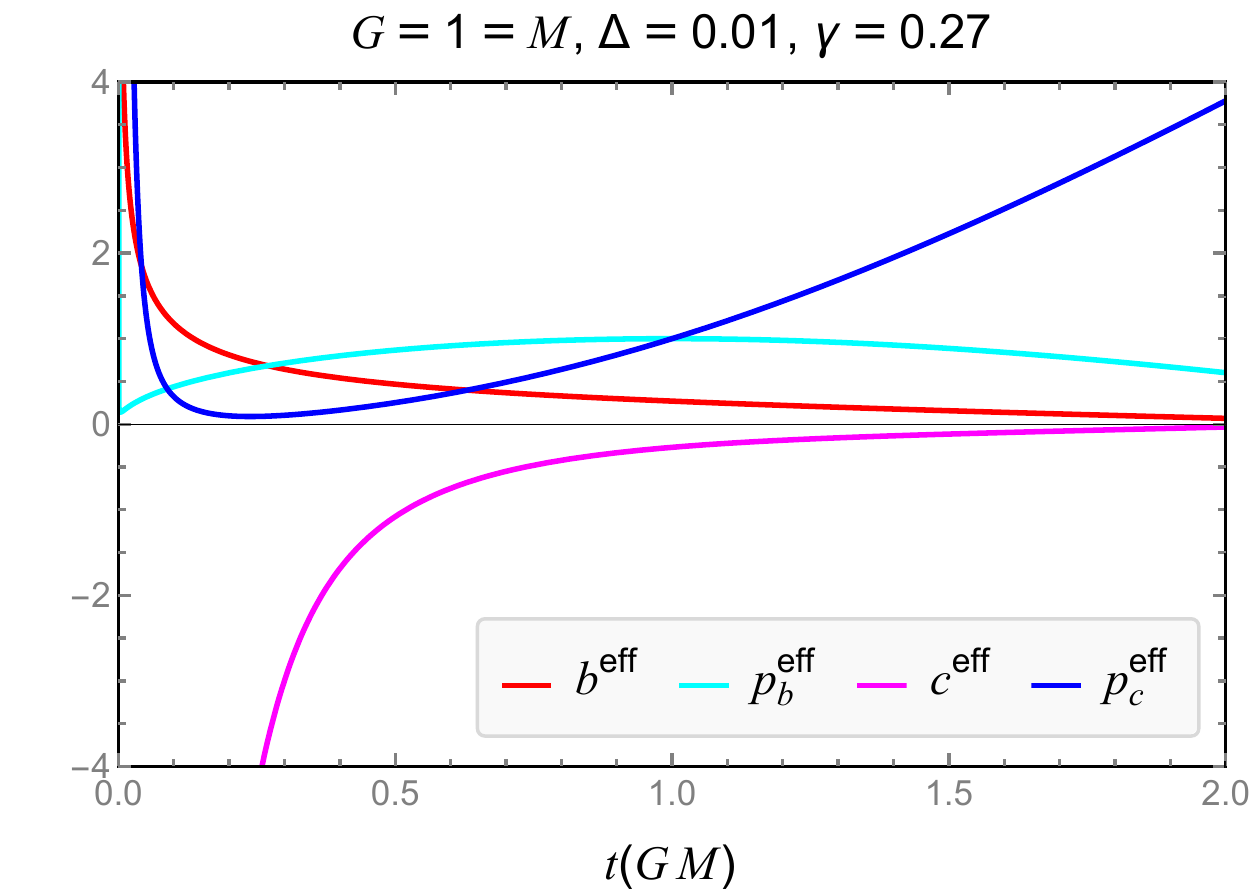}
		\par\end{centering}
	\caption{Solutions to the equations of motion in the $\bar{\mu}$ case as a function
		of the Schwarzschild time $t$. Once again, $p_{c}$ never vanishes.
		\label{fig:Sol-LQG-mu-bar}}
\end{figure}

In this scheme $\bar{\mu}_{b},\,\bar{\mu}_{c}$ are assumed to depend
on the triad components as 
\begin{align}
	\bar{\mu}_{b}= & \sqrt{\frac{\Delta}{p_{b}}},\label{eq:mu-bar-b}\\
	\bar{\mu}_{c}= & \sqrt{\frac{\Delta}{p_{c}}},\label{eq:mu-bar-c}
\end{align}
where $\Delta$ is related to the minimum area in loop quantum gravity.
Using the same lapse as \eqref{eq:laps-eff} but keeping in mind the
above dependence of $\bar{\mu}_{b},\,\bar{\mu}_{c}$, one can easily
obtain the equations of motion as 
\begin{align}
	\frac{db}{dT}= & \frac{1}{4}\left(b\cos\left(\bar{\mu}_{b}b\right)-3\frac{\sin\left(\bar{\mu}_{b}b\right)}{\bar{\mu}_{b}}-\gamma^{2}\frac{\bar{\mu}_{b}}{\sin\left(\bar{\mu}_{b}b\right)}\left[1+b\cos\left(\bar{\mu}_{b}b\right)\frac{\bar{\mu}_{b}}{\sin\left(\bar{\mu}_{b}b\right)}\right]\right),\\
	\frac{dp_{b}}{dT}= & \frac{1}{2}p_{b}\cos\left(\bar{\mu}_{b}b\right)\left[1-\gamma^{2}\frac{\bar{\mu}_{b}^{2}}{\sin^{2}\left(\bar{\mu}_{b}b\right)}\right],\\
	\frac{dc}{dT}= & c\cos\left(\bar{\mu}_{c}c\right)-3\frac{\sin\left(\bar{\mu}_{c}c\right)}{\bar{\mu}_{c}},\\
	\frac{dp_{c}}{dT}= & 2p_{c}\cos\left(\bar{\mu}_{c}c\right).
\end{align}
The solutions to these equations can be derived numerically by demanding
the solutions match the classical ones very close to the horizon at
$t=2GM$. These are plotted in Fig. \ref{fig:Sol-LQG-mu-bar}

Analytical solutions to these equations are hard to obtain, but one
can numerically solve them. The constants of integration are determined
by matching these solution with the classical ones very close to the
horizon for $\text{\ensuremath{\Delta}\ensuremath{\to0}}$. The expressions
for timelike $\theta_{(\bar{\mu})}^{\mathrm{TL}}$ and $\frac{d\theta_{(\bar{\mu})}^{\mathrm{TL}}}{d\tau}$
turn out to be

\begin{align}
	\theta_{(\bar{\mu})}^{\mathrm{TL}}= & \pm\frac{1}{2\gamma\sqrt{p_{c}}}\left\{ 2\frac{\sin\left(\bar{\mu}_{b}b\right)}{\bar{\mu}_{b}}\cos\left(\bar{\mu}_{c}c\right)+\left[\frac{\sin\left(\bar{\mu}_{b}b\right)}{\bar{\mu}_{b}}-\frac{\bar{\mu}_{b}}{\sin\left(\bar{\mu}_{b}b\right)}\gamma^{2}\right]\cos\left(\bar{\mu}_{b}b\right)\right\} ,\\
	\frac{d\theta_{(\bar{\mu})}^{\mathrm{TL}}}{d\tau}= & \frac{1}{8p_{c}}\left\{ -3+\frac{\sin^{2}\left(\bar{\mu}_{b}b\right)}{\bar{\mu}_{b}^{2}}\left[\frac{17}{2\gamma^{2}}-\frac{8}{\gamma^{2}}\cos\left(\bar{\mu}_{b}b\right)\cos\left(\bar{\mu}_{c}c\right)-\frac{16}{\gamma^{2}}\cos\left(2\bar{\mu}_{c}c\right)\right]\right.\nonumber \\
	& \left.-\frac{3\gamma^{2}}{2}\frac{\bar{\mu}_{b}^{2}}{\sin^{2}\left(\bar{\mu}_{b}b\right)}-\left[\frac{5}{2\gamma^{2}}\frac{\sin^{2}\left(\bar{\mu}_{b}b\right)}{\bar{\mu}_{b}^{2}}+\frac{\gamma^{2}}{2}\frac{\bar{\mu}_{b}^{2}}{\sin^{2}\left(\bar{\mu}_{b}b\right)}+1\right]\cos\left(2\bar{\mu}_{b}b\right)\right\} .
\end{align}

\begin{figure}
	\subfloat[Left: Classical vs timlike $\theta$ in the $\bar{\mu}$ scheme as
	a function of the Schwarzschild time $t$. The effective expansion
	$\theta_{(\bar{\mu})}^{\mathrm{TL}}$ goes to zero as $t\to0$. Right:
	Close up of the left figure close to $t=0$.]{%
		\begin{minipage}[t]{0.45\textwidth}%
			\begin{center}
				\includegraphics[scale=0.5]{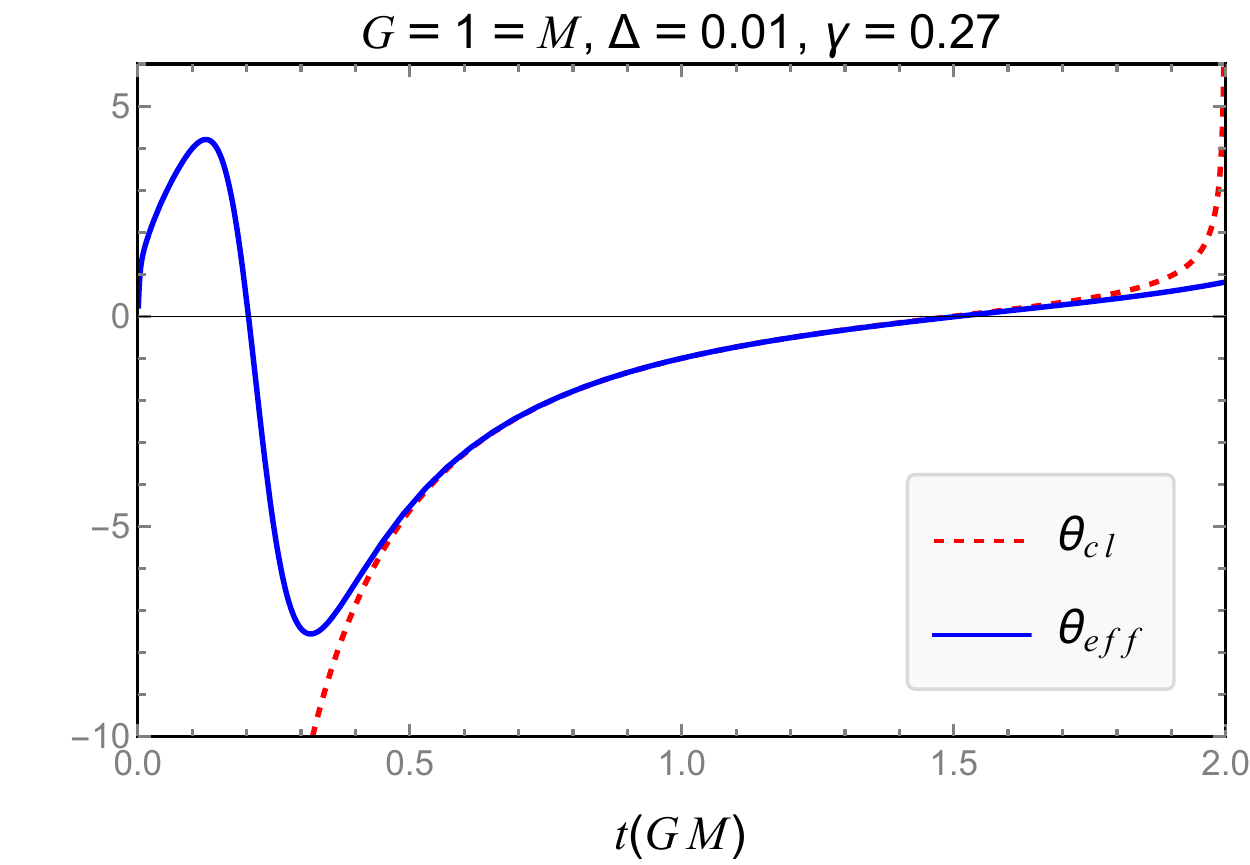}
				\par\end{center}%
		\end{minipage}\hfill{}%
		\begin{minipage}[t]{0.45\textwidth}%
			\begin{center}
				\includegraphics[scale=0.5]{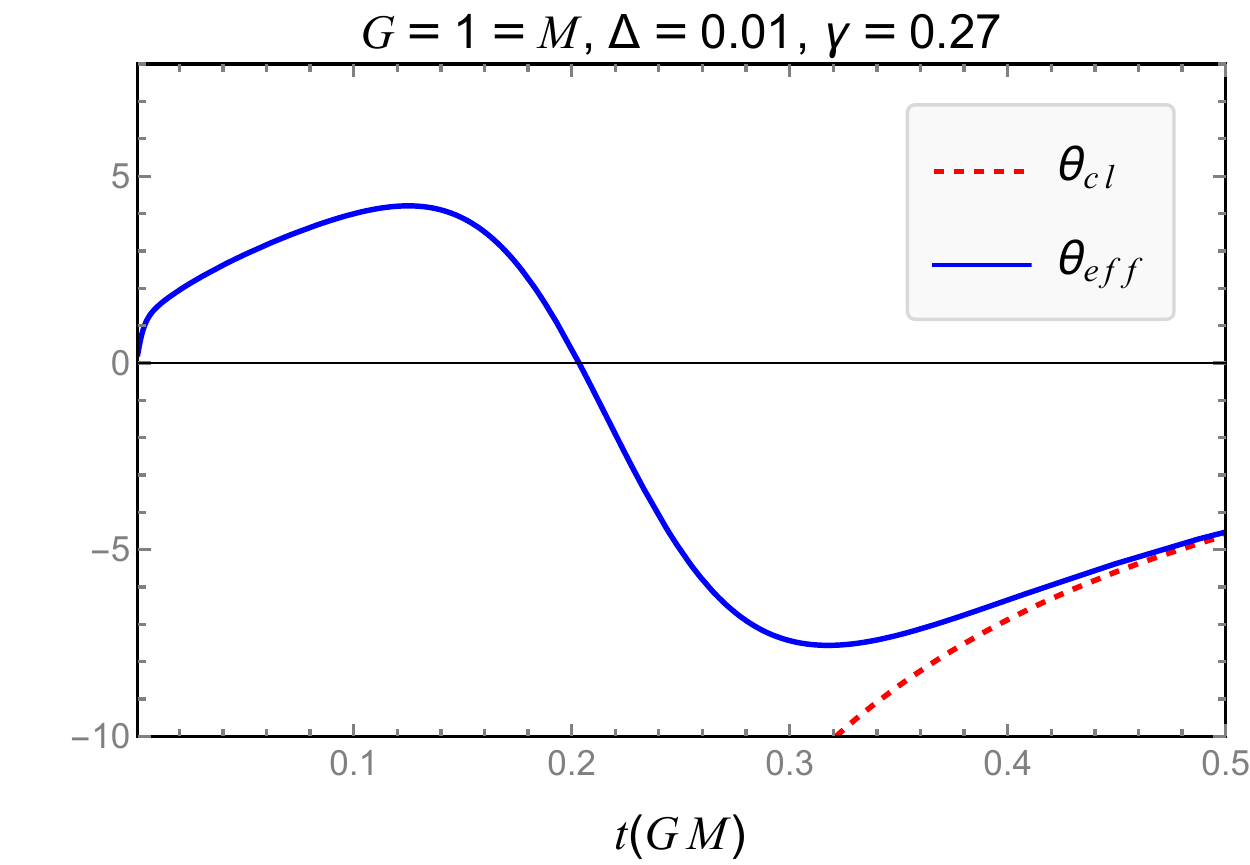}
				\par\end{center}%
		\end{minipage}
		
	}\hfill{}\subfloat[Left: Classical vs timlike $\frac{d\theta}{d\tau}$ in the $\bar{\mu}$
	scheme as a function of the Schwarzschild time $t$. The effective
	$\frac{d\theta_{(\bar{\mu})}^{\mathrm{TL}}}{d\tau}$ goes to zero
	as $t\to0$. Right: Close up of the left figure close to $t=0$.]{%
		\begin{minipage}[t]{0.45\textwidth}%
			\begin{center}
				\includegraphics[scale=0.50]{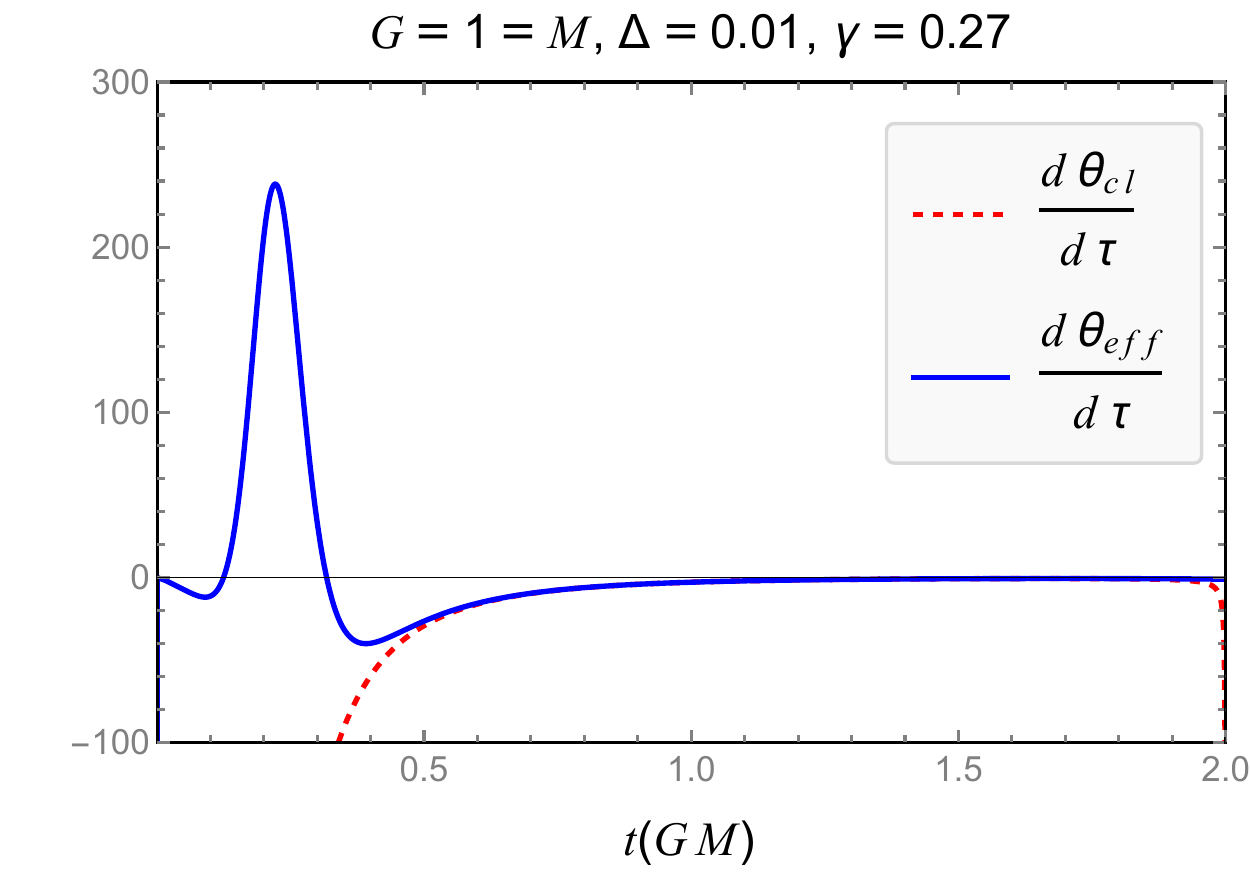}
				\par\end{center}%
		\end{minipage}\hfill{}%
		\begin{minipage}[t]{0.45\textwidth}%
			\begin{center}
				\includegraphics[scale=0.50]{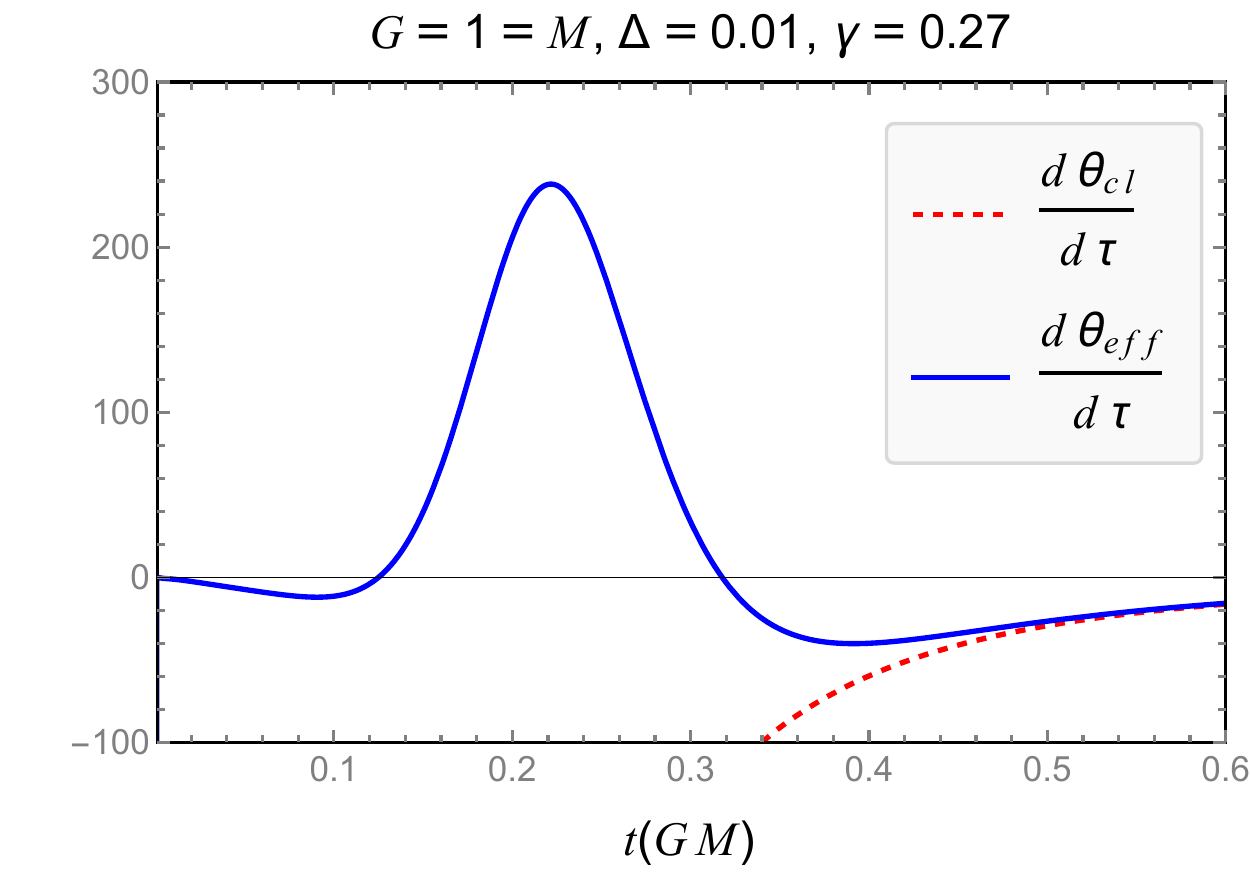}
				\par\end{center}%
		\end{minipage}
		
	}
	
	\caption{$\theta_{(\bar{\mu})}^{\mathrm{TL}}$ and $\frac{d\theta_{(\bar{\mu})}^{\mathrm{TL}}}{d\tau}$\label{fig:theta-RE-TL-mubar-vs-classic}}
\end{figure}

Notice that these expressions should in fact be thought of being in
terms of $\Delta$. Also note that the form of $\theta_{(\bar{\mu})}^{\mathrm{TL}}$
is exactly the same as the $\mathring{\mu}$ case but with $\mathring{\mu}$
replaced by $\bar{\mu}$. The perturbative expansion of these expressions
for small $\Delta$ become (negative branch for the expansion scalar)
\begin{align}
	\theta_{(\bar{\mu})}^{\mathrm{TL}}\approx & -\frac{1}{2\gamma\sqrt{p_{c}}}\left[3b-\frac{\gamma^{2}}{b}-\left(\frac{b^{3}}{p_{b}}+\frac{bc^{2}}{p_{c}}-\frac{b\gamma^{2}}{3p_{b}}\right)\Delta\right]+\mathcal{O}\left(\Delta^{2}\right),\\
	\frac{d\theta_{(\bar{\mu})}^{\mathrm{TL}}}{d\tau}\approx & -\frac{1}{2p_{c}}\left[1+\frac{9b^{2}}{2\gamma^{2}}+\frac{\gamma^{2}}{2b^{2}}-\left[\frac{15b^{4}}{4\gamma^{2}p_{b}}+\frac{9b^{2}c^{2}}{\gamma^{2}p_{c}}+\frac{b^{2}}{2p_{b}}+\frac{\gamma^{2}}{12p_{b}}\right]\Delta\right]+\mathcal{O}\left(\Delta^{2}\right).
\end{align}
The behavior of the solution to the canonical variables is such that
the combination of the effective terms above are not only positive,
but also become very large and balance the classical terms and lead
to cancellation of focusing of the congruence at $t=0$. This can
be seen from the full nonperturbative behavior of $\theta_{(\bar{\mu})}^{\mathrm{TL}}$
and $\frac{d\theta_{(\bar{\mu})}^{\mathrm{TL}}}{d\tau}$ depicted
in Fig. \ref{fig:theta-RE-TL-mubar-vs-classic}.

\begin{figure}
	\subfloat[Left: Classical vs null $\theta$ in the $\bar{\mu}$ scheme as a
	function of the Schwarzschild time $t$. The effective expansion $\theta_{(\bar{\mu})}^{\mathrm{NL}}$
	goes to zero as $t\to0$. Right: Close up of the left figure close
	to $t=0$.]{%
		\begin{minipage}[t]{0.45\textwidth}%
			\begin{center}
				\includegraphics[scale=0.5]{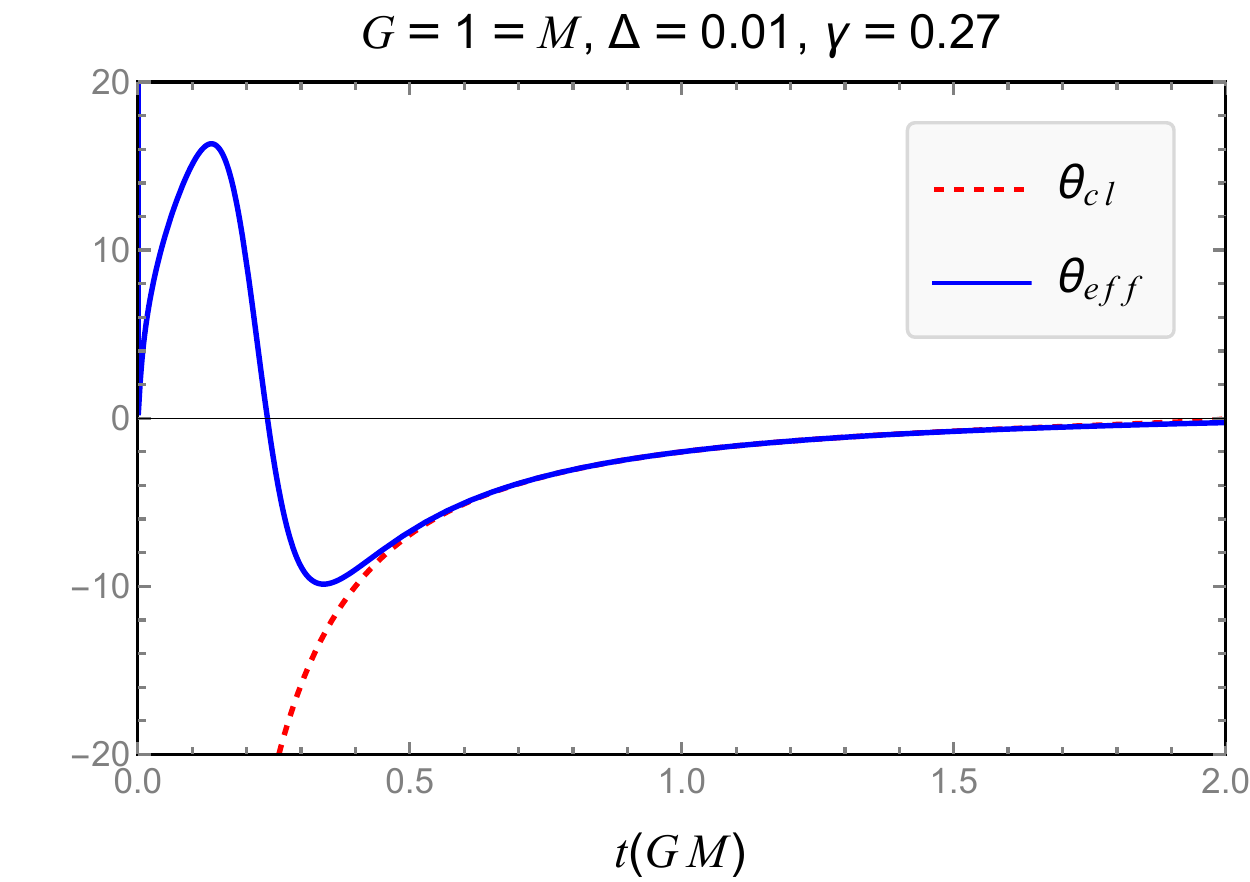}
				\par\end{center}%
		\end{minipage}\hfill{}%
		\begin{minipage}[t]{0.45\textwidth}%
			\begin{center}
				\includegraphics[scale=0.5]{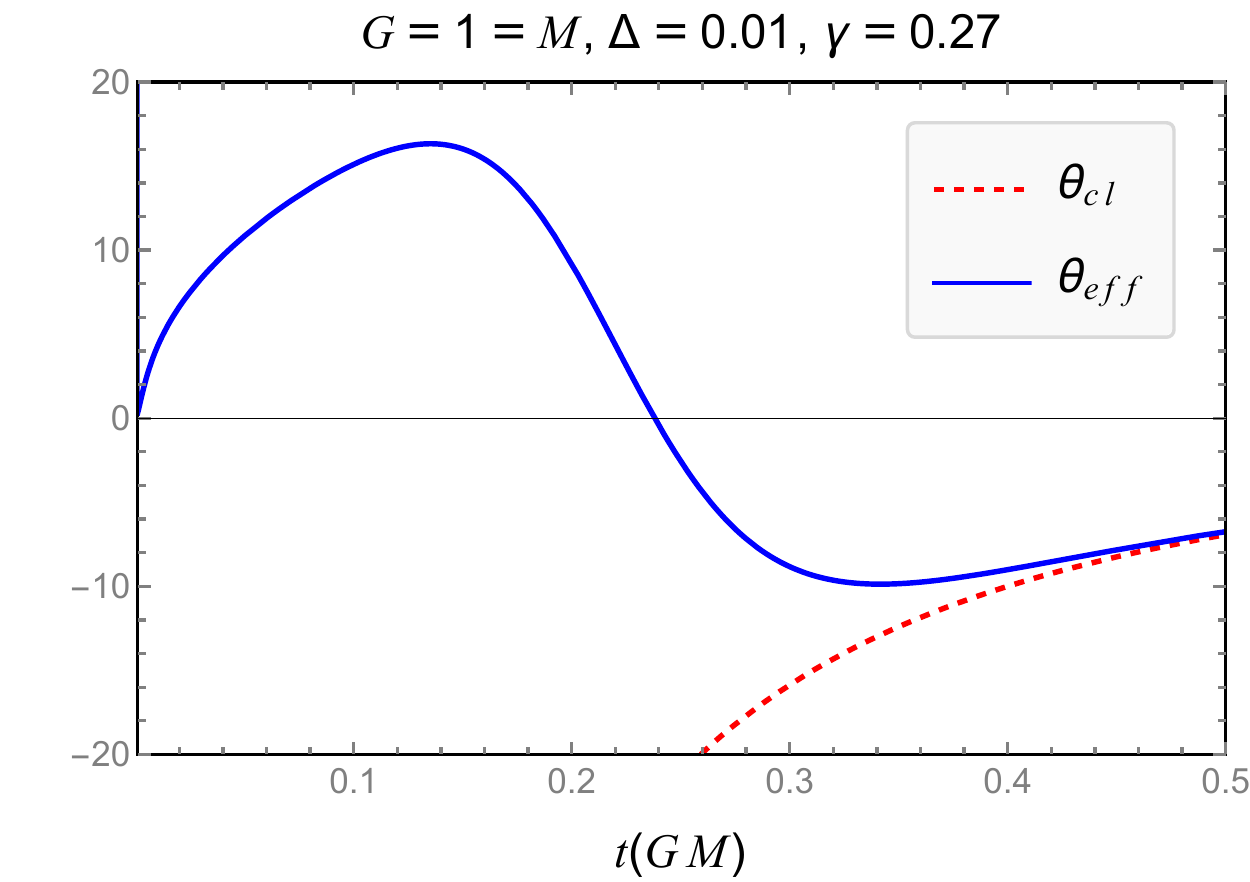}
				\par\end{center}%
		\end{minipage}
		
	}\hfill{}\subfloat[Left: Classical vs null $\frac{d\theta}{d\lambda}$ in the $\bar{\mu}$
	scheme as a function of the Schwarzschild time $t$. The effective
	$\frac{d\theta_{(\bar{\mu})}^{\mathrm{NL}}}{d\lambda}$ goes to zero
	as $t\to0$. Right: Close up of the left figure close to $t=0$.]{%
		\begin{minipage}[t]{0.45\textwidth}%
			\begin{center}
				\includegraphics[scale=0.5]{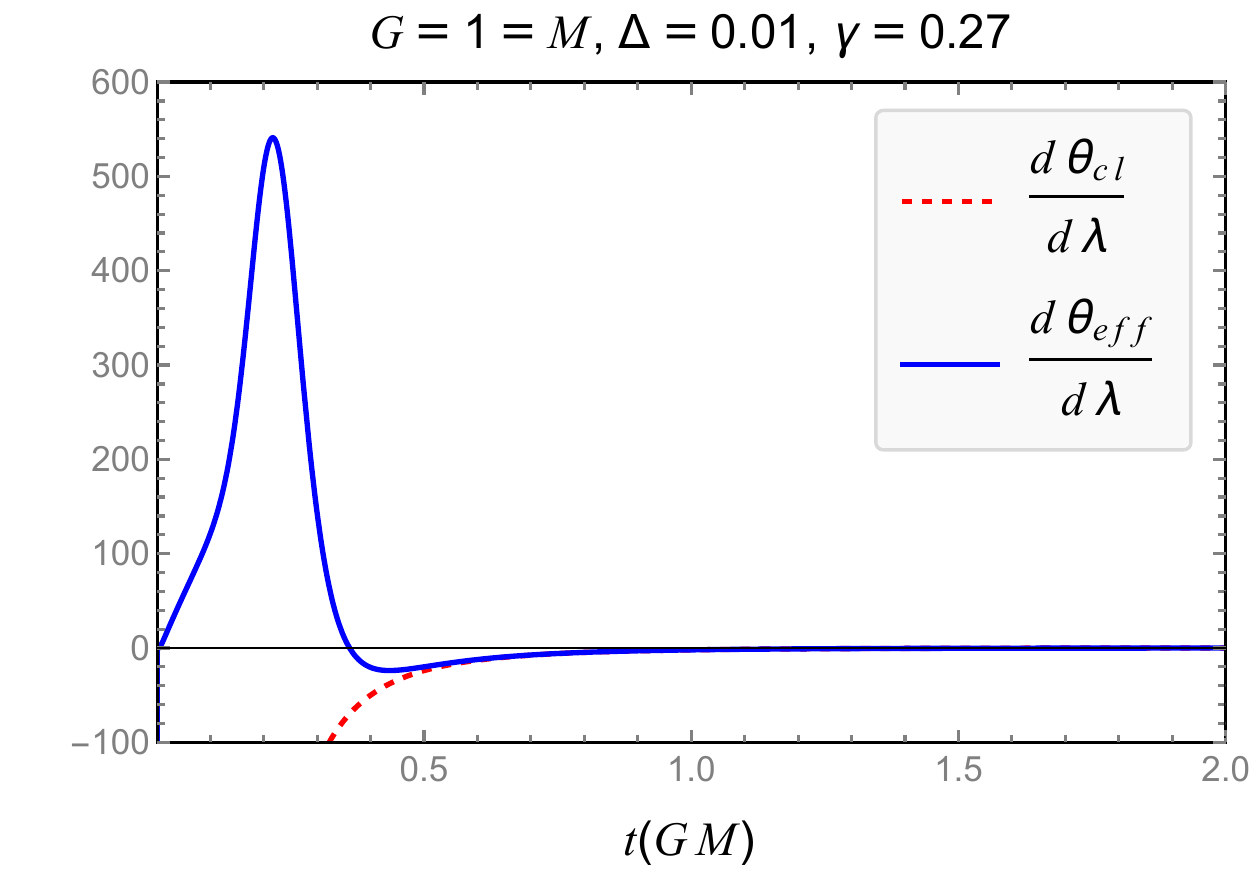}
				\par\end{center}%
		\end{minipage}\hfill{}%
		\begin{minipage}[t]{0.45\textwidth}%
			\begin{center}
				\includegraphics[scale=0.5]{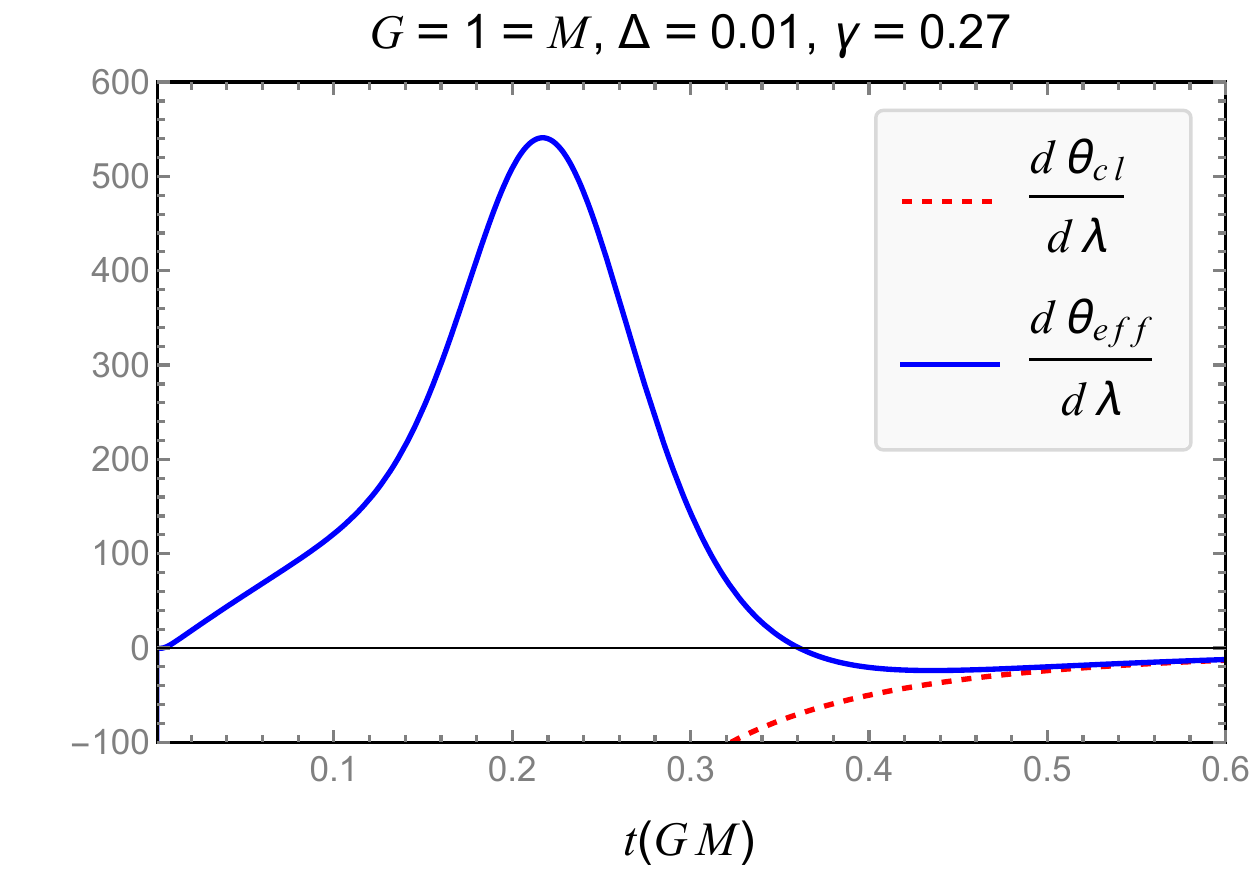}
				\par\end{center}%
		\end{minipage}
		
	}
	
	\caption{$\theta_{(\bar{\mu})}^{\mathrm{NL}}$ and $\frac{d\theta_{(\bar{\mu})}^{\mathrm{NL}}}{d\lambda}$\label{fig:theta-RE-NL-mubar-vs-classic}}
\end{figure}

In the same way, we can compute the null expansion scalar and Raychaudhuri
equation using \eqref{eq:expansion-pbpc-gen-null} and \eqref{eq:RE-pbpc-gen-null}
and the equations of motion of this scheme. Doing so, one obtains
\begin{align}
	\theta_{(\bar{\mu})}^{\mathrm{NL}}= & \pm\frac{2}{\gamma\sqrt{p_{c}}}\frac{\sin\left(\bar{\mu}_{b}b\right)}{\bar{\mu}_{b}}\cos\left(\bar{\mu}_{c}c\right),\\
	\frac{d\theta_{(\bar{\mu})}^{\mathrm{NL}}}{d\lambda}= & \frac{2}{\gamma^{2}p_{c}}\left\{ \frac{\sin^{2}\left(\bar{\mu}_{b}b\right)}{\bar{\mu}_{b}^{2}}\left[\frac{3}{2}-\frac{3}{2}\cos\left(2\bar{\mu}_{c}c\right)-\cos\left(\bar{\mu}_{b}b\right)\cos\left(\bar{\mu}_{c}c\right)\right]\right\} ,\nonumber 
\end{align}
with perturbative forms (negative branch for the expansion scalar)
\begin{align}
	\theta_{(\bar{\mu})}^{\mathrm{NL}}\approx & -\frac{1}{\gamma\sqrt{p_{c}}}\left[2b-\left(\frac{b^{3}}{3p_{b}}+\frac{bc^{2}}{p_{c}}\right)\Delta\right]+\mathcal{O}\left(\Delta^{2}\right),\\
	\frac{d\theta_{(\bar{\mu})}^{\mathrm{NL}}}{d\lambda}\approx & \frac{1}{\gamma^{2}p_{c}}\left[-2b^{2}+\left(\frac{5b^{4}}{3p_{b}}+7b^{2}c^{2}\right)\Delta\right]+\mathcal{O}\left(\Delta^{2}\right).
\end{align}
One can make similar observation about these expressions by noting
that the effective terms are positive and take over close to $t=0$.
These observations are confirmed by plotting the full nonperturbative
expressions in Fig. \eqref{fig:theta-RE-NL-mubar-vs-classic}, where
it is seen that nowhere inside the black hole does either $\theta_{(\bar{\mu})}^{\mathrm{NL}}$
or $\frac{d\theta_{(\bar{\mu})}^{\mathrm{NL}}}{d\lambda}$ blow up.

\begin{figure}
	\subfloat[The Kretschmann scalar $K$ for the $\bar{\mu}$ case as a function
	of the Schwarzschild time $t$.]{\begin{centering}
			\includegraphics[scale=0.51]{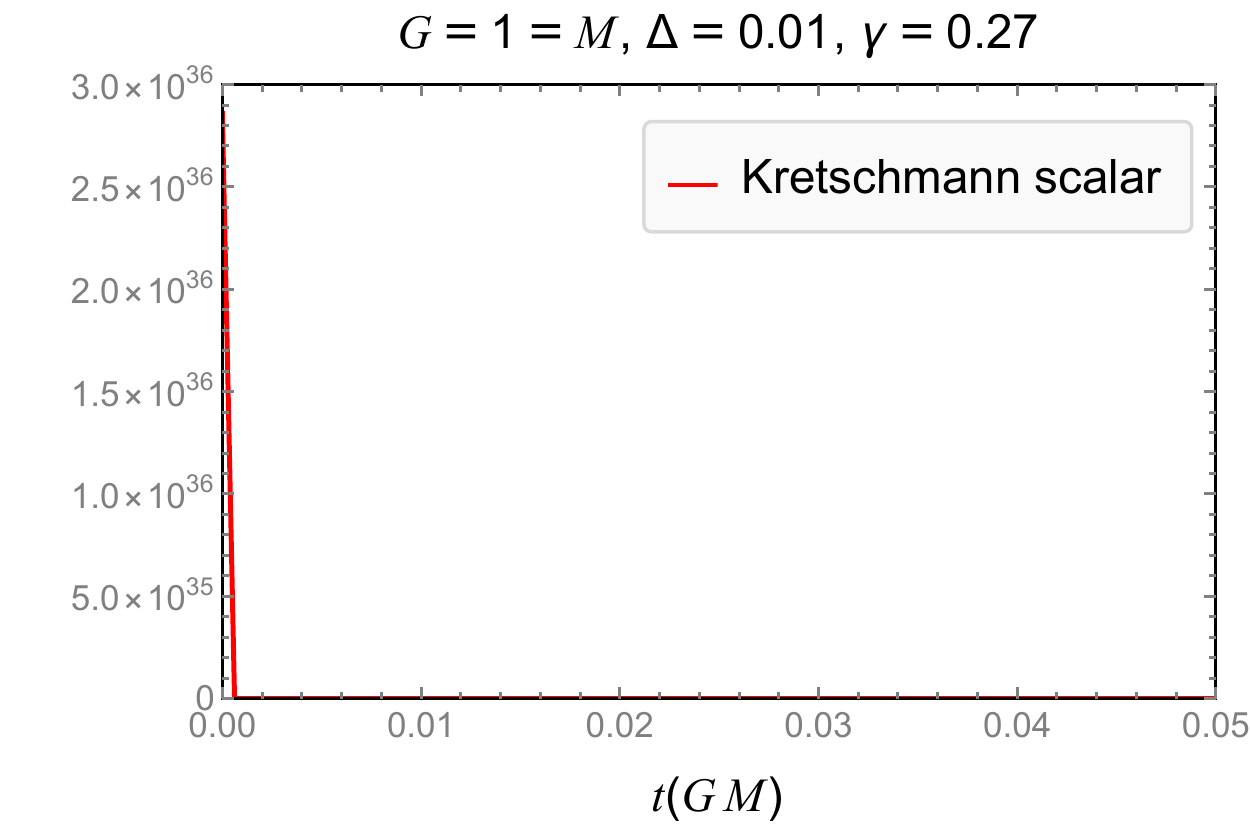}
			\par\end{centering}
	}\hfill{}\subfloat[Close up of $K$ close to $t=0$. It is seen that $K$ remains finite
	everywhere in the interior although it does not vanishes for $t=0$.]{\begin{centering}
			\includegraphics[scale=0.51]{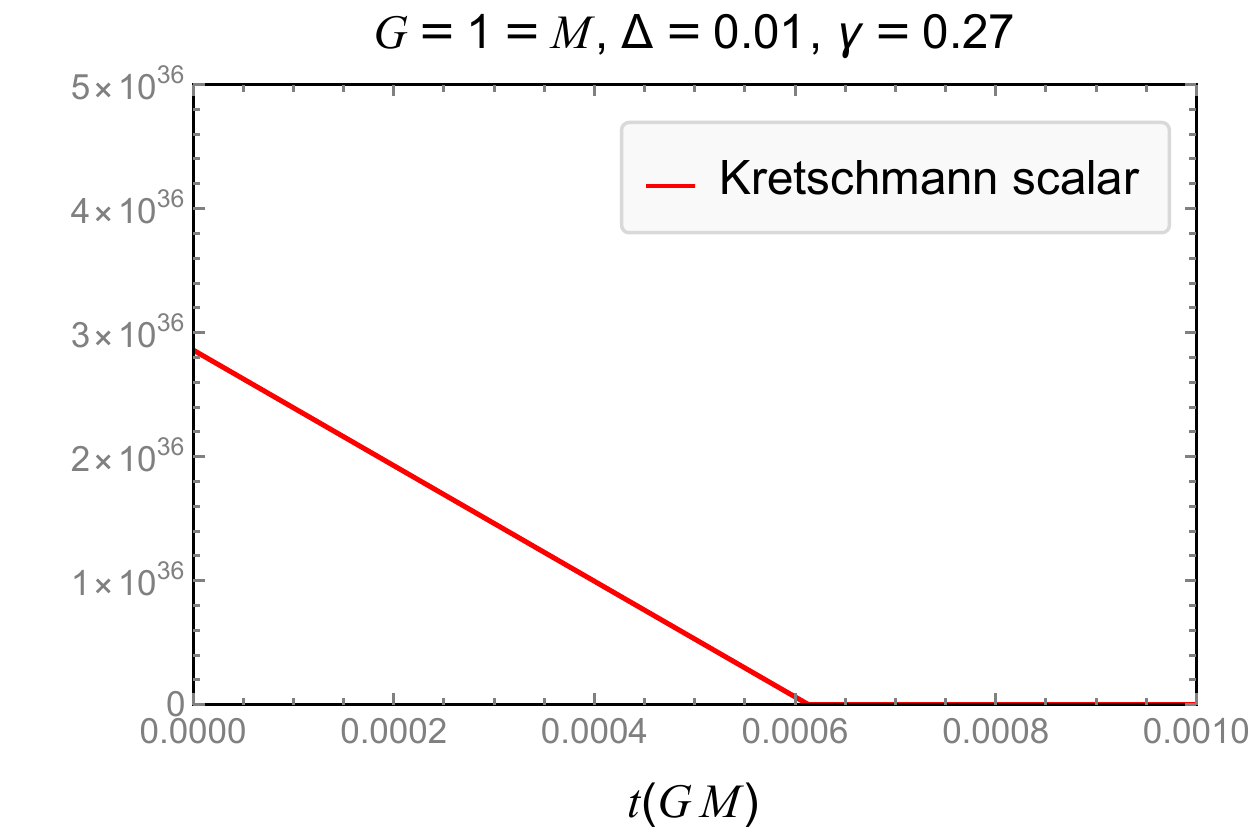}
			\par\end{centering}
	}
	
	\caption{$K$ in the $\bar{\mu}$ case \label{fig:K-LQG-mu-bar}}
\end{figure}

The Kretschmann scalar in this case has a similar behavior to the
previous case. However, while it does not diverge, it becomes quite
large at $t\to0$. By looking at its profile in Fig. \ref{fig:K-LQG-mu-bar},
we see that not only it does not diverge anywhere in the interior,
but also it becomes zero as $t\to0$. Since the full expression for
$K$ in this case is also very large, let us first check the perturbative
expression up to $\Delta$,
\begin{align}
	K= & \frac{1}{p_{c}^{2}}\left[12+\frac{12b^{4}}{\gamma^{4}}+\frac{24b^{2}}{\gamma^{2}}\right.\nonumber \\
	& -\Delta\left(\frac{b^{6}}{\gamma^{4}p_{b}}+\frac{76b^{4}c^{2}}{\gamma^{4}p_{c}}+\frac{7b^{4}}{\gamma^{2}p_{b}}+\frac{88b^{2}c^{2}}{\gamma^{2}p_{c}}+\frac{7b^{2}}{p_{b}}+\frac{12c^{2}}{p_{c}}+\frac{\gamma^{2}}{p_{b}}\right)\nonumber \\
	& +\mathcal{O}\left(\Delta^{2}\right).
\end{align}
The same pattern emerges here too similar to the previous case where
the correction terms are all negative, given the behavior of the solutions
to the equations of motion, particularly of $p_{b},\,p_{c}$. These
effective terms will counter the classical terms and become large
close to the region used to be the singularity, thus stopping $K$
from diverging. However, in this model, $K$ does not vanish at $t=0$.
Based on the value we chose for $\Delta$, we obtain $K\approx3\times10^{36}$
for $t=0$. This is seen from the full numerical nonperturbative behavior
of $K$ as a function of the Schwarzschild time $t$ in Fig. \ref{fig:K-LQG-mu-bar}. 

\subsubsection{$\bar{\mu}^{\prime}$ scheme\label{subsec:mu-bar-prime-scheme}}

\begin{figure}
	\subfloat[Solutions to the equations of motion in the $\bar{\mu}^{\prime}$ case
	as a function of the Schwarzschild time $t$.]{\begin{centering}
			\includegraphics[scale=0.51]{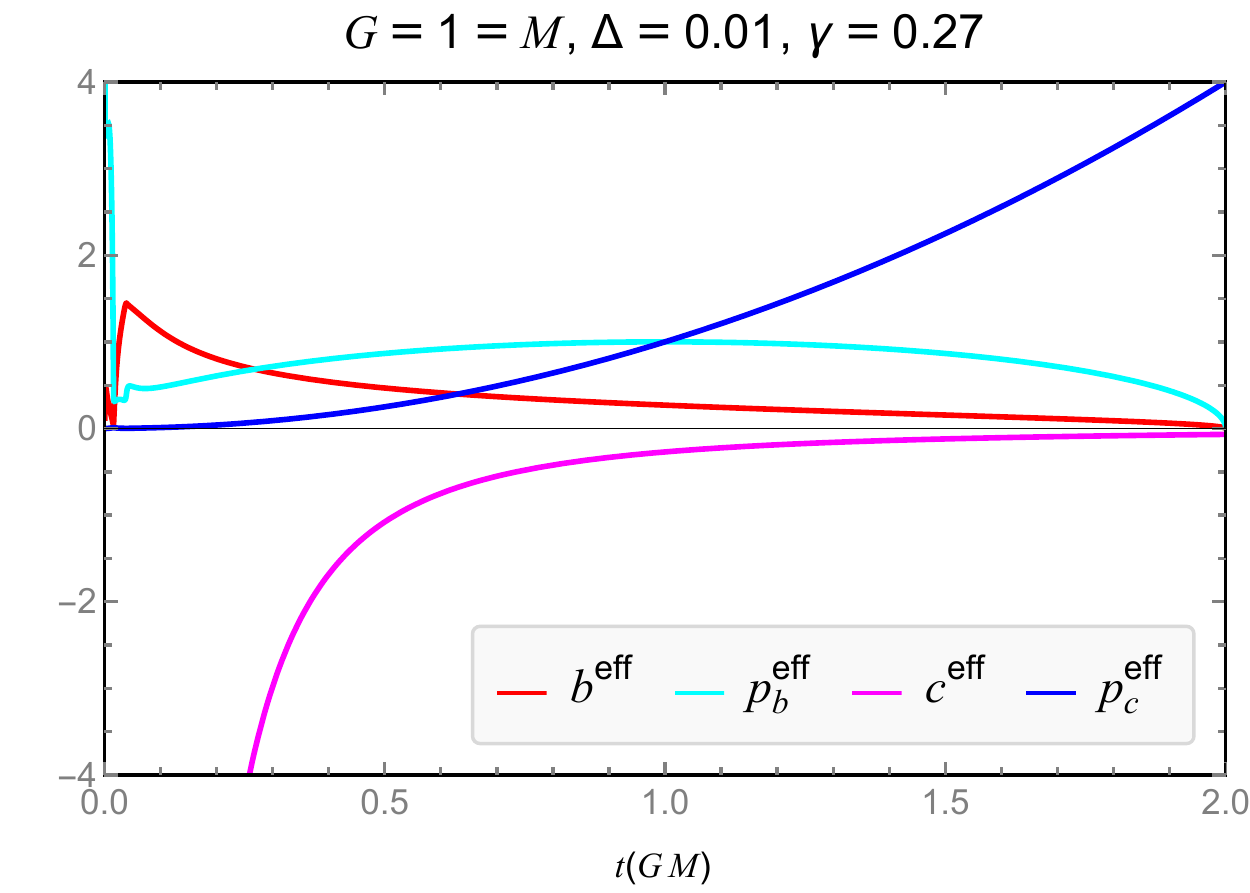}
			\par\end{centering}
	}\hfill{}\subfloat[Close up of the left figure close to $t=0$. Although $p_{c}$ behaves
	rather erratically close to $t=0$, it never vanishes in this case
	either.]{\begin{centering}
			\includegraphics[scale=0.51]{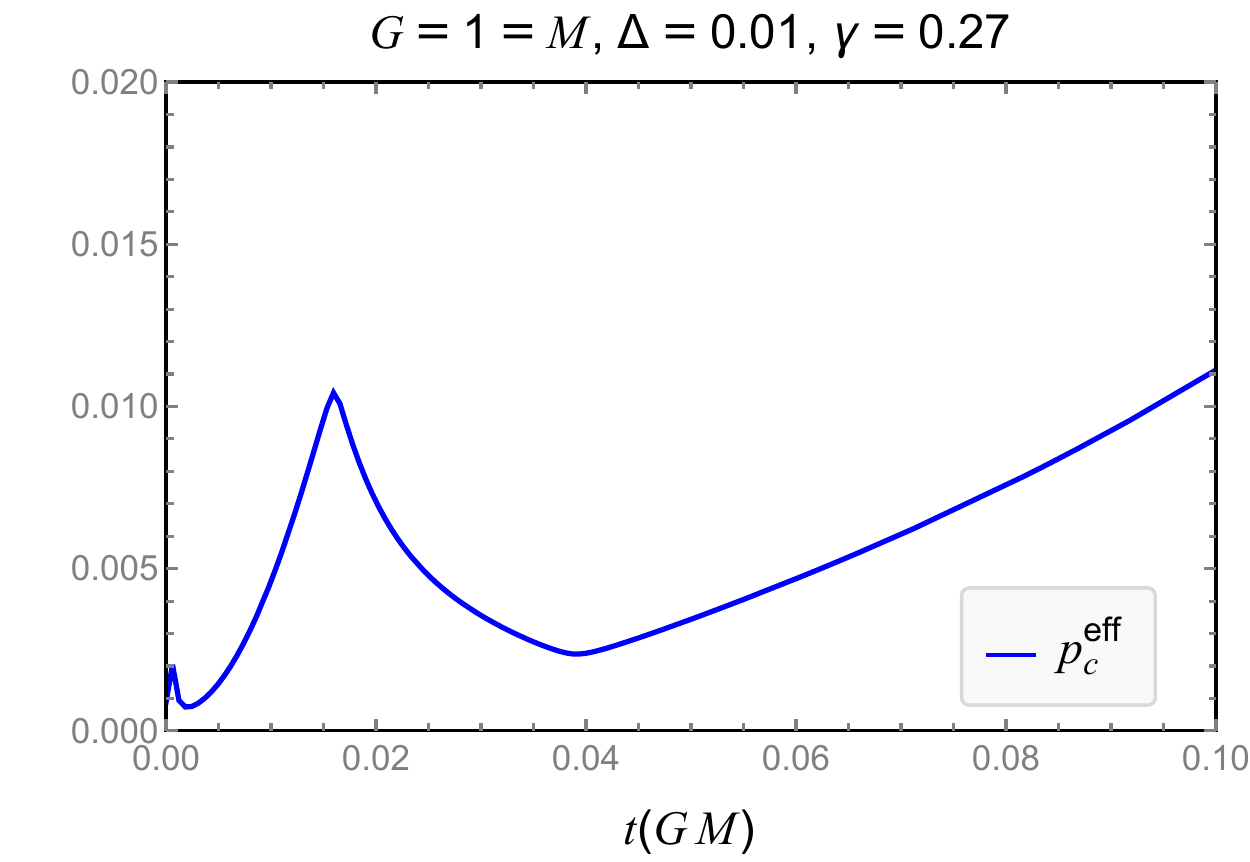}
			\par\end{centering}
	}
	
	\caption{Solutions of the EoM of the $\bar{\mu}$ case \label{fig:Sol-LQG-mu-bar-prime}}
\end{figure}

Here $\bar{\mu}_{b}^{\prime},\,\bar{\mu}_{c}^{\prime}$ have the following
dependence on the triad components, 
\begin{align}
	\bar{\mu}_{b}^{\prime}= & \sqrt{\frac{\Delta}{p_{c}}},\label{eq:mu-bar-b-prime}\\
	\bar{\mu}_{c}^{\prime}= & \frac{\sqrt{p_{c}\Delta}}{p_{b}},\label{eq:mu-bar-c-prime}
\end{align}
and the equations of motion in this case are 
\begin{align}
	\frac{db}{dT}= & -\frac{1}{2}\gamma^{2}\frac{\bar{\mu}_{b}^{\prime}}{\sin\left(\bar{\mu}_{b}^{\prime}b\right)}-\frac{1}{2}\frac{\sin\left(\bar{\mu}_{b}^{\prime}b\right)}{\bar{\mu}_{b}^{\prime}}-\frac{p_{c}}{p_{b}}\left[\frac{\sin\left(\bar{\mu}_{c}^{\prime}c\right)}{\bar{\mu}_{c}^{\prime}}+c\cos\left(\bar{\mu}_{c}^{\prime}c\right)\right],\\
	\frac{dp_{b}}{dT}= & \frac{1}{2}p_{b}\cos\left(\bar{\mu}_{b}^{\prime}b\right)\left[1-\gamma^{2}\frac{\bar{\mu}_{b}^{\prime2}}{\sin^{2}\left(\bar{\mu}_{b}^{\prime}b\right)}\right],\\
	\frac{dc}{dT}= & \frac{p_{b}}{2p_{c}}\left[\gamma^{2}\frac{\bar{\mu}_{b}^{\prime}}{\sin\left(\bar{\mu}_{b}^{\prime}b\right)}\left[1-\frac{\bar{\mu}_{b}^{\prime}}{\sin\left(\bar{\mu}_{b}^{\prime}b\right)}b\cos\left(\bar{\mu}_{b}^{\prime}b\right)\right]-\frac{\sin\left(\bar{\mu}_{b}^{\prime}b\right)}{\bar{\mu}_{b}^{\prime}}\right]\nonumber \\
	& +\frac{bp_{b}\cos\left(\bar{\mu}_{b}^{\prime}b\right)}{2p_{c}}-\frac{\sin\left(\bar{\mu}_{c}^{\prime}c\right)}{\bar{\mu}_{c}^{\prime}}-c\cos\left(\bar{\mu}_{c}^{\prime}c\right),\\
	\frac{dp_{c}}{dT}= & 2p_{c}\cos\left(\bar{\mu}_{c}^{\prime}c\right).
\end{align}
These equations can also be solved numerically as before. By demanding
the solutions match the classical ones at the horizon $t\to2GM$,
we obtain the behavior of the canonical variables as depicted in Fig.
\ref{fig:Sol-LQG-mu-bar-prime}.

With the help of these equations and their solutions together with
\eqref{eq:expansion-pbpc-gen-timelike} and \eqref{eq:RE-pbpc-gen-timelike},
we can obtain the following expressions for the expansion scalar and
the Raychaudhuri equation
\begin{align}
	\theta_{(\bar{\mu}^{\prime})}^{\mathrm{TL}}= & \pm\frac{1}{2\gamma\sqrt{p_{c}}}\left\{ 2\frac{\sin\left(\bar{\mu}_{b}^{\prime}b\right)}{\bar{\mu}_{b}^{\prime}}\cos\left(\bar{\mu}_{c}^{\prime}c\right)+\left[\frac{\sin\left(\bar{\mu}_{b}^{\prime}b\right)}{\bar{\mu}_{b}^{\prime}}-\gamma^{2}\frac{\bar{\mu}_{b}^{\prime}}{\sin\left(\bar{\mu}_{b}^{\prime}b\right)}\right]\cos\left(\bar{\mu}_{b}^{\prime}b\right)\right\} ,\\
	\frac{d\theta_{(\bar{\mu}^{\prime})}^{\mathrm{TL}}}{d\tau}= & \frac{1}{8p_{c}}+\frac{\bar{\mu}_{b}^{\prime}\cos\left(\bar{\mu}_{b}^{\prime}b\right)\sin\left(\bar{\mu}_{c}^{\prime}c\right)}{2}\left(\frac{b}{p_{c}}-\frac{c}{2p}\right)-\frac{1}{4p_{c}}\cos\left(2\bar{\mu}_{b}^{\prime}b\right)+\frac{\cos\left(\bar{\mu}_{b}^{\prime}b\right)\cos\left(\bar{\mu}_{c}^{\prime}c\right)}{2p_{c}}\nonumber \\
	& +\frac{c}{2p_{b}\gamma^{2}}\frac{\sin\left(\bar{\mu}_{b}^{\prime}b\right)}{\bar{\mu}_{b}^{\prime}}\left[-\frac{\cos\left(\bar{\mu}_{c}^{\prime}c\right)}{2}\left[1-\cos\left(2\bar{\mu}_{b}^{\prime}b\right)\right]+\cos^{2}\left(\bar{\mu}_{b}^{\prime}b\right)\cos\left(\bar{\mu}_{c}^{\prime}c\right)\right.\nonumber \\
	& \left.+2\cos\left(\bar{\mu}_{b}^{\prime}b\right)\cos^{2}\left(\bar{\mu}_{c}^{\prime}c\right)-\frac{\cos\left(\bar{\mu}_{b}^{\prime}b\right)}{c}\frac{\sin\left(2\bar{\mu}_{c}^{\prime}c\right)}{2\bar{\mu}_{c}^{\prime}}\right]\nonumber \\
	& +\frac{b}{2\gamma^{2}p_{c}}\frac{\sin\left(\bar{\mu}_{b}^{\prime}b\right)}{\bar{\mu}_{b}^{\prime}}\left[\frac{\cos\left(\bar{\mu}_{c}^{\prime}c\right)}{2}\left[1-\cos\left(2\bar{\mu}_{b}^{\prime}b\right)\right]-\cos^{2}\left(\bar{\mu}_{b}^{\prime}b\right)\cos\left(\bar{\mu}_{c}^{\prime}c\right)\right.\nonumber \\
	& \left.-2\cos\left(\bar{\mu}_{b}^{\prime}b\right)\cos^{2}\left(\bar{\mu}_{c}^{\prime}c\right)-\frac{\cos\left(\bar{\mu}_{b}^{\prime}b\right)}{b}\frac{\sin\left(2\bar{\mu}_{c}^{\prime}c\right)}{2\bar{\mu}_{c}^{\prime}}\right]\nonumber \\
	& +\frac{b}{4\gamma^{2}p_{b}}\frac{\sin\left(\bar{\mu}_{c}^{\prime}c\right)}{\bar{\mu}_{c}^{\prime}}\left[-\cos\left(\bar{\mu}_{b}^{\prime}b\right)\left[1-\cos\left(2\bar{\mu}_{b}^{\prime}b\right)\right]\right]\nonumber \\
	& +\frac{cp_{c}}{4\gamma^{2}p_{b}^{2}}\frac{\sin\left(\bar{\mu}_{c}^{\prime}c\right)}{\bar{\mu}_{c}^{\prime}}\left[\cos\left(\bar{\mu}_{b}^{\prime}b\right)\left[1-\cos\left(2\bar{\mu}_{b}^{\prime}b\right)\right]\right]\nonumber \\
	& +\frac{1}{2\gamma^{2}p_{c}}\frac{\sin^{2}\left(\bar{\mu}_{b}^{\prime}b\right)}{\bar{\mu}_{b}^{\prime2}}\left[1-\frac{3\cos\left(2\bar{\mu}_{b}^{\prime}b\right)}{2}-\cos\left(\bar{\mu}_{b}^{\prime}b\right)\cos\left(\bar{\mu}_{c}^{\prime}c\right)\right]\nonumber \\
	& +\frac{1}{2\gamma^{2}p_{b}}\frac{\sin\left(\bar{\mu}_{b}^{\prime}b\right)}{\bar{\mu}_{b}^{\prime}}\frac{\sin\left(\bar{\mu}_{c}^{\prime}c\right)}{\bar{\mu}_{c}^{\prime}}\left[1-\cos^{2}\left(\bar{\mu}_{b}^{\prime}b\right)-\cos\left(2\bar{\mu}_{b}^{\prime}b\right)-\gamma^{2}\bar{\mu}_{b}^{\prime}\bar{\mu}_{c}^{\prime}\frac{p_{b}}{p_{c}}\right]\nonumber \\
	& +\frac{1}{2}\frac{\bar{\mu}_{b}^{\prime}}{\sin\left(\bar{\mu}_{b}^{\prime}b\right)}\left[\left(\frac{c}{p_{b}}-\frac{b}{p_{c}}\right)\cos\left(\bar{\mu}_{c}^{\prime}c\right)-\frac{1}{p_{b}}\frac{\sin\left(\bar{\mu}_{c}^{\prime}c\right)}{\bar{\mu}_{c}^{\prime}}-\frac{\gamma^{2}}{2p_{c}}\frac{\bar{\mu}_{b}^{\prime}}{\sin\left(\bar{\mu}_{b}^{\prime}b\right)}\right].
\end{align}

\begin{figure}
	\subfloat[Left: Classical vs timlike $\theta$ in the $\bar{\mu}^{\prime}$
	scheme as a function of the Schwarzschild time $t$. The effective
	$\theta_{(\bar{\mu}^{\prime})}^{\mathrm{TL}}$ goes to zero as $t\to0$.
	Right: Close up of the left figure close to $t=0$.]{%
		\begin{minipage}[t]{0.45\textwidth}%
			\begin{center}
				\includegraphics[scale=0.5]{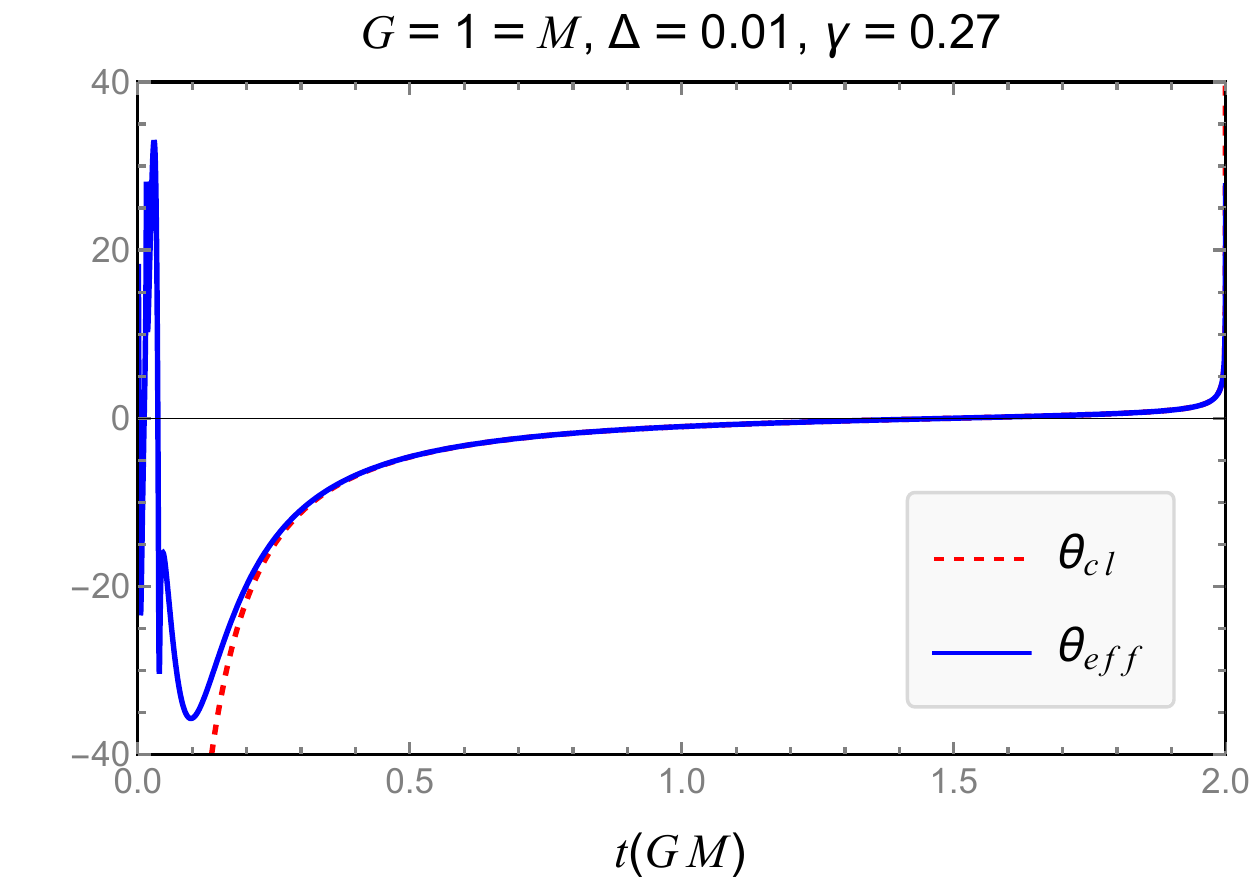}
				\par\end{center}%
		\end{minipage}\hfill{}%
		\begin{minipage}[t]{0.45\textwidth}%
			\begin{center}
				\includegraphics[scale=0.5]{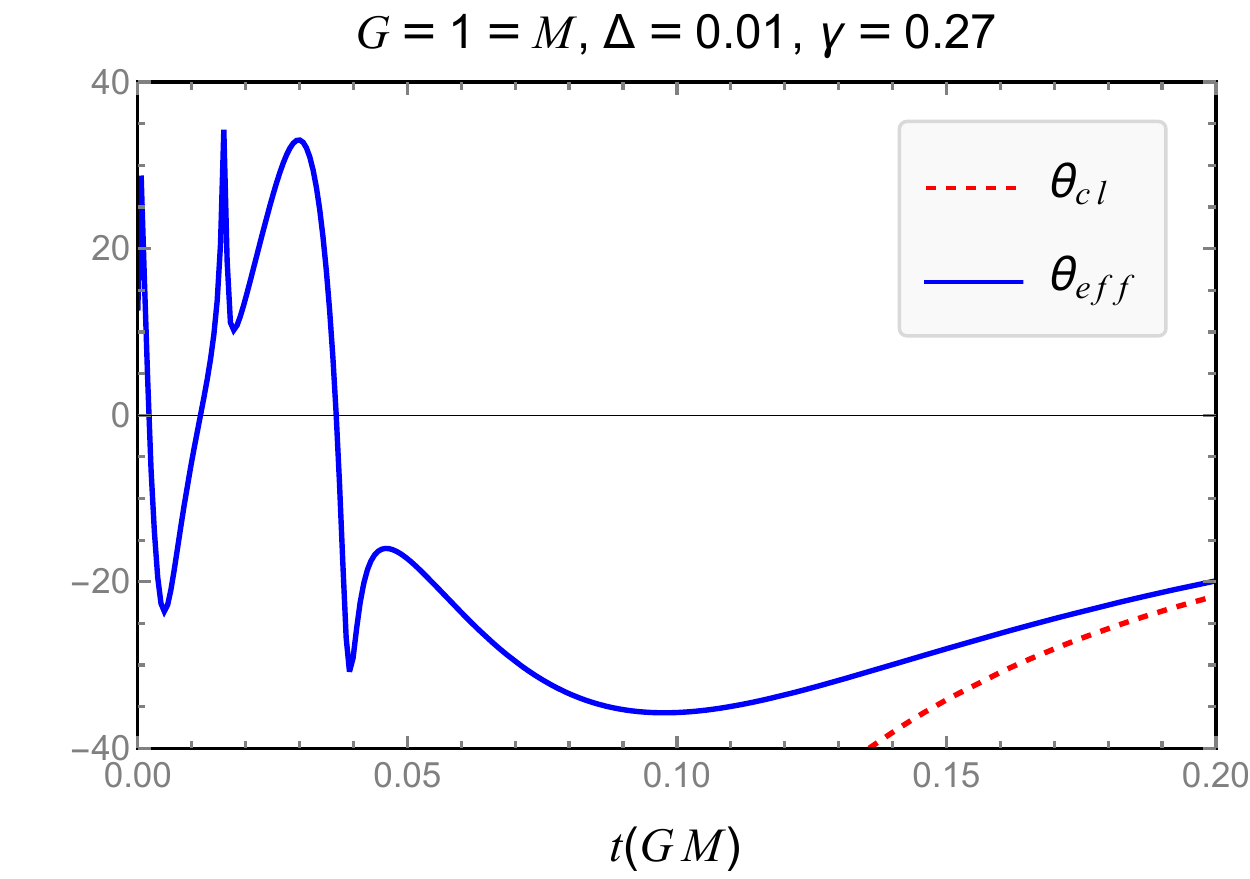}
				\par\end{center}%
		\end{minipage}
		
	}\hfill{}\subfloat[Left: Classical vs timlike $\frac{d\theta}{d\tau}$ in the $\bar{\mu}^{\prime}$
	scheme as a function of the Schwarzschild time $t$. The effective
	$\frac{d\theta_{(\bar{\mu}^{\prime})}^{\mathrm{TL}}}{d\tau}$ goes
	to zero as $t\to0$. Right: Close up of the left figure close
	to $t=0$.]{%
		\begin{minipage}[t]{0.45\textwidth}%
			\begin{center}
				\includegraphics[scale=0.5]{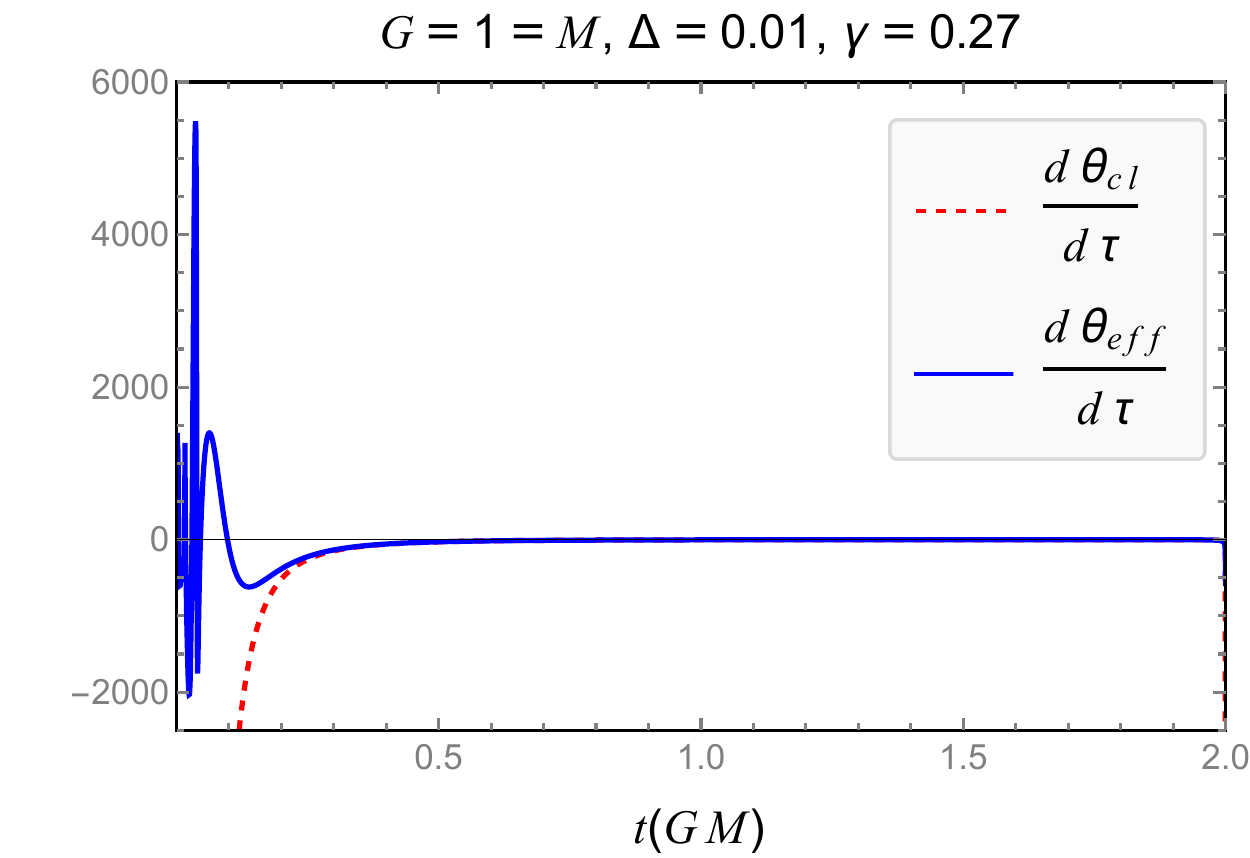}
				\par\end{center}%
		\end{minipage}\hfill{}%
		\begin{minipage}[t]{0.45\textwidth}%
			\begin{center}
				\includegraphics[scale=0.5]{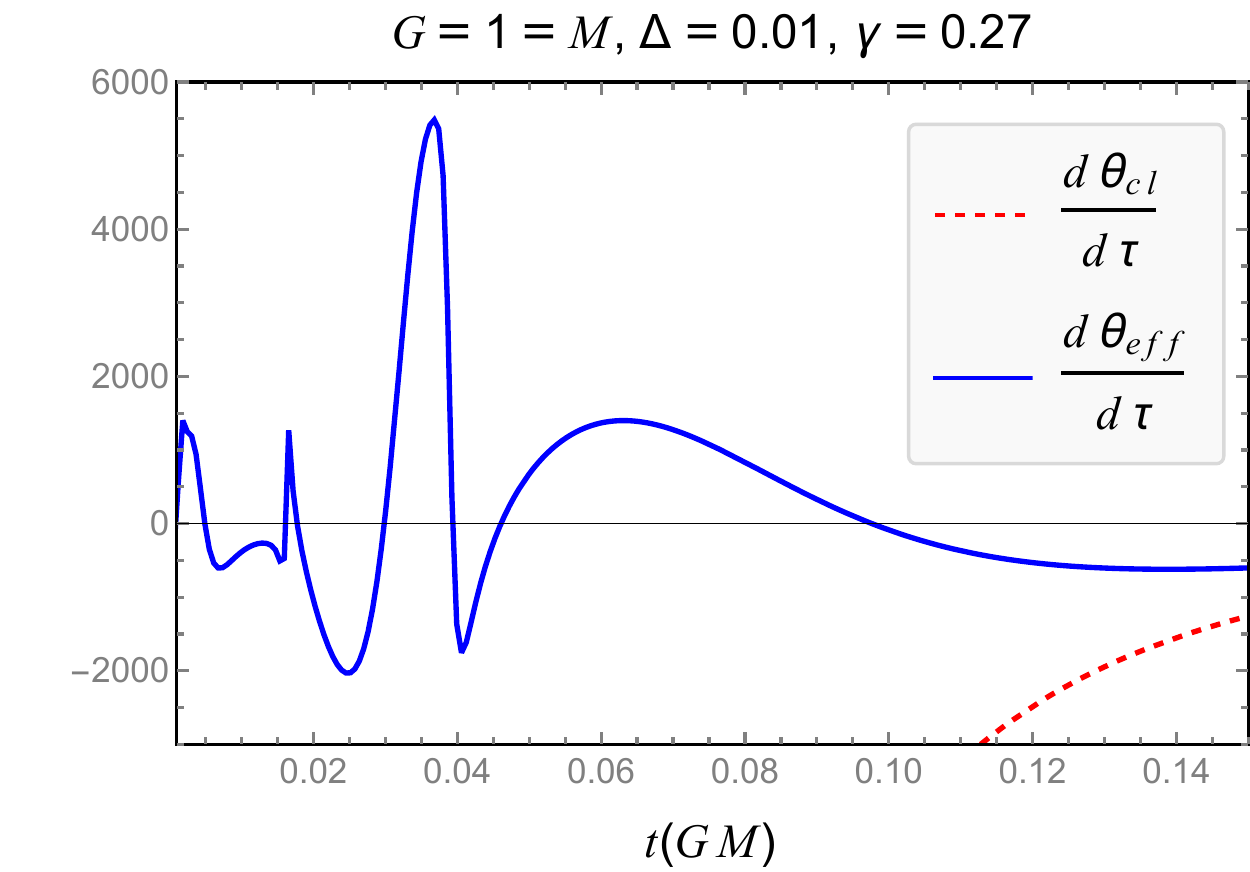}
				\par\end{center}%
		\end{minipage}
		
	}
	
	\caption{$\theta_{(\bar{\mu}^{\prime})}^{\mathrm{TL}}$ and $\frac{d\theta_{(\bar{\mu}^{\prime})}^{\mathrm{TL}}}{d\tau}$\label{fig:theta-RE-TL-mubarprime-vs-classic}}
\end{figure}

Here also the form of $\theta_{(\bar{\mu}^{\prime})}^{\mathrm{TL}}$
is exactly the same as the $\mathring{\mu}$ and $\bar{\mu}$ cases
but with $\mathring{\mu}$ and $\bar{\mu}$ replaced by $\bar{\mu}^{\prime}$.
The perturbative expressions corresponding to these results for $\text{\ensuremath{\Delta}\ensuremath{\to0}}$
are 
\begin{align}
	\theta_{(\bar{\mu}^{\prime})}^{\mathrm{TL}}\approx & -\frac{1}{2\gamma\sqrt{p_{c}}}\left[3b-\frac{\gamma^{2}}{b}-\left(\frac{b^{3}}{p_{c}^{2}}+\frac{bc^{2}p_{c}}{p_{b}^{2}}-\frac{b\gamma^{2}}{3p_{c}^{2}}\right)\Delta\right]+\mathcal{O}\left(\Delta^{2}\right),\\
	\frac{d\theta_{(\bar{\mu}^{\prime})}^{\mathrm{TL}}}{d\tau}\approx & -\frac{1}{2p_{c}}\left\{ \frac{9b^{2}}{2\gamma^{2}}+\frac{\gamma^{2}}{2b^{2}}+\right.\nonumber \\
	& \left.+\Delta\left[\frac{\gamma^{2}}{6p_{c}}-\frac{b^{2}}{3p_{c}}-\frac{11b^{4}}{2\gamma^{2}p_{c}}+\frac{c^{2}p_{c}}{p_{b}^{2}}\left(1-\frac{6b^{2}}{\gamma^{2}}\right)+\frac{c^{3}p_{c}^{2}}{p_{b}^{3}}\left(\frac{b}{\gamma^{2}}+\frac{1}{3b}\right)\right]\right\} \nonumber \\
	& +\mathcal{O}\left(\Delta^{2}\right).
\end{align}
It is not very clear from $\frac{d\theta_{(\bar{\mu}^{\prime})}^{\mathrm{TL}}}{d\tau}$
that whether the effective terms are overall positive or negative
since they contain a mixture of positive and negative terms. However,
due to the behavior of the solutions of the equations of motion, indeed
their overall sign close to $t\to0$ is positive and they almost cancel
out the classical focusing terms. In fact $\theta_{(\bar{\mu}^{\prime})}^{\mathrm{TL}}$
at $t\to0$ is a small positive number (in the figure its value is
$13.6$) while $\frac{d\theta_{(\bar{\mu}^{\prime})}^{\mathrm{TL}}}{d\tau}\to0$
for $t\to0$. These can be better seen in Fig. \ref{fig:theta-RE-TL-mubarprime-vs-classic},
which depicts the full nonperturbative behavior of these terms. 

\begin{figure}
	\subfloat[Left: Classical vs null $\theta$ in the $\bar{\mu}^{\prime}$ scheme
	as a function of the Schwarzschild time $t$. The effective expansion
	$\theta_{(\bar{\mu}^{\prime})}^{\mathrm{NL}}$ goes to zero as $t\to0$.
	Right: Close up of the left figure close to $t=0$.]{%
		\begin{minipage}[t]{0.45\textwidth}%
			\begin{center}
				\includegraphics[scale=0.5]{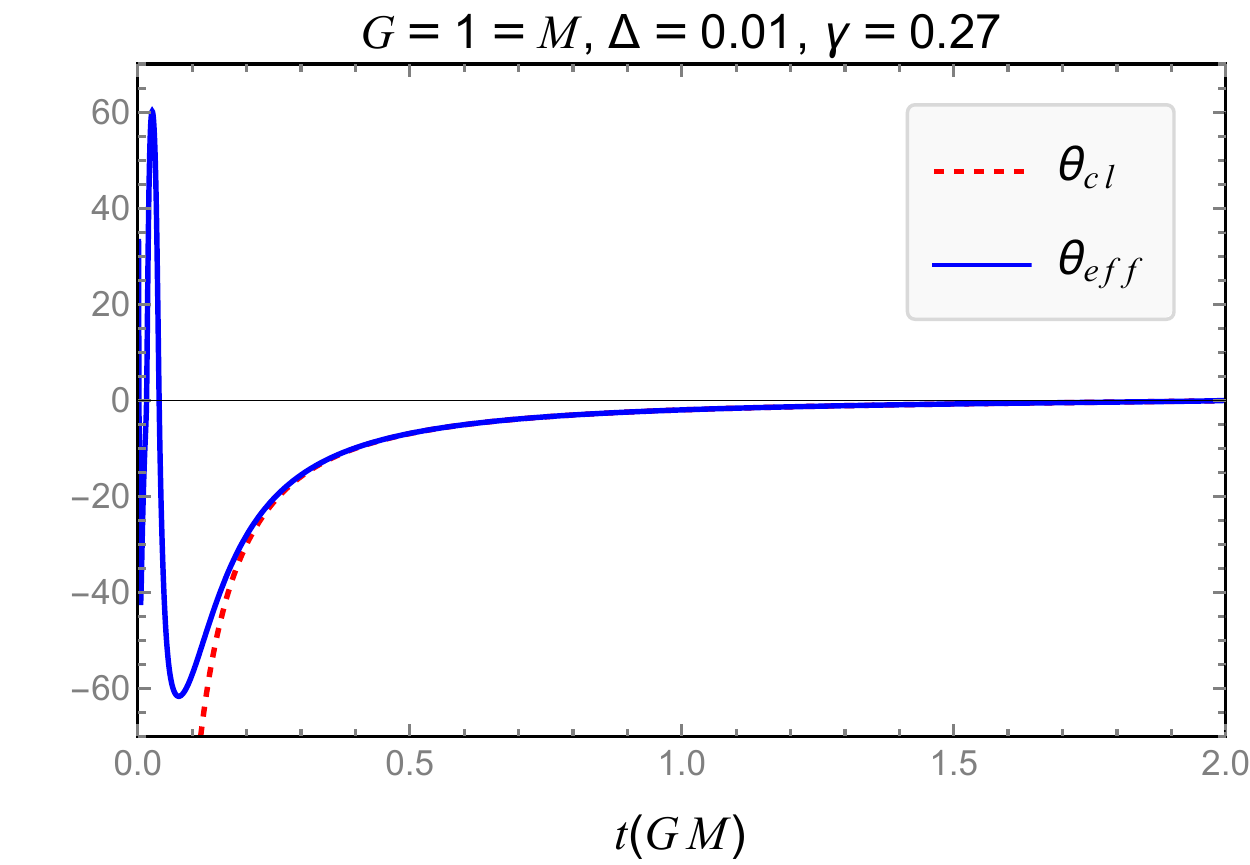}
				\par\end{center}%
		\end{minipage}\hfill{}%
		\begin{minipage}[t]{0.45\textwidth}%
			\begin{center}
				\includegraphics[scale=0.5]{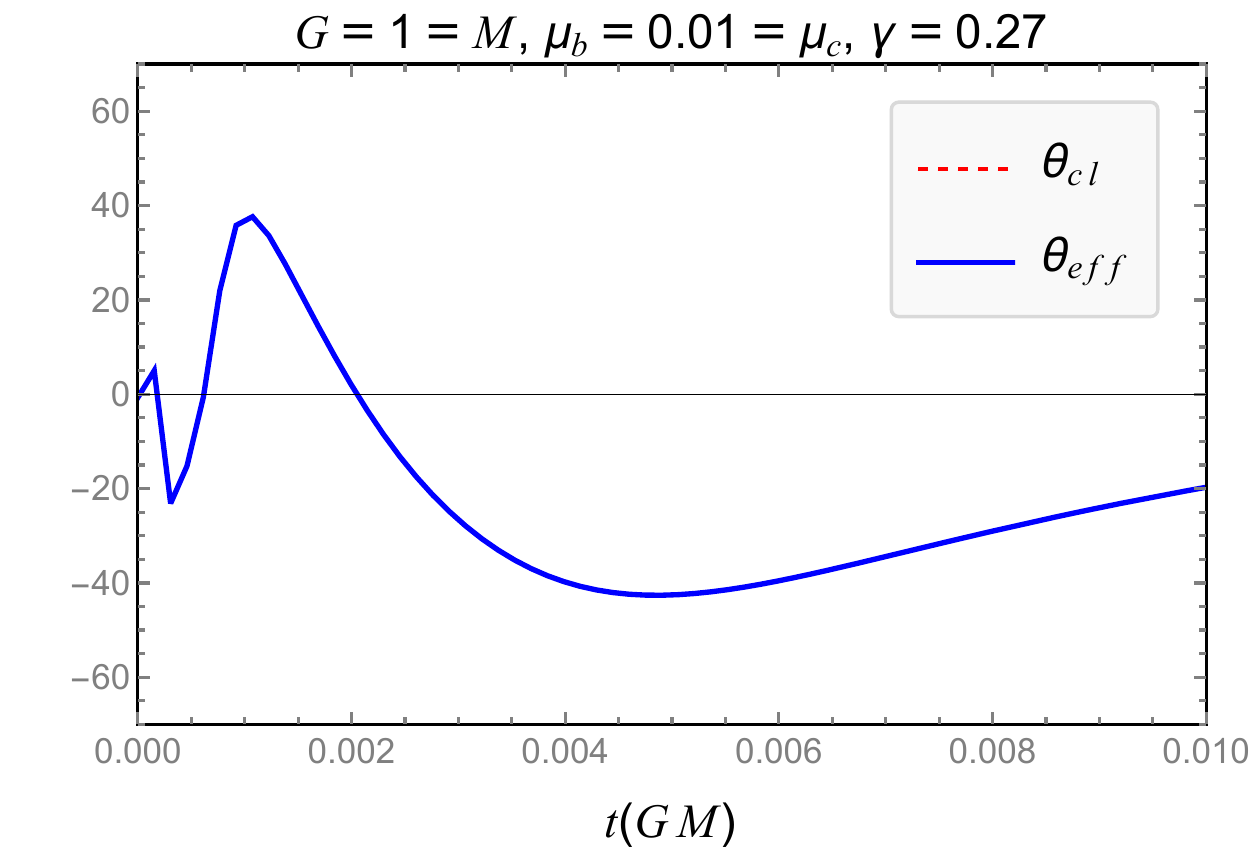}
				\par\end{center}%
		\end{minipage}
		
	}\hfill{}\subfloat[Left: Classical vs null $\frac{d\theta}{d\lambda}$ in the $\bar{\mu}^{\prime}$
	scheme as a function of the Schwarzschild time $t$. The effective
	$\frac{d\theta_{(\bar{\mu}^{\prime})}^{\mathrm{NL}}}{d\lambda}$ goes
	to zero as $t\to0$. Right: Close up of the left figure close
	to $t=0$.]{%
		\begin{minipage}[t]{0.45\textwidth}%
			\begin{center}
				\includegraphics[scale=0.5]{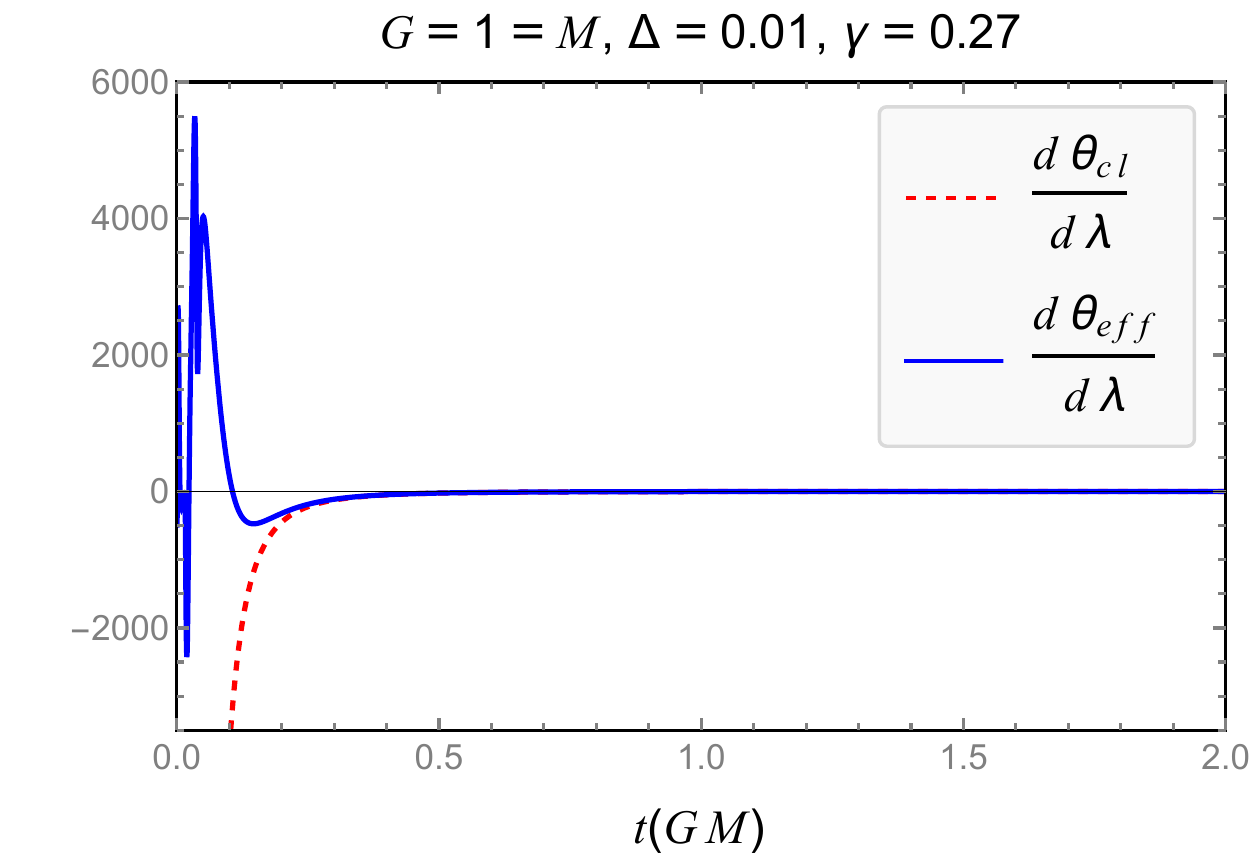}
				\par\end{center}%
		\end{minipage}\hfill{}%
		\begin{minipage}[t]{0.45\textwidth}%
			\begin{center}
				\includegraphics[scale=0.5]{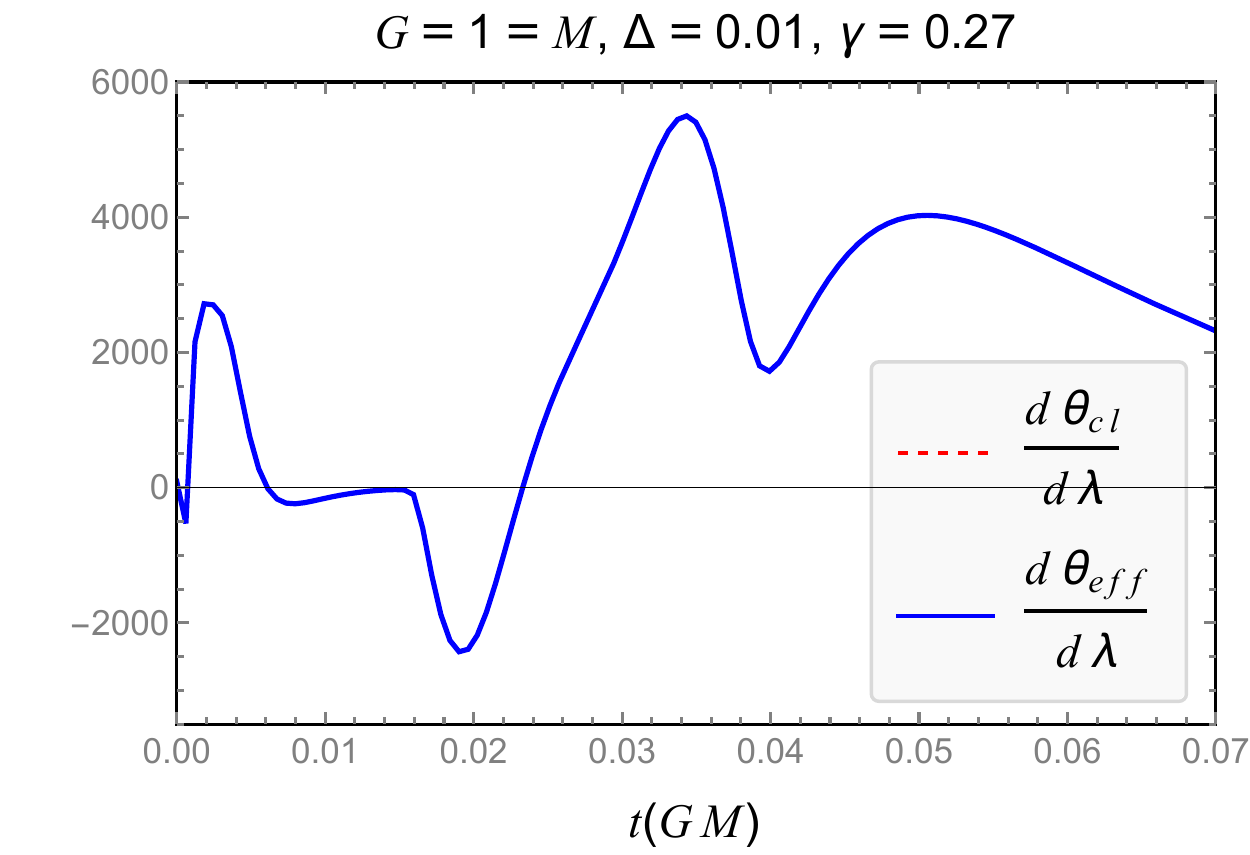}
				\par\end{center}%
		\end{minipage}
		
	}
	
	\caption{$\theta_{(\bar{\mu}^{\prime})}^{\mathrm{NL}}$ and $\frac{d\theta_{(\bar{\mu}^{\prime})}^{\mathrm{NL}}}{d\lambda}$\label{fig:theta-RE-NL-mubarprime-vs-classic}}
\end{figure}

By replacing our lapse \eqref{eq:laps-eff} and differential equations
of motion for this case into \eqref{eq:expansion-pbpc-gen-null} and
\eqref{eq:RE-pbpc-gen-null}, we obtain the null expansion scalar
and the Raychaudhuri equation as follows
\begin{align}
	\theta_{(\bar{\mu}^{\prime})}^{\mathrm{NL}}= & \pm\frac{2}{\gamma\sqrt{p_{c}}}\frac{\sin\left(\bar{\mu}_{b}^{\prime}b\right)}{\bar{\mu}_{b}^{\prime}}\cos\left(\bar{\mu}_{c}^{\prime}c\right)\\
	\frac{d\theta_{(\bar{\mu}^{\prime})}^{\mathrm{NL}}}{d\lambda}= & \frac{\bar{\mu}_{c}^{\prime}\cos\left(\bar{\mu}_{b}^{\prime}b\right)\sin\left(\bar{\mu}_{c}^{\prime}c\right)}{p_{c}^{2}}\left(bp_{b}-cp_{c}\right)\nonumber \\
	& +\frac{2}{\gamma^{2}}\frac{\sin\left(2\bar{\mu}_{b}^{\prime}b\right)}{2\bar{\mu}_{b}^{\prime}}\cos^{2}\left(\bar{\mu}_{c}^{\prime}c\right)\left[\frac{c}{p_{b}}-\frac{b}{p_{c}}\right]\nonumber \\
	& +\frac{1}{2}\frac{cp_{c}\cos\left(\bar{\mu}_{b}^{\prime}b\right)}{\gamma^{2}p_{b}^{2}}\frac{\sin\left(\bar{\mu}_{c}^{\prime}c\right)}{\bar{\mu}_{c}^{\prime}}\left[1-\cos\left(\bar{\mu}_{b}^{\prime}b\right)\right]\nonumber \\
	& +\frac{1}{2\gamma^{2}p_{b}}\frac{\sin\left(\bar{\mu}_{b}^{\prime}b\right)}{\bar{\mu}_{b}^{\prime}}\frac{\sin\left(\bar{\mu}_{c}^{\prime}c\right)}{\bar{\mu}_{c}^{\prime}}\left[1-\cos\left(\bar{\mu}_{b}^{\prime}b\right)\right]\nonumber \\
	& +\frac{1}{\gamma^{2}p_{c}}\frac{\sin^{2}\left(\bar{\mu}_{b}^{\prime}b\right)}{\bar{\mu}_{b}^{\prime2}}\left[1-\cos\left(\bar{\mu}_{c}^{\prime}c\right)\left(1+2\cos\left(\bar{\mu}_{b}^{\prime}b\right)-2\cos\left(\bar{\mu}_{c}^{\prime}c\right)\right)\right]\nonumber \\
	& -\frac{\sin\left(2\bar{\mu}_{b}^{\prime}b\right)}{2\bar{\mu}_{b}^{\prime}\gamma^{2}p_{b}}\left[b\bar{\mu}_{b}^{\prime}\sin\left(\bar{\mu}_{b}^{\prime}b\right)\frac{\sin\left(\bar{\mu}_{c}^{\prime}c\right)}{\bar{\mu}_{c}^{\prime}}+\frac{\sin\left(2\bar{\mu}_{c}^{\prime}c\right)}{\bar{\mu}_{c}^{\prime}}\right]\nonumber \\
	& -\frac{\sin\left(\bar{\mu}_{b}^{\prime}b\right)\sin\left(\bar{\mu}_{c}^{\prime}c\right)}{p_{c}}.
\end{align}
which are a bit simpler expressions compared to the timelike case.
These have perturbative forms (negative branch for the expansion scalar)
\begin{align}
	\theta_{(\bar{\mu}^{\prime})}^{\mathrm{NL}}\approx & -\frac{1}{\gamma\sqrt{p_{c}}}\left[2b-\left(\frac{b^{3}}{3p_{c}}+\frac{bc^{2}p_{c}}{p_{b}^{2}}\right)\Delta\right]+\mathcal{O}\left(\Delta^{2}\right),\\
	\frac{d\theta_{(\bar{\mu}^{\prime})}^{\mathrm{NL}}}{d\lambda}\approx & -\frac{2b^{2}}{\gamma^{2}p_{c}}+\left(\frac{7b^{4}}{3\gamma^{2}p_{c}^{2}}+\frac{4b^{2}c^{2}}{\gamma^{2}p_{b}^{2}}-\frac{2bc^{3}p_{c}}{3\gamma^{2}p_{b}^{3}}-\frac{c^{2}}{p_{b}^{2}}\right)\Delta+\mathcal{O}\left(\Delta^{2}\right).
\end{align}
For $\theta_{(\bar{\mu}^{\prime})}^{\mathrm{NL}}$ in this approximation,
it is clear that the effective terms are positive, and in fact they
become significant near $t=0$. In case of $\frac{d\theta_{(\bar{\mu}^{\prime})}^{\mathrm{NL}}}{d\lambda}$,
it is not quite clear whether the combination of the effective terms
is positive or negative, but due to the behavior of the solutions
to the equations of motion, these turn out to be positive close to
$t=0$ and become quite significant there. The full nonperturbative
behavior is plotted in Fig. \ref{fig:theta-RE-NL-mubarprime-vs-classic}
and it is seen that both $\theta_{(\bar{\mu}^{\prime})}^{\mathrm{NL}}$
and $\frac{d\theta_{(\bar{\mu}^{\prime})}^{\mathrm{NL}}}{d\lambda}$
go to zero as $t\to0$.

\begin{figure}
	\subfloat[The Kretschmann scalar $K$ for the $\bar{\mu}^{\prime}$ case as
	a function of the Schwarzschild time $t$.]{\begin{centering}
			\includegraphics[scale=0.51]{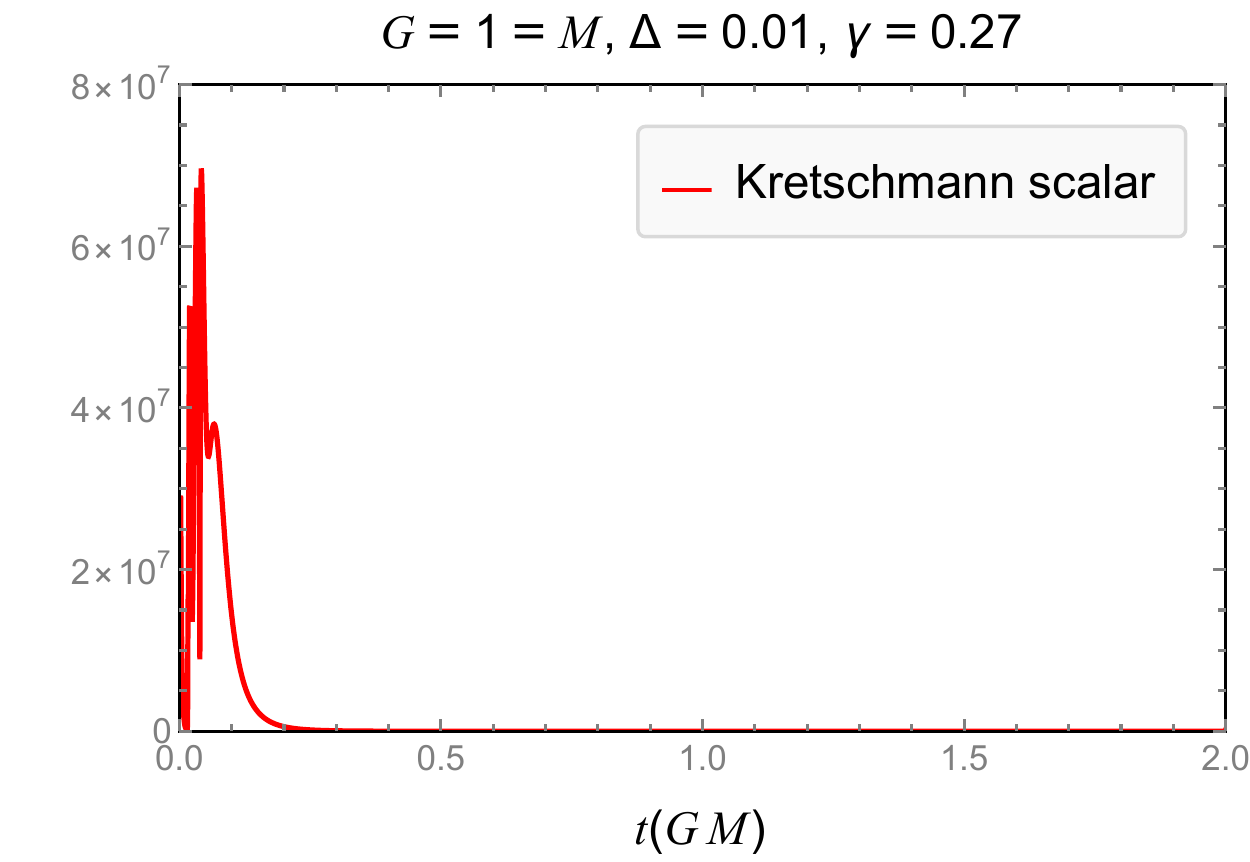}
			\par\end{centering}
	}\hfill{}\subfloat[Close up of $K$ close to $t=0$. It is seen that $K$ remains finite
	everywhere in the interior and vanishes for $t=0$.]{\begin{centering}
			\includegraphics[scale=0.51]{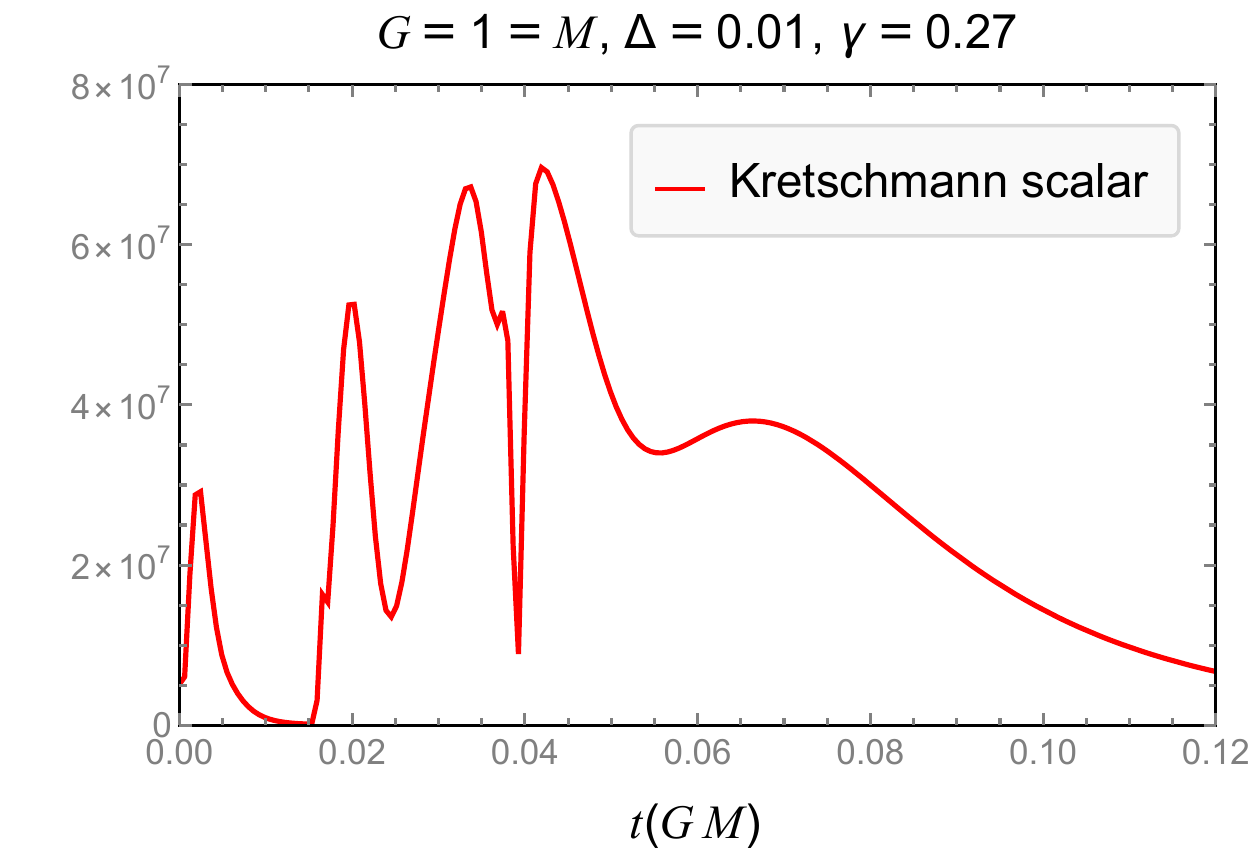}
			\par\end{centering}
	}
	
	\caption{$K$ in the $\bar{\mu}^{\prime}$ case \label{fig:K-LQG-mu-bar-prime}}
\end{figure}

The Kretschmann scalar always in this remains finite just as the two
previous cases. Due to the erratic behavior of the solutions to the
equations of motion, $K$ starts oscillating close to $t=0$, but
nevertheless it always remains finite. Let us first check the perturbative
expression of $K$ up to $\Delta$,
\begin{align}
	K= & \frac{1}{p_{c}^{2}}\left[12+\frac{24b^{2}}{\gamma^{2}}+\frac{12b^{4}}{\gamma^{4}}\right.\nonumber \\
	& -\Delta\left(\frac{2b^{6}}{\gamma^{4}p_{c}}+\frac{52b^{4}p_{c}c^{2}}{\gamma^{4}p_{b}^{2}}+\frac{26b^{4}}{3\gamma^{2}p_{c}}+\frac{56p_{c}b^{2}c^{2}}{\gamma^{2}p_{b}^{2}}+\frac{26b^{2}}{3p_{c}}\right.\nonumber \\
	& \left.\left.+\frac{4p_{c}c^{3}}{3bp_{b}^{3}}+\frac{4p_{c}c^{2}}{p_{b}^{2}}+\frac{2\gamma^{2}}{p_{c}}-\frac{4b^{3}p_{c}^{2}c^{3}}{\gamma^{4}p_{b}^{3}}-\frac{8p_{c}bc^{3}}{3\gamma^{2}p_{b}^{3}}\right)\right]\nonumber \\
	& +\mathcal{O}\left(\Delta^{2}\right).
\end{align}
The correction terms in this case are not all negative, although most
of them are. It turns out that the full $K$ behaves in the desired
way. It stays finite everywhere and goes to zero for $t=0$. This
can be seen from Fig. \ref{fig:K-LQG-mu-bar-prime}. 

\subsection{Generalized uncertainty principle\label{subsec:Generalized-uncertainty-principle}}

In the case of the generalized uncertainty principle (GUP), the standard
Heisenberg algebra of the system is effectively modified. Inspired
by the above, and the fact that a corrected quantum algebra also implies
suitable modifications of the corresponding Poisson algebra, one can
assume that the fundamental Poisson brackets between the canonical
variables also get modified as

\begin{align}
	\{b,p_{b}\} & =G\gamma F_{1}\left(b,c,p_{b},p_{c},\beta_{b},\beta_{c}\right),\label{eq:GUP-Gen-Modif-PB-1}\\
	\{c,p_{c}\} & =2G\gamma F_{2}\left(b,c,p_{b},p_{c},\beta_{b},\beta_{c}\right),\label{eq:GUP-Gen-Modif-PB-2}
\end{align}
where the modifications are encoded entirely in $F_{1}$ and $F_{2}$,
and hence the non-deformed classical limit is obtained by setting
$F_{1}=1=F_{2}$. Here, $\beta_{i}$ refer to the parameters by which
the algebra is modified. Such modification will result in the effective
equations of motion 
\begin{align}
	\frac{db}{dT} & =\{b,H\}=-\frac{1}{2}\left(b+\frac{\gamma^{2}}{b}\right)F_{1},\label{eq:EoM-gen-b}\\
	\frac{dp_{b}}{dT} & =\{p_{b},H\}=\frac{p_{b}}{2}\left(1-\frac{\gamma^{2}}{b^{2}}\right)F_{1},\label{eq:EoM-gen-pb}\\
	\frac{dc}{dT} & =\{c,H\}=-2cF_{2},\label{eq:EoM-gen-c}\\
	\frac{dp_{c}}{dT} & =\{p_{c},H\}=2p_{c}F_{2}.\label{eq:EoM-gen-pc}
\end{align}
which should also be supplemented by weakly vanishing of the Hamiltonian
constraint \eqref{eq:H-class-1}. Notice that in this approach the
Hamiltonian does not get modified and the effective modifications
to the equations of motion come from the modifications to the Poisson
algebra.

\begin{table}
	\noindent \begin{centering}
		\begin{tabular}{|c|c|c|}
			\hline 
			Model & Dependence of GUP modifications on & Expansion and RE finite for\tabularnewline
			\hline 
			1 & Configuration & $\beta_{b}<0$ and $\beta_{c}<0$\tabularnewline
			\hline 
			2 & Momenta & No values of $\alpha_{b}^{\prime}$ and $\alpha_{c}^{\prime}$\tabularnewline
			\hline 
			3 & Configuration & $\alpha_{b}<0$ and $\alpha_{c}>0$\tabularnewline
			\hline 
			4 & Momenta & No values of $\beta_{b}^{\prime}$ and $\beta_{c}^{\prime}$\tabularnewline
			\hline 
		\end{tabular}
		\par\end{centering}
	\caption{Comparison of GUP models with regard to the possibility of singularity
		resolution.\label{tab:Comparison-of-GUP-Models}}
\end{table}

The above equations of motion \eqref{eq:EoM-gen-b}--\eqref{eq:EoM-gen-pc}
can now be substituted into the timelike expansion \eqref{eq:expansion-pbpc-gen-timelike}
and Raychaudhuri equation \eqref{eq:RE-pbpc-gen-timelike} to yield
(with $N=\frac{\gamma\sqrt{p_{c}}}{b}$ as before): 
\begin{equation}
	\theta_{\mathrm{GUP}}^{(\mathrm{TL})}=\pm\frac{1}{2\gamma\sqrt{p_{c}}}\left(bF_{1}-\frac{\gamma^{2}F_{1}}{b}+2bF_{2}\right),\label{eq:theta-GUP-Timelike-factored}
\end{equation}
and
\begin{equation}
	\frac{d\theta_{\mathrm{GUP}}^{(\mathrm{TL})}}{d\tau}=\frac{1}{2\gamma^{2}p_{c}}\left[\left(b^{2}-\gamma^{2}\right)\dot{F}_{1}+2b^{2}\dot{F}_{2}-F_{1}^{2}\left(\frac{b^{2}}{2}+\frac{\gamma^{4}}{2b^{2}}+\gamma^{2}\right)-2b^{2}F_{1}F_{2}-2b^{2}F_{2}^{2}\right]\label{eq:RE-GUP-Timelike}
\end{equation}
in terms of the canonical variables. For the null case, using \eqref{eq:expansion-pbpc-gen-null}
and \eqref{eq:RE-pbpc-gen-null} we get
\begin{align}
	\theta_{\mathrm{GUP}}^{(\mathrm{NL})}= & \frac{2bF_{2}}{\gamma\sqrt{p_{c}}},\label{eq:theta-GUP-Null}\\
	\frac{d\theta_{\mathrm{GUP}}^{(\mathrm{NL})}}{d\lambda}= & \frac{2b^{2}}{\gamma^{2}p_{c}}\left(\dot{F}_{2}-F_{1}F_{2}\right).\label{eq:RE-GUP-Null}
\end{align}
As can be seen from \eqref{eq:GUP-Gen-Modif-PB-1} and \eqref{eq:GUP-Gen-Modif-PB-2},
the modifications to the Poisson algebra is controlled by functions
$F_{1}$ and $F_{2}$. A generic class of modifications, containing
terms up to second order in canonical variables and with no cross
terms, can be written as
\begin{equation}
	F_{i}=1+\alpha_{b}^{(i)}b+\alpha_{c}^{(i)}c+\beta_{b}^{(i)}b^{2}+\beta_{c}^{(i)}c^{2}+\alpha_{b}^{\prime(i)}p_{b}+\alpha_{c}^{\prime(i)}p_{c}+\beta_{b}^{\prime(i)}p_{b}^{2}+\beta_{c}^{\prime(i)}p_{c}^{2},
\end{equation}
where $\alpha_{l},\,\beta_{l},\,\alpha_{l}^{\prime},\,\beta_{l}^{\prime}$
parameters (with $l=b,\,c$) encode the quantum gravity effects associated
to the noncommutativity of the model, and $i=1,2$. We will consider
the four most common cases appearing in the literature,
\begin{align}
	\mathrm{Model\,1:} &  & F_{1}= & 1+\beta_{b}b^{2}, & F_{2}= & 1+\beta_{c}c^{2},\label{eq:GUP-Model-1}\\
	\mathrm{Model\,2:} &  & F_{1}= & 1+\alpha_{b}^{\prime}p_{b}, & F_{2}= & 1+\alpha_{c}^{\prime}p_{c},\label{eq:GUP-Model-2}\\
	\mathrm{Model\,3:} &  & F_{1}= & 1+\alpha_{b}b, & F_{2}= & 1+\alpha_{c}c,\label{eq:GUP-Model-3}\\
	\mathrm{Model\,4:} &  & F_{1}= & 1+\beta_{b}^{\prime}p_{b}^{2}, & F_{2}= & 1+\beta_{c}^{\prime}p_{c}^{2}.\label{eq:GUP-Model-4}
\end{align}
In each model, we can consider the parameters $\alpha_{l},\,\beta_{l},\,\alpha_{l}^{\prime},\,\beta_{l}^{\prime}$
to be positive, negative or zero. It turns out that only in configuration-dependent
cases models 1 and 3 and just for certain signs of $\alpha_{l},\,\beta_{l},\,\alpha_{l}^{\prime},\,\beta_{l}^{\prime}$
will both $\theta$ and the Raychaudhuri equation remain finite everywhere
in the interior. In fact, finding the solutions to the equations of
motion \eqref{eq:EoM-gen-b}--\eqref{eq:EoM-gen-pc} by using either
of the momentum-dependent case \eqref{eq:GUP-Model-2} or \eqref{eq:GUP-Model-4}
and replacing the solutions into any of the equations \eqref{eq:theta-GUP-Timelike-factored}--\eqref{eq:RE-GUP-Null},
one can never obtain a finite value for them as $t\to0$ for any value
of $\alpha_{b}^{\prime},\,\alpha_{c}^{\prime},\,\beta_{b}^{\prime},\,\beta_{c}^{\prime}$.
This means that the momentum-dependent cases are not viable for modeling
GUP if we demand the singularity of the black hole model we are using
is resolved. On the other hand configuration-dependent cases do in
fact allow for such a possibility. This is summarized in table \ref{tab:Comparison-of-GUP-Models}.

\begin{figure}
	\subfloat[Solutions to the effective equations of motion for model 1 as a function
	of the Schwarzschild time $t$.]{\begin{centering}
			\includegraphics[scale=0.51]{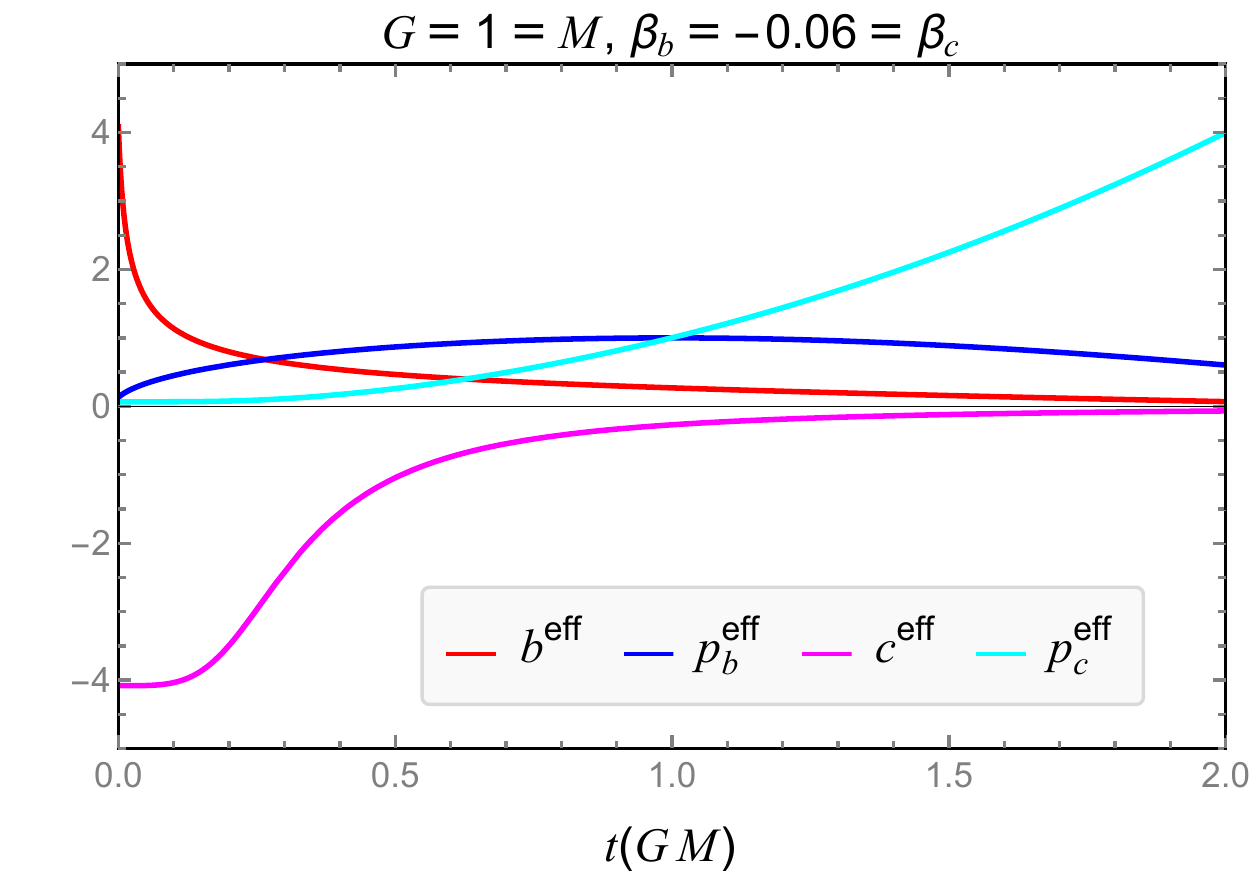}
			\par\end{centering}
	}\hfill{}\subfloat[Close up of the left figure close to $t=0$. Note that none of the
	variables diverge or vanish in the interior, and everywhere in the
	interior $b,p_{b},p_{c}>0$ while $c<0$.]{\begin{centering}
			\includegraphics[scale=0.51]{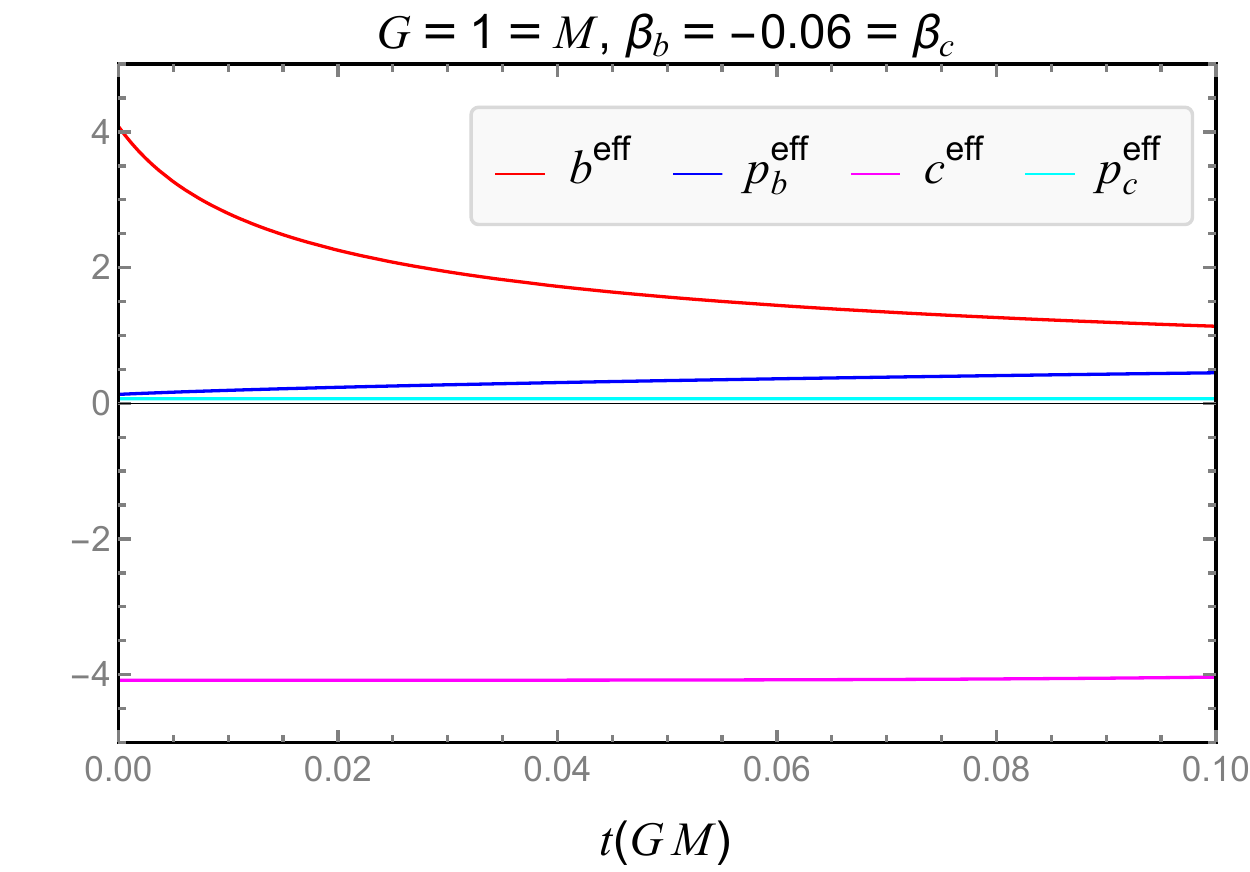}
			\par\end{centering}
	}
	
	\caption{Solutions of the EoM of GUP model 1 \label{fig:GUP-EoM-Model-1}}
\end{figure}

Let us first consider model 1. The solution to the equations of motion
in this model are plotted in Fig. \ref{fig:GUP-EoM-Model-1}. From
this figure we see that none of the variables diverge or vanish in
the interior, and everywhere in the interior $b,p_{b},p_{c}>0$ while
$c<0$. For the chosen values of $\beta_{b},\,\beta_{c}$, for $t\to0$
we obtain $b\approx4$ and $c\approx-4$.

\begin{figure}
	\subfloat[Classical vs timelike $\theta$ in GUP model 1 as a function of the
	Schwarzschild time $t$.]{%
		\noindent\begin{minipage}[t]{1\columnwidth}%
			\begin{center}
				\includegraphics[scale=0.7]{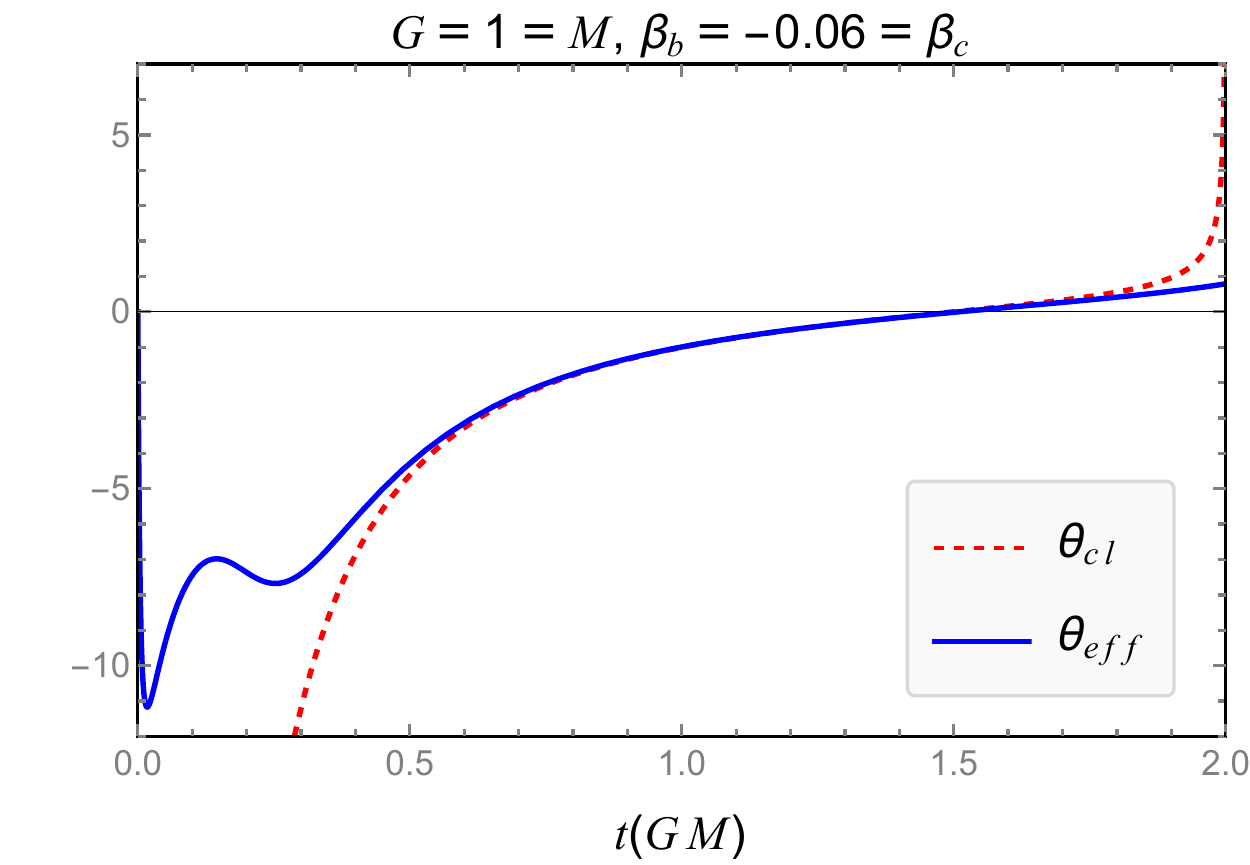}
				\par\end{center}%
		\end{minipage}
		
	}\hfill{}\subfloat[Left: Classical vs timelike $\frac{d\theta}{d\tau}$ in GUP model
	1 as a function of the Schwarzschild time $t$. Both the effective
	effective expansion $\theta_{\mathrm{GUP}(1)}^{(\mathrm{TL})}$ and
	its rate of change $\frac{d\theta_{\mathrm{GUP}(1)}^{(\mathrm{TL})}}{d\tau}$
	go to zero as $t\to0$. Right: Close up of the left figure
	close to $t=0$.]{%
		\begin{minipage}[t]{0.45\textwidth}%
			\begin{center}
				\includegraphics[scale=0.5]{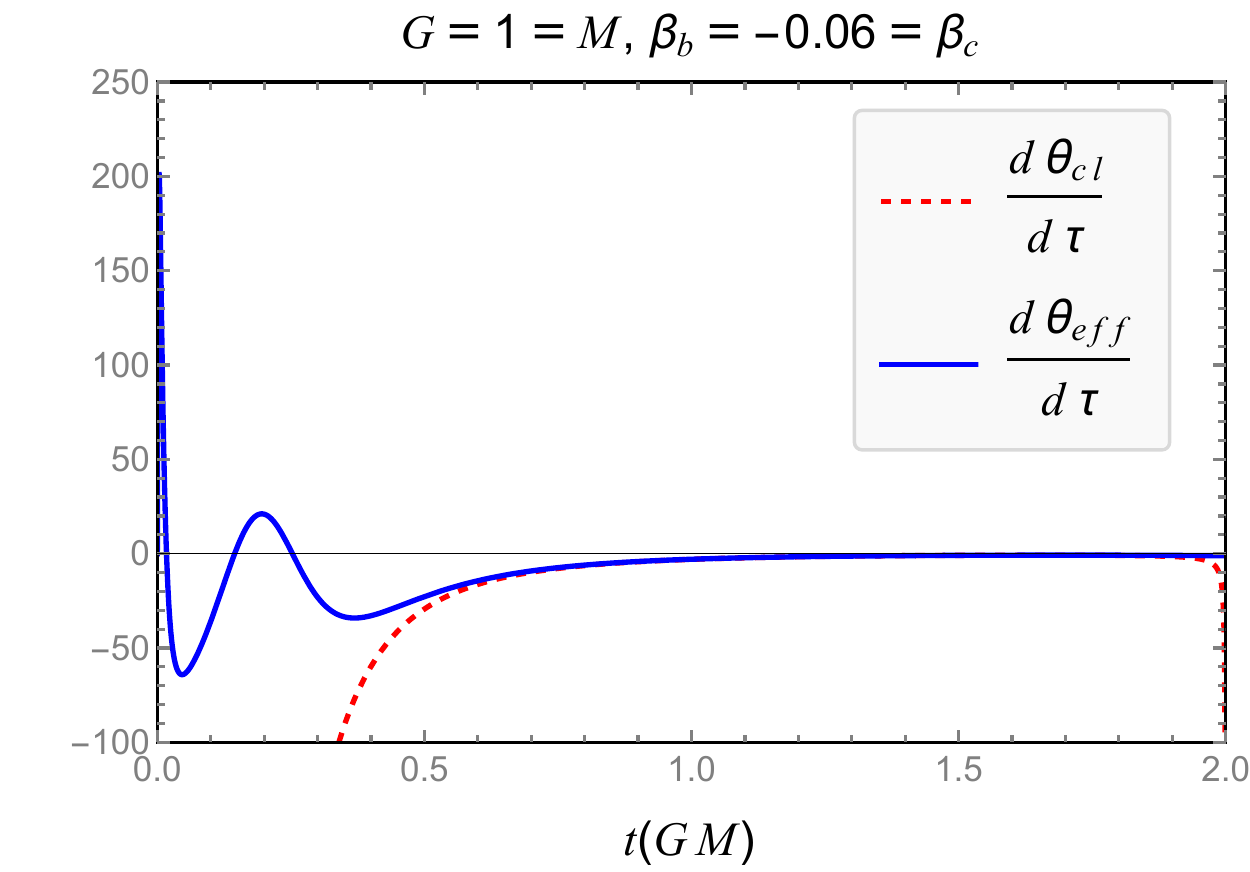}
				\par\end{center}%
		\end{minipage}\hfill{}%
		\begin{minipage}[t]{0.45\textwidth}%
			\begin{center}
				\includegraphics[scale=0.5]{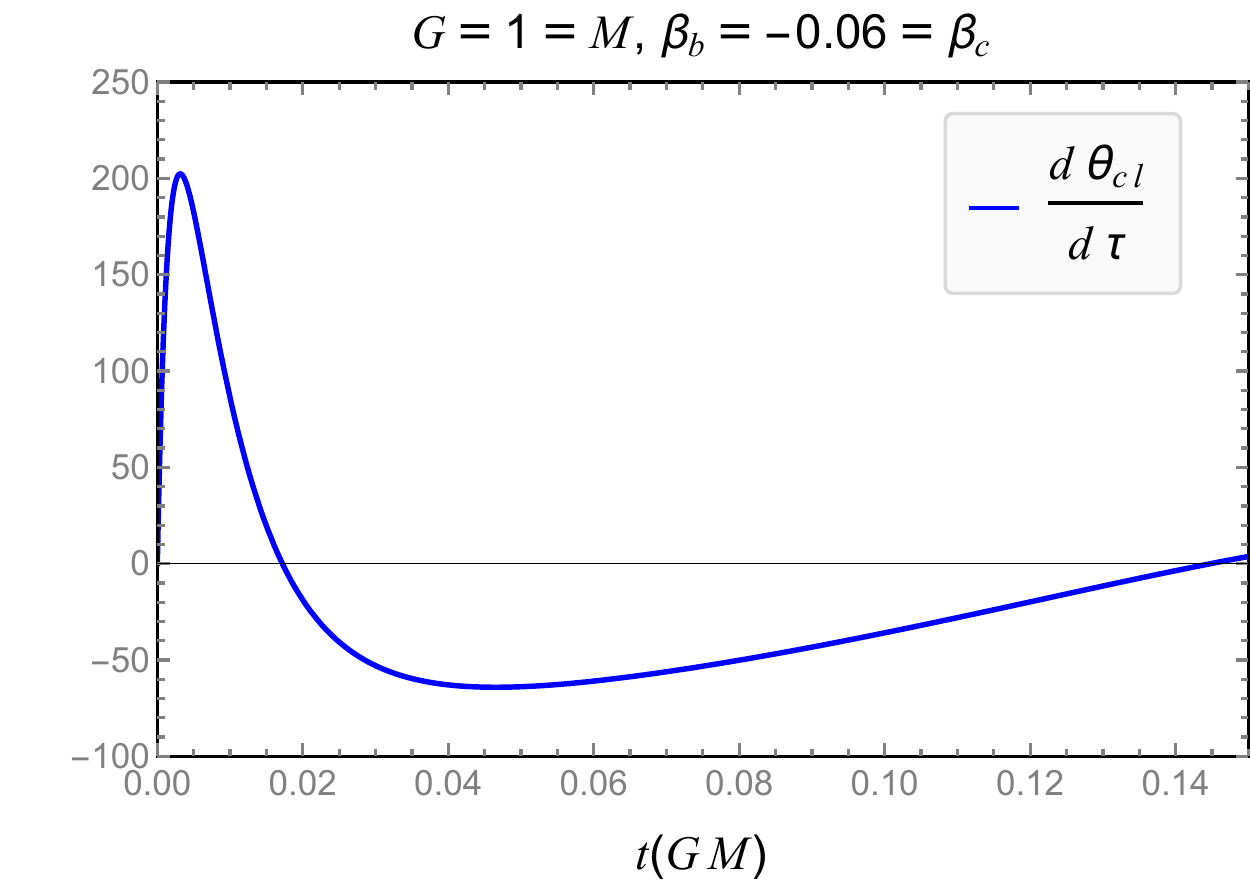}
				\par\end{center}%
		\end{minipage}
		
	}
	
	\caption{$\theta_{\mathrm{GUP}(1)}^{(\mathrm{TL})}$ and $\frac{d\theta_{\mathrm{GUP}(1)}^{(\mathrm{TL})}}{d\tau}$
		\label{fig:theta-RE-TL-GUP-vs-classic-Model-1}}
\end{figure}

For this model we obtain the expressions for the timelike case
\begin{align}
	\theta_{\mathrm{GUP(1)}}^{(\mathrm{TL})}= & \pm\frac{b}{2\gamma\sqrt{p_{c}}}\left[3-\frac{\gamma^{2}}{b^{2}}+2\beta_{c}c^{2}+\beta_{b}\left(b^{2}-\gamma^{2}\right)\right],\label{eq:theta-TL-GUP-1-in-canon-vars}\\
	\frac{d\theta_{\mathrm{GUP}(1)}^{(\mathrm{TL})}}{d\tau}= & \frac{1}{\gamma^{2}p_{c}}\left[-\frac{9b^{2}}{4}-\frac{\gamma^{4}}{4b^{2}}-\frac{\gamma^{2}}{2}\right.\nonumber \\
	& \left.-\beta_{b}b^{2}\left(2b^{2}+\ensuremath{\gamma^{2}}\right)+\frac{\beta_{b}^{2}b^{2}}{4}\left(\gamma^{4}-3b^{4}-2\gamma^{2}b^{2}\right)-\beta_{c}^{2}c^{2}b^{2}\left(5c^{2}+7\right)-\beta_{b}\beta_{c}b^{2}c^{2}\right]\label{eq:RE-TL-GUP-1-in-canon-vars}
\end{align}
Solving the equations of motion \eqref{eq:EoM-gen-b}--\eqref{eq:EoM-gen-pc}
with \eqref{eq:GUP-Model-1} and replacing the solutions in the above
two expressions, we obtain the behavior of $\theta_{\mathrm{GUP}}^{(\mathrm{TL})}$
and $\frac{d\theta_{\mathrm{GUP}}^{(\mathrm{TL})}}{d\tau}$ as a function
of the Schwarzschild time $t$. It turns out that only the case with
both $\beta_{b}<0$ and $\beta_{c}<0$ will yield expressions that
are always finite in the interior. With such choice of $\beta$'s
we can numerically find the expressions for \eqref{eq:theta-TL-GUP-1-in-canon-vars}
and \eqref{eq:RE-TL-GUP-1-in-canon-vars}, which are plotted in Fig.
\ref{fig:theta-RE-TL-GUP-vs-classic-Model-1}. Interestingly, although
the expansion scalar $\theta_{\mathrm{GUP}}^{(\mathrm{TL})}$ dips towrads the negative values
and its rate of change $\frac{d\theta_{\mathrm{GUP}}^{(\mathrm{TL})}}{d\tau}$
peaks towards positive values when we get closer to $t\to 0$, at some points quantum effects take over
and turn them both back towards zero. The qualitative behavior is similar
to the cases in LQG but the difference is that in those cases, the
expansion scalar in some regions actually becomes positive and then
goes to zero, while here the expansion remain negative (after initially
being positive) and it goes to zero from below.

To get an analytical sense of the expressions, we see that for the
negative branch of the expansion scalar, Eq. \eqref{eq:theta-TL-GUP-1-in-canon-vars},
the correction terms are all positive (note that $\beta_{b}<0$ and
$\beta_{c}<0$) and it is these terms that overcome the classical
negative terms close to $t\to0$ and turn the curve around. In the
same way, in Eq. \eqref{eq:RE-TL-GUP-1-in-canon-vars} and up to the first order in $\beta$'s, the correction
term $-\beta_{b}b^{2}\left(2b^{2}+\ensuremath{\gamma^{2}}\right)$
is positive and has the same effect. 

\begin{figure}
	\subfloat[Classical vs null $\theta$ in GUP model 1 as a function of the Schwarzschild
	time $t$.]{\begin{centering}
			\includegraphics[scale=0.51]{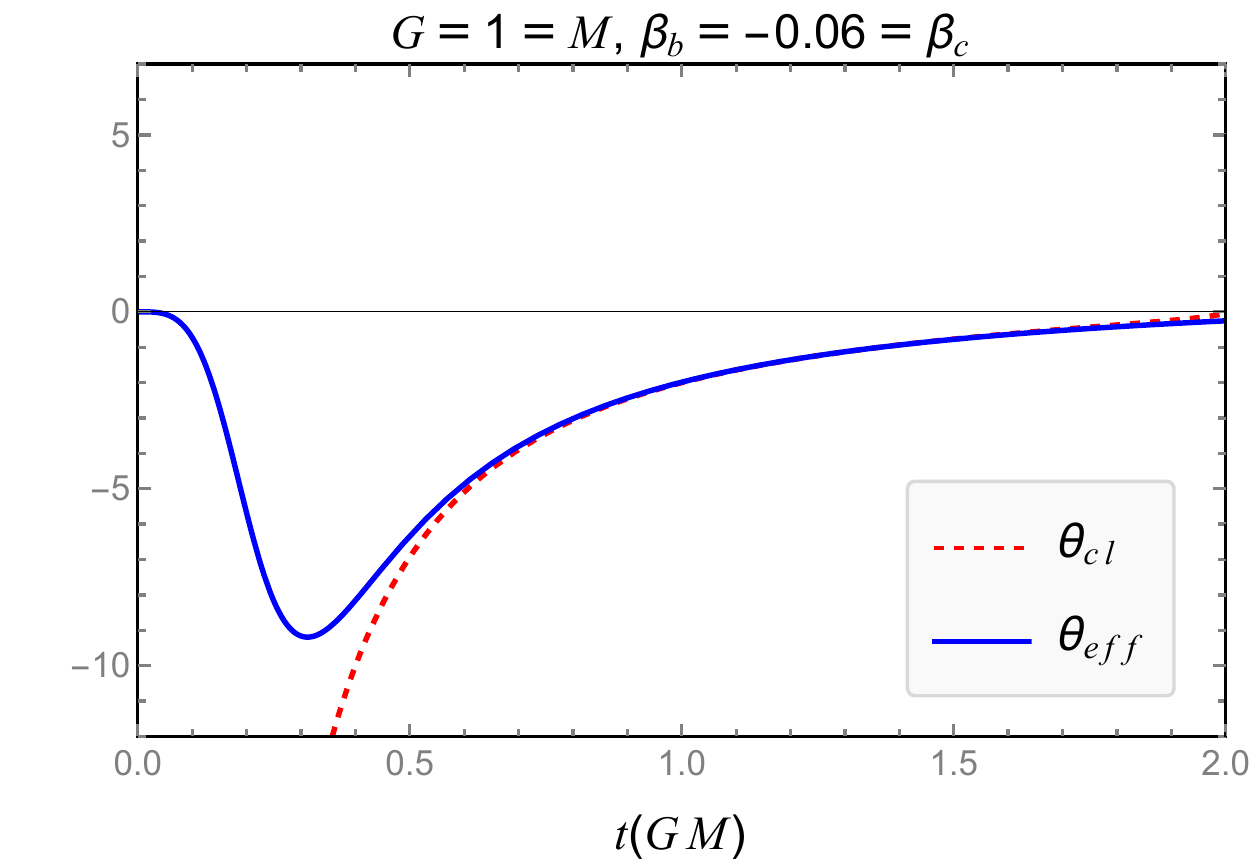}
			\par\end{centering}
	}\hfill{}\subfloat[Classical vs null $\frac{d\theta}{d\tau}$ in GUP model 1 as a function
	of the Schwarzschild time $t$. Both the effective expansion
	$\theta_{\mathrm{GUP}(1)}^{(\mathrm{NL})}$ and its rate of change
	$\frac{d\theta_{\mathrm{GUP}(1)}^{(\mathrm{NL})}}{d\lambda}$ go to
	zero as $t\to0$.]{\begin{centering}
			\includegraphics[scale=0.51]{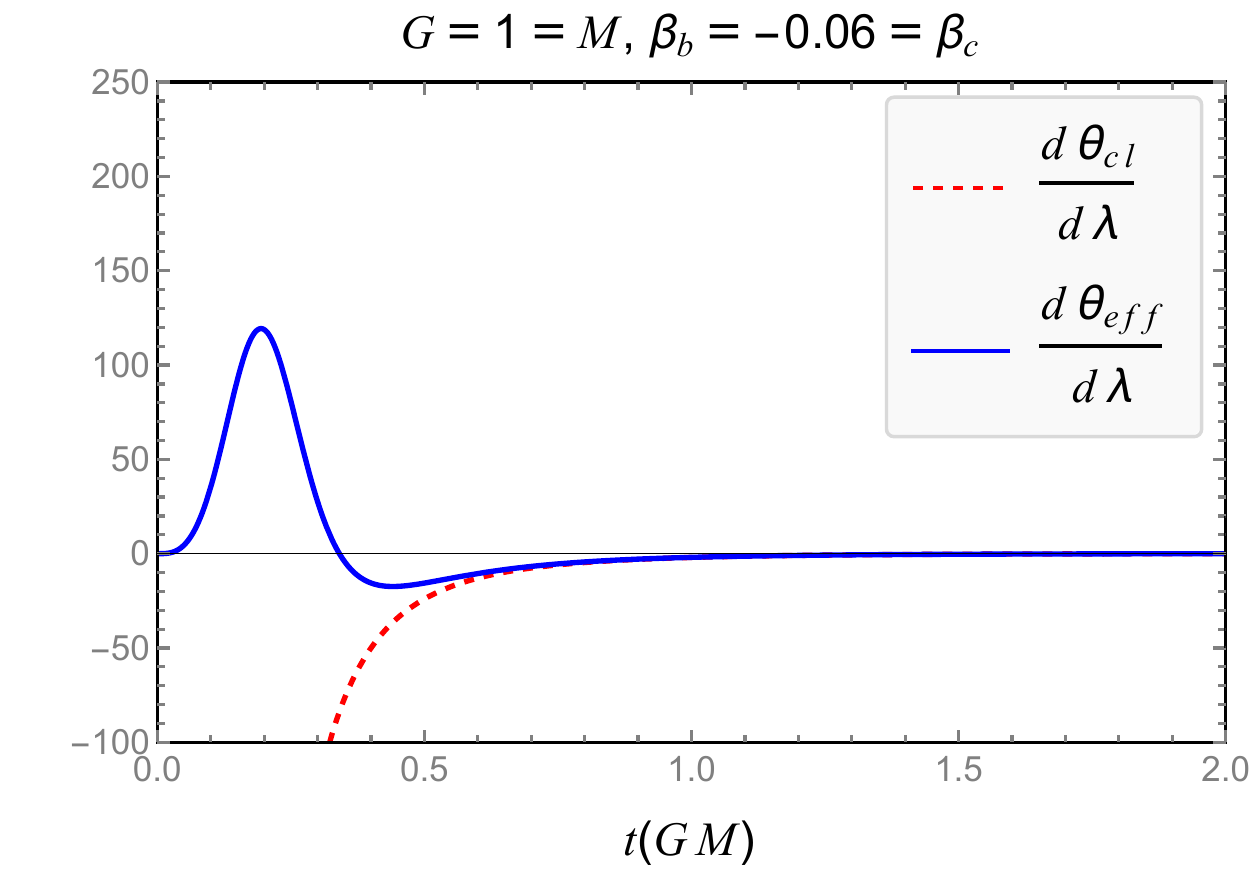}
			\par\end{centering}
	}
	
	\caption{$\theta_{\mathrm{GUP}(1)}^{(\mathrm{NL})}$ and $\frac{d\theta_{\mathrm{GUP}(1)}^{(\mathrm{NL})}}{d\lambda}$
		\label{fig:theta-RE-NL-GUP-vs-classic-Model-1}}
\end{figure}

The expressions for the null case for this model become
\begin{align}
	\theta_{\mathrm{GUP(1)}}^{(\mathrm{NL})}= & \pm\frac{2b}{\gamma\sqrt{p_{c}}}\left(1+\beta_{c}c^{2}\right)=\pm\frac{2b}{\gamma\sqrt{p_{c}}}F_{2},\\
	\frac{d\theta_{\mathrm{GUP}(1)}^{(\mathrm{NL})}}{d\lambda}= & \frac{1}{\gamma^{2}p_{c}}\left[-2b^{2}-2\beta_{b}b^{4}-2\beta_{c}c^{2}b^{2}\left(4\beta_{c}c^{2}+5\right)-2\beta_{b}\beta_{c}b^{4}c^{2}\right].
\end{align}
Clearly, with negative $\beta_{b}$ and $\beta_{c}$, the correction
term in the negative branch of the expansion scalar is positive, and
counters the negative classical term. Up to the first order in $\beta$,
the same is true in the Raychaudhuri equation. In fact the full numerical
results once again show that both $\theta_{\mathrm{GUP(1)}}^{(\mathrm{NL})}$
and $\frac{d\theta_{\mathrm{GUP}(1)}^{(\mathrm{NL})}}{d\tau}$ remain
finite everywhere inside the black hole as can be seen in Fig. \ref{fig:theta-RE-NL-GUP-vs-classic-Model-1}.

\begin{figure}
	\subfloat[The Kretschmann scalar $K$ for model 1 as a function of the Schwarzschild
	time $t$.]{\begin{centering}
			\includegraphics[scale=0.51]{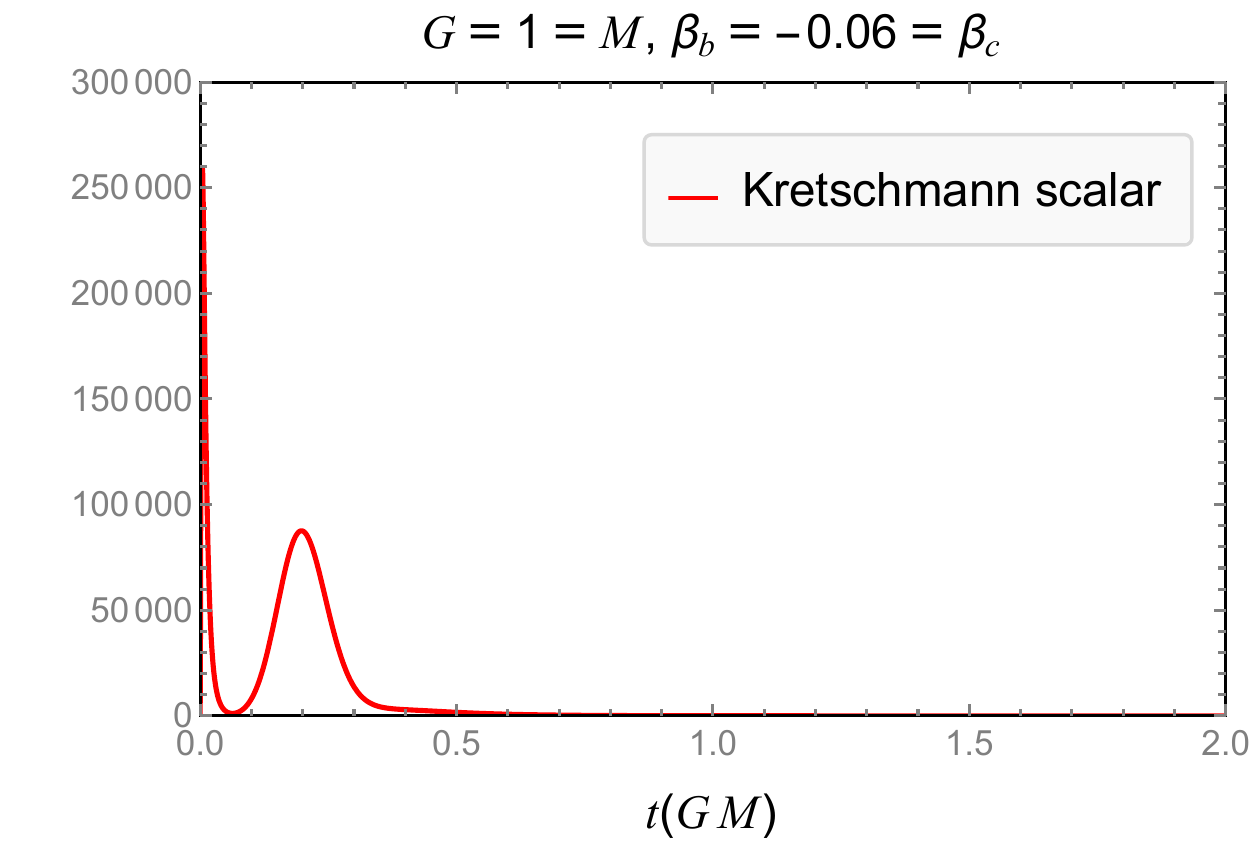}
			\par\end{centering}
	}\hfill{}\subfloat[Close up of $K$ close to $t=0$. It is seen that $K$ remains finite
	everywhere in the interior and vanishes for $t=0$.]{\begin{centering}
			\includegraphics[scale=0.51]{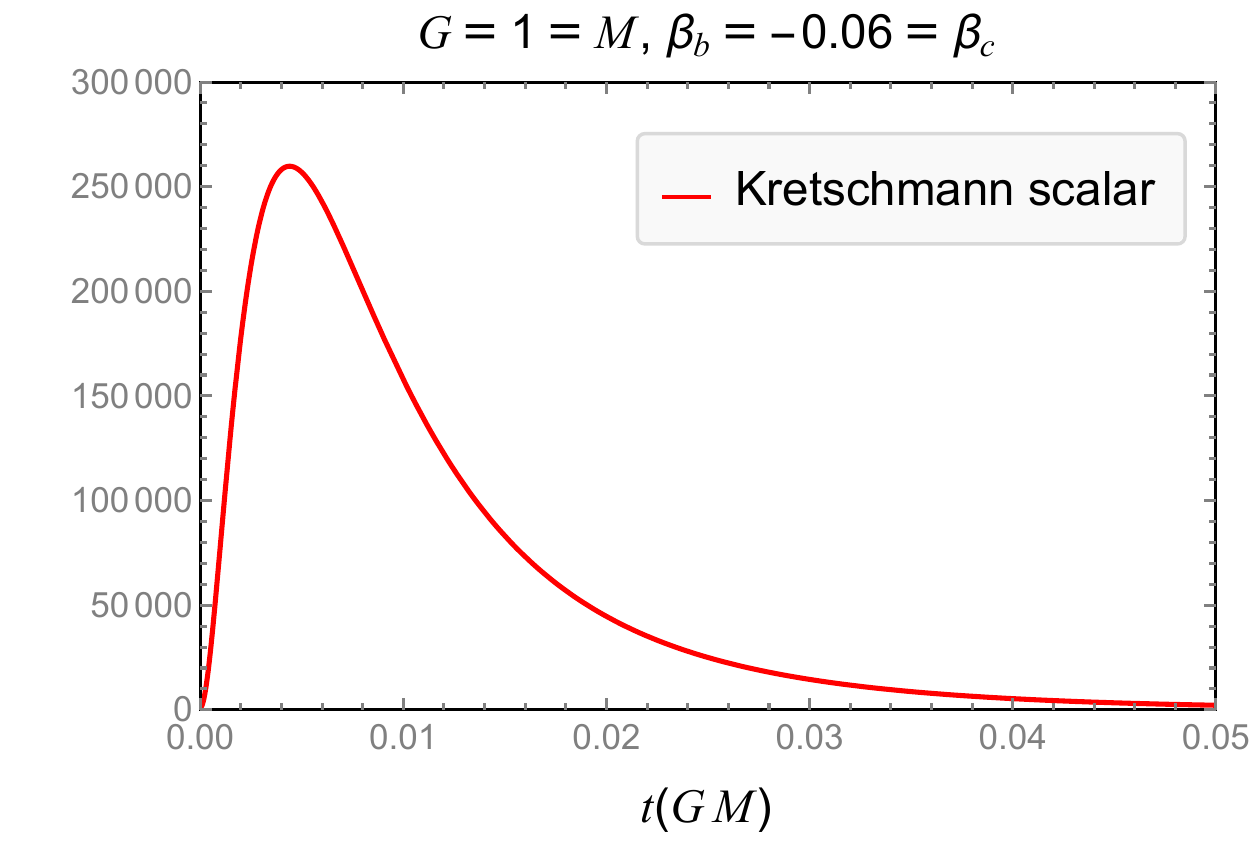}
			\par\end{centering}
	}
	
	\caption{$K$ in GUP model 1 \label{fig:K-GUP-Model-1}}
\end{figure}

We can also compute the expression for the effective Kretschmann scalar
$K$ in this case. This expression is quite large and we do not write
it down here. However, the plot of the $K$ in this case is presented
in Fig. \ref{fig:K-GUP-Model-1}. We can clearly see that $K$ remains
finite everywhere in the interior particularly for $t\to0$, although
it has a big bump close to this time. This confirms that the quantum
effects take over very close to $t=0$ and keep the curvature finite.
In fact $K\to0$ as $t\to0$.

\begin{figure}
	\subfloat[Left: Classical vs timelike $\theta$ in GUP model 3 as a function
	of the Schwarzschild time $t$. Right: Close up of the left figure
	close to $t=0$.]{%
		\begin{minipage}[t]{0.45\textwidth}%
			\begin{center}
				\includegraphics[scale=0.5]{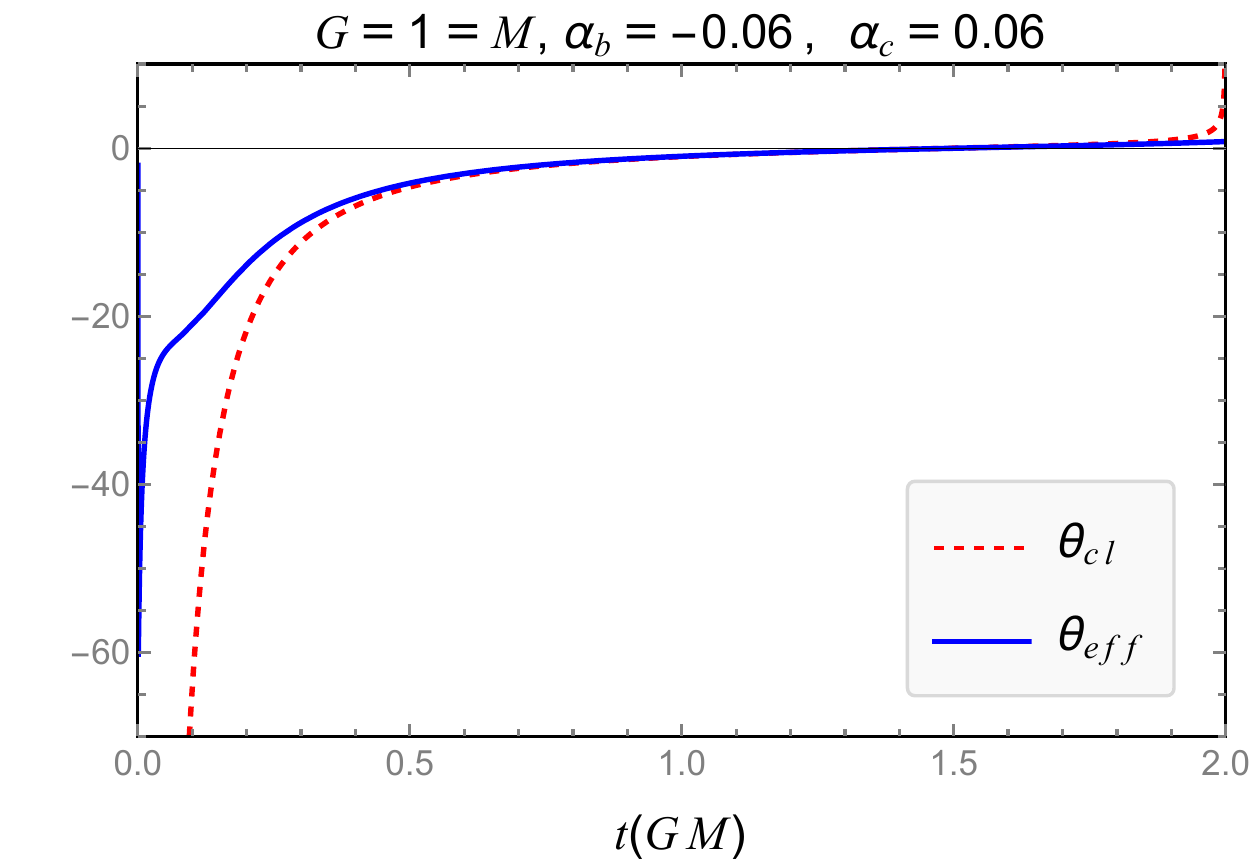}
				\par\end{center}%
		\end{minipage}\hfill{}%
		\begin{minipage}[t]{0.45\textwidth}%
			\begin{center}
				\includegraphics[scale=0.5]{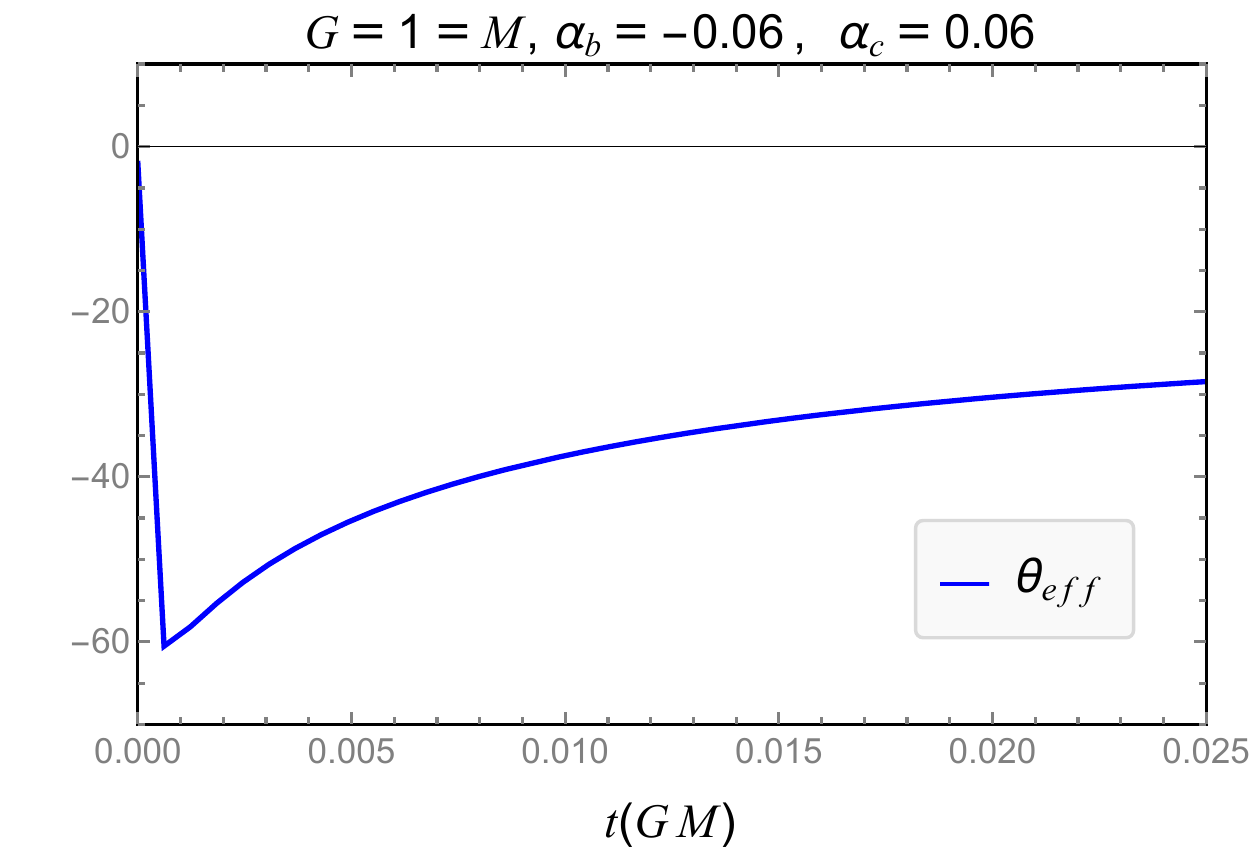}
				\par\end{center}%
		\end{minipage}
		
	}\hfill{}\subfloat[Left: Classical vs timelike $\frac{d\theta}{d\tau}$ in GUP model
	3 as a function of the Schwarzschild time $t$. Right: Close
	up of the left figure close to $t=0$.]{%
		\begin{minipage}[t]{0.45\textwidth}%
			\begin{center}
				\includegraphics[scale=0.5]{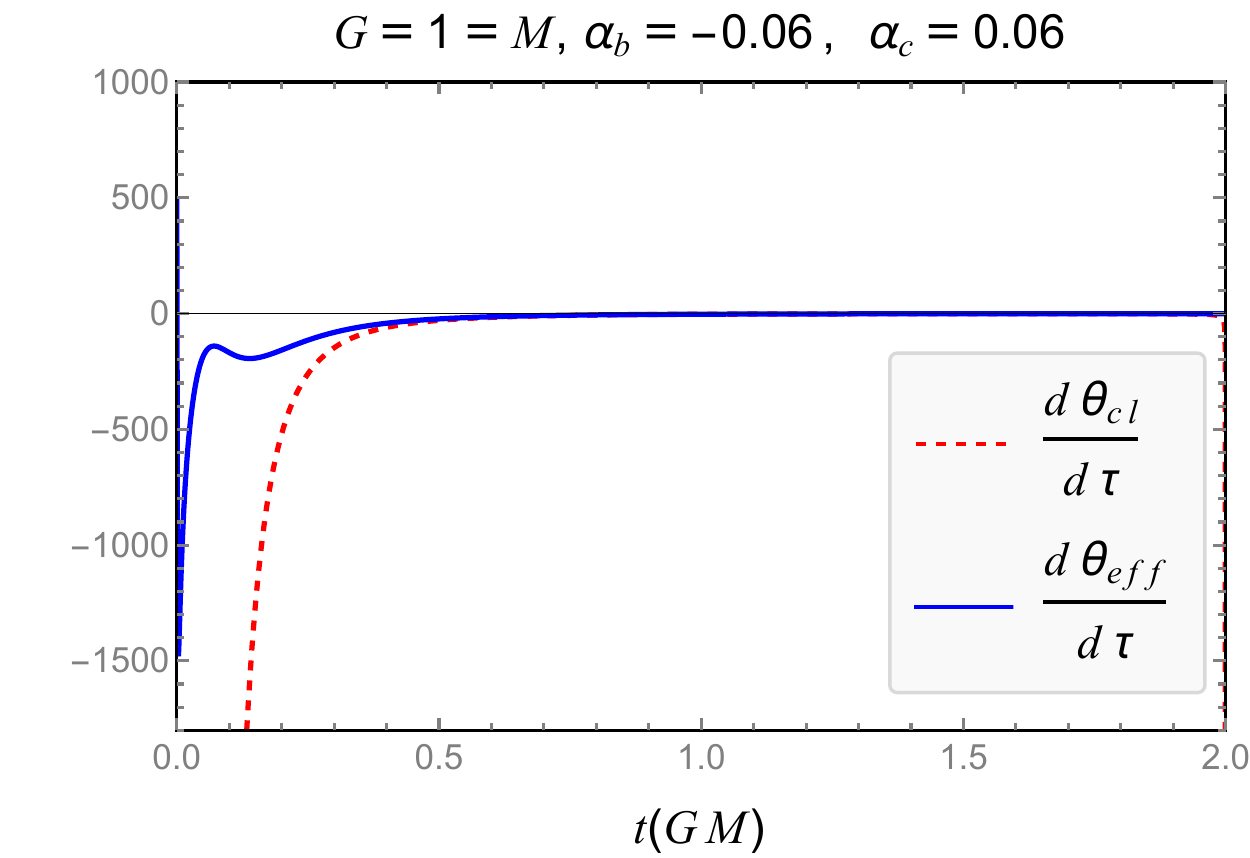}
				\par\end{center}%
		\end{minipage}\hfill{}%
		\begin{minipage}[t]{0.45\textwidth}%
			\begin{center}
				\includegraphics[scale=0.5]{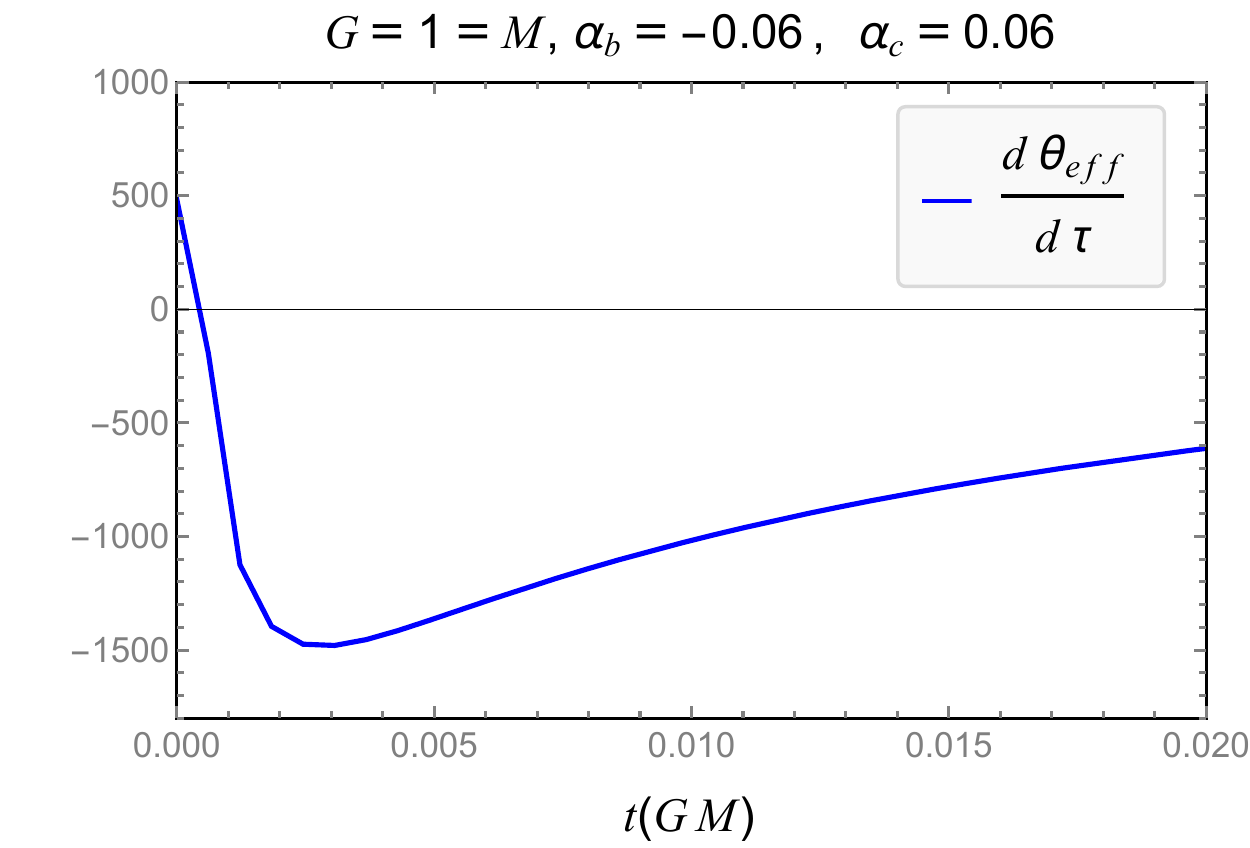}
				\par\end{center}%
		\end{minipage}
		
	}
	
	\caption{$\theta_{\mathrm{GUP}(3)}^{(\mathrm{TL})}$ and $\frac{d\theta_{\mathrm{GUP}(3)}^{(\mathrm{TL})}}{d\tau}$.
		Both effective expressions go to finite values as $t\to0$.
		\label{fig:theta-RE-TL-GUP-vs-classic-Model-3}}
\end{figure}

Another model for which it is possible to resolve the singularity
is model 3 which is the other configuration-dependent case. For the
timelike congruence in this model we obtain
\begin{align}
	\theta_{\mathrm{GUP(3)}}^{(\mathrm{TL})}= & \pm\frac{1}{2\gamma\sqrt{p_{c}}}\left[3b-\frac{\gamma^{2}}{b}+\alpha_{b}\left(b^{2}-\gamma^{2}\right)+2\alpha_{c}cb\right],\label{eq:theta-TL-GUP-3-in-canon-vars}\\
	\frac{d\theta_{\mathrm{GUP}(3)}^{(\mathrm{TL})}}{d\tau}= & \frac{1}{2\gamma^{2}p_{c}}\left[-\frac{9b^{2}}{2}-\frac{\gamma^{4}}{2b^{2}}-\gamma^{2}-10\alpha_{c}cb^{2}\right.\nonumber \\
	& \left.-\alpha_{b}b\left(\frac{7b^{2}}{2}+\frac{\gamma^{4}}{2b^{2}}+2\gamma^{2}\right)-\alpha_{b}^{2}b^{4}-\alpha_{b}^{2}b^{2}\left(\gamma^{2}+6c^{2}\right)-2\alpha_{b}\alpha_{c}b^{3}c\right].\label{eq:RE-TL-GUP-3-in-canon-vars}
\end{align}
It turns out that, as mentioned before, in this model the only case
where both the expansion scalar and its rate of change are finite
in the interior is when $\alpha_{b}<0$ and $\alpha_{c}>0$. This
can be checked by finding the solutions to the equations of motion
for all the cases of the signs of $\alpha_{b},\,\alpha_{c}$ and checking
the behavior of the expansion scalar and Raychaudhuri equation based
on them. Having this in mind, we notice that up to the first order
in $\alpha$'s, the correction term proportional to $\alpha_{b}$
in the negative branch of $\theta_{\mathrm{GUP(3)}}^{(\mathrm{TL})}$is
positive for $b^{2}>\gamma^{2}$ which is always the case when we
get close to $t=0$. The other correction term $\theta_{\mathrm{GUP(3)}}^{(\mathrm{TL})}$
proportional to $\alpha_{c}$ is positive for $bc<0$ which is guaranteed
to always hold in the interior based on the solutions to the equations
of motions in this model. Up to the same order, the correction term
in $\frac{d\theta_{\mathrm{GUP}(3)}^{(\mathrm{TL})}}{d\tau}$ proportional
to $\alpha_{b}$ is clearly always positive. The term proportional
to $\alpha_{c}$ is positive for $c<0$ which is also guaranteed to
hold based on the solutions to the equations of motion. Hence, the
first order correction terms contribute to defocusing of the geodesics.
This is indeed the case if we also consider the full form of the expressions
as can be seen from Fig. \eqref{fig:theta-RE-TL-GUP-vs-classic-Model-3}.

We see a difference in this model compared to model 1. Unlike model
1, in model 3 neither $\theta_{\mathrm{GUP}(3)}^{(\mathrm{TL})}$
nor $\frac{d\theta_{\mathrm{GUP}(3)}^{(\mathrm{TL})}}{d\tau}$ become
zero at $t=0$. They both take a finite values (negative and positive, respectively) as can be seen
from Fig. \eqref{fig:theta-RE-TL-GUP-vs-classic-Model-3}.

\begin{figure}
	\subfloat[Classical vs null $\theta$ in GUP model 3 as a function of the Schwarzschild
	time $t$.]{\begin{centering}
			\includegraphics[scale=0.51]{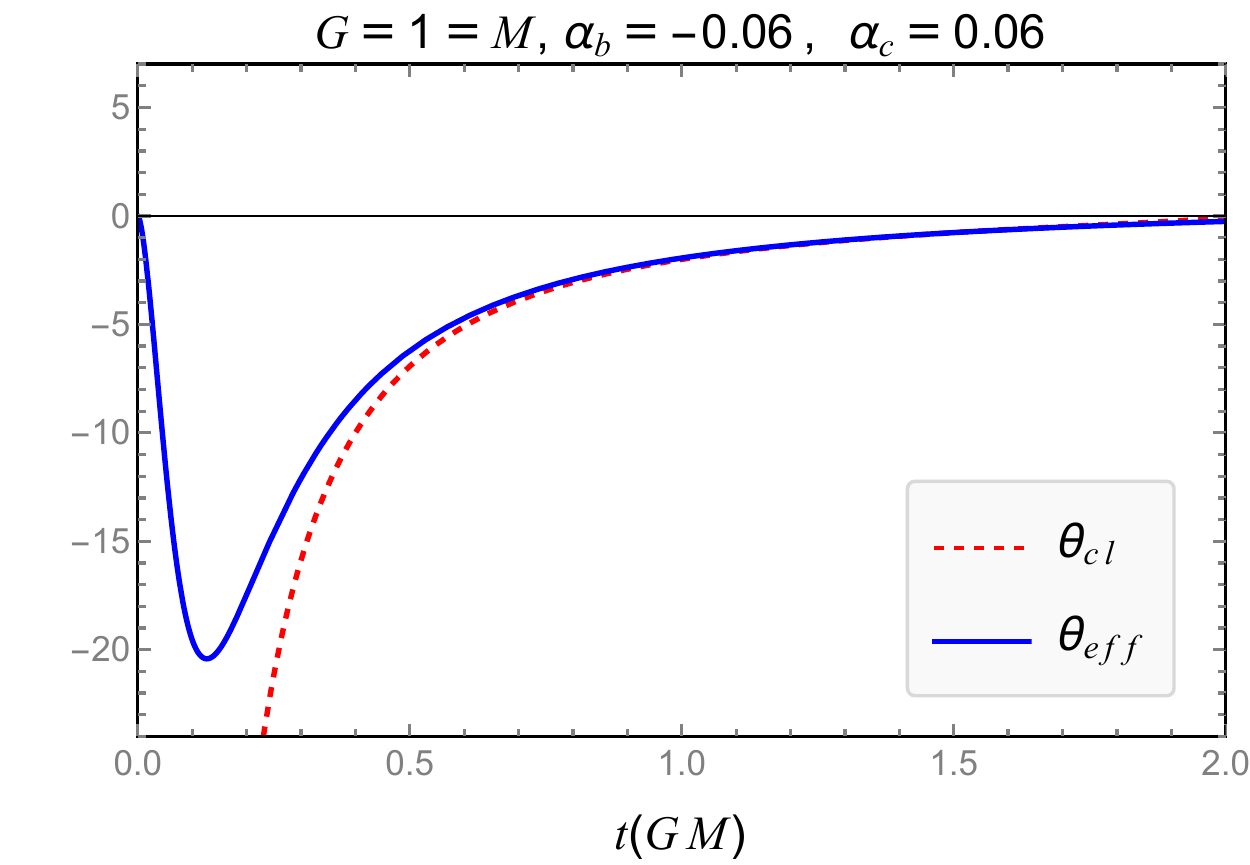}
			\par\end{centering}
	}\hfill{}\subfloat[Classical vs null $\frac{d\theta}{d\tau}$ in GUP model 3 as a function
	of the Schwarzschild time $t$.]{\begin{centering}
			\includegraphics[scale=0.51]{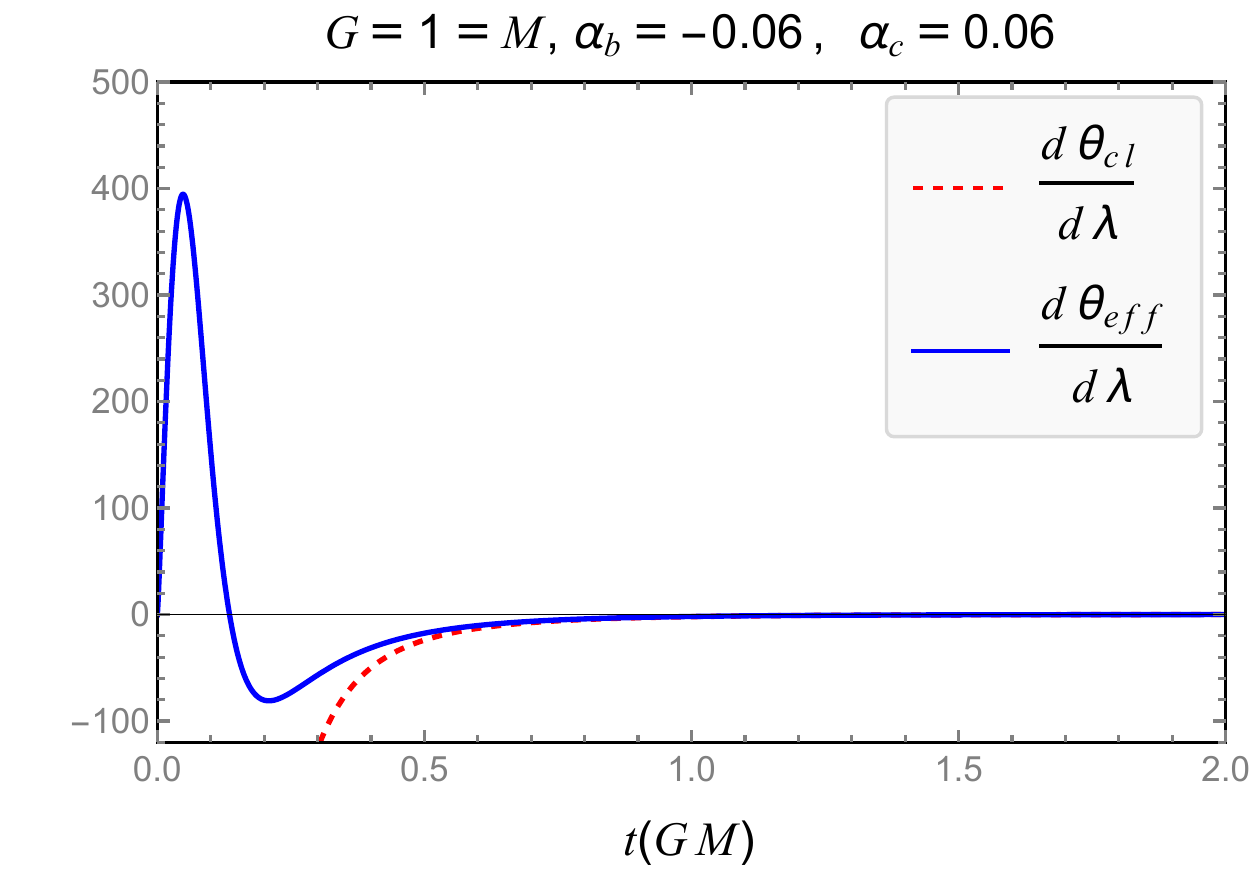}
			\par\end{centering}
	}
	
	\caption{$\theta_{\mathrm{GUP}(3)}^{(\mathrm{NL})}$ and $\frac{d\theta_{\mathrm{GUP}(3)}^{(\mathrm{NL})}}{d\lambda}$.
		Both effective expressions vanish as $t\to0$.
		\label{fig:theta-RE-NL-GUP-vs-classic-Model-3}}
\end{figure}

The behavior of null congruence for this model results in the following
expressions, 
\begin{align}
	\theta_{\mathrm{GUP(3)}}^{(\mathrm{NL})}= & \pm\frac{2b}{\gamma\sqrt{p_{c}}}\left(1+\alpha_{c}c\right)=\pm\frac{2b}{\gamma\sqrt{p_{c}}}F_{2},\\
	\frac{d\theta_{\mathrm{GUP}(3)}^{(\mathrm{NL})}}{d\lambda}= & -\frac{2}{\gamma^{2}p_{c}}\left[b^{2}+\alpha_{b}b^{3}+3\alpha_{c}b^{2}c+\alpha_{b}\alpha_{c}b^{3}c+2\alpha_{c}^{2}b^{2}c^{2}\right].
\end{align}
Given the behavior of the equations of motion and the signs of $\alpha_{b},\,\alpha_{c}$,
clearly the correction term up to the first order in $\alpha$'s in
the negative branch of $\theta_{\mathrm{GUP(3)}}^{(\mathrm{NL})}$
is positive. The same is true for similar terms in $\frac{d\theta_{\mathrm{GUP}(3)}^{(\mathrm{NL})}}{d\lambda}$.
As is expected the full expressions behave in a way that both $\theta_{\mathrm{GUP(3)}}^{(\mathrm{NL})}$
and $\frac{d\theta_{\mathrm{GUP}(3)}^{(\mathrm{NL})}}{d\lambda}$
remain finite everywhere in the interior. This is shown in Fig. \ref{fig:theta-RE-NL-GUP-vs-classic-Model-3}

We see that unlike the timelike case, the null case, $\theta_{\mathrm{GUP}(3)}^{(\mathrm{NL})}$
and $\frac{d\theta_{\mathrm{GUP}(3)}^{(\mathrm{NL})}}{d\lambda}$actually
end up being zero for $t=0$.

\begin{figure}
	\subfloat[The Kretschmann scalar $K$ for model 3 as a function of the Schwarzschild
	time $t$.]{\begin{centering}
			\includegraphics[scale=0.51]{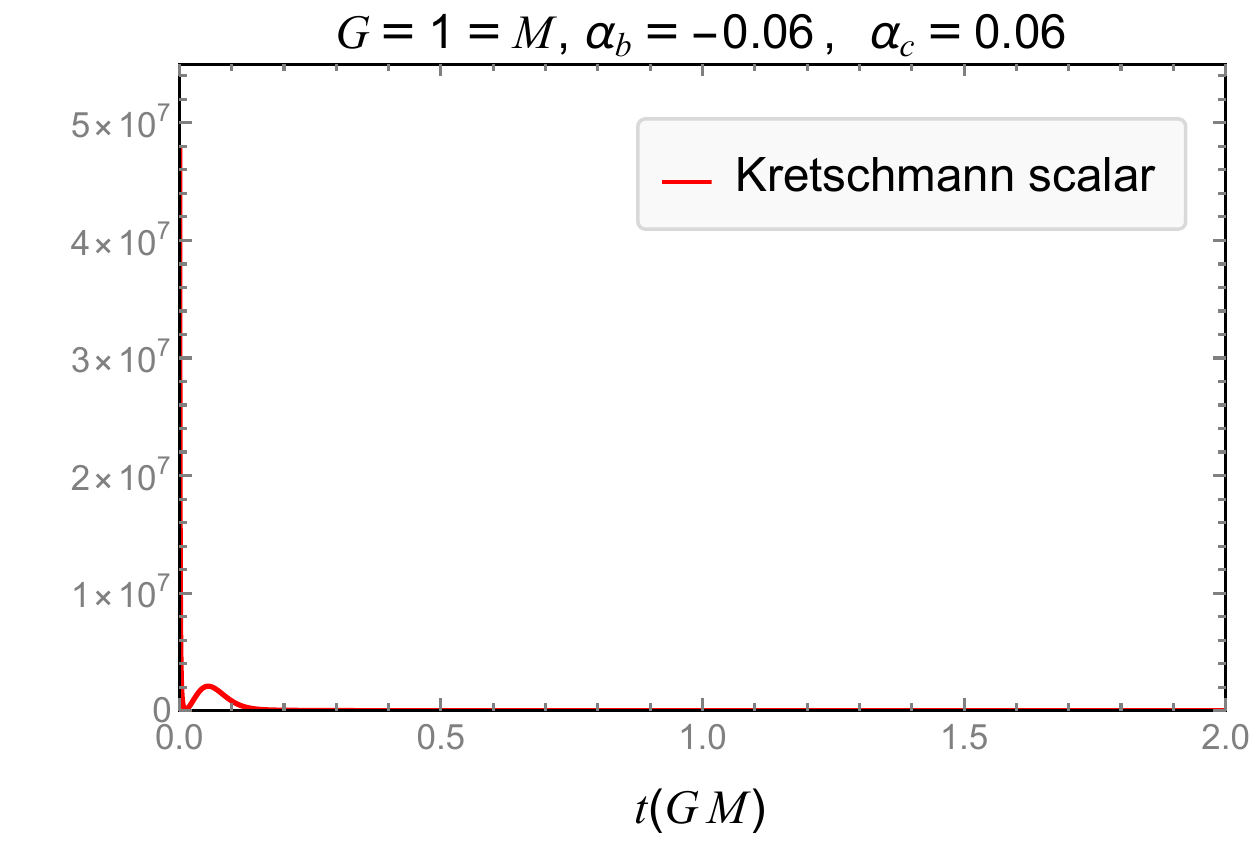}
			\par\end{centering}
	}\hfill{}\subfloat[Close up of $K$ close to $t=0$. It is seen that $K$ remains finite
	everywhere in the interior and vanishes for $t=0$.]{\begin{centering}
			\includegraphics[scale=0.51]{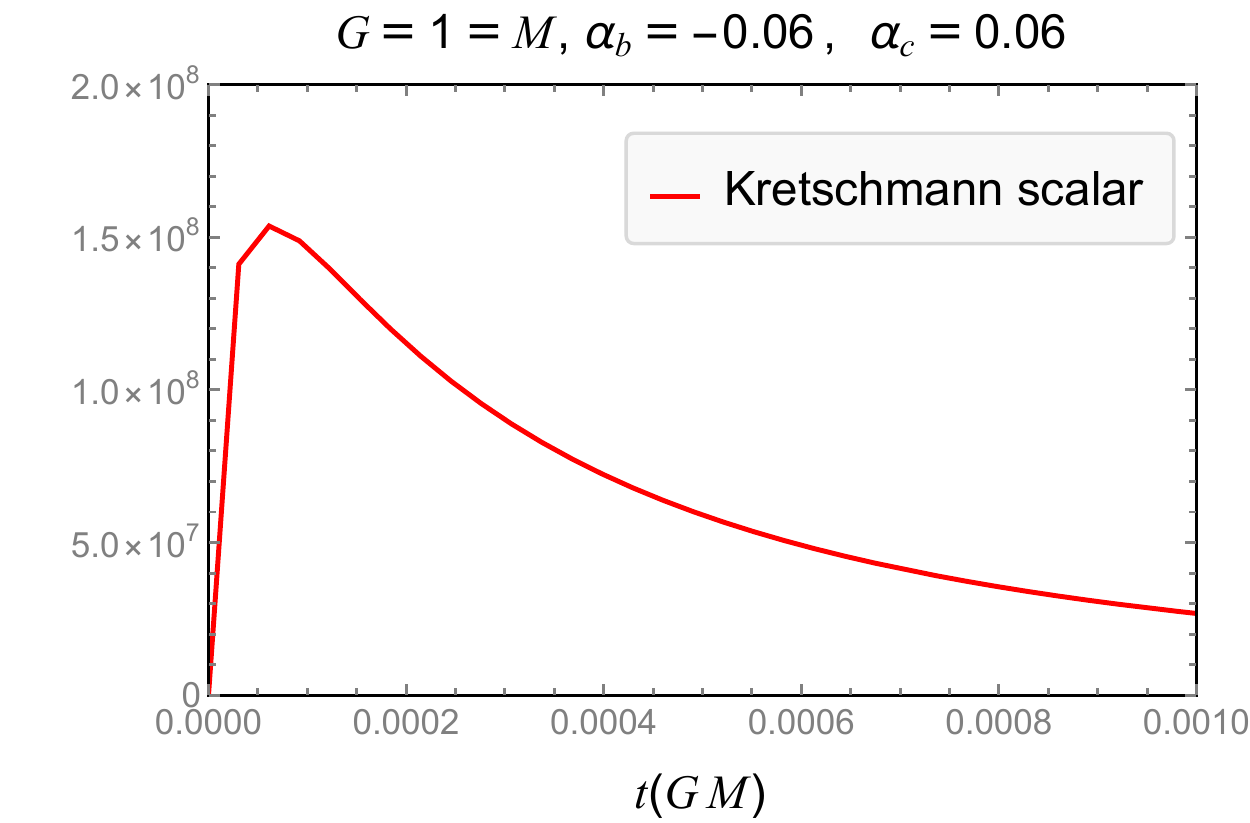}
			\par\end{centering}
	}
	
	\caption{$K$ in GUP model 3 \label{fig:K-GUP-Model-3}}
\end{figure}

Finally we can derive the Kretschmann scalar $K$ for this model and
plot is against $t$. this is depicted in Fig. \ref{fig:K-GUP-Model-3}.
Once again we see that although there is a big bump in $K$ close
to $t=0$, the quantum effects take over close to that time and turn
the curve around such that $k\to0$ as $t\to0$. This again confirms
that the singularity is resolved due to quantum effects.

\section{Discussion\label{sec:Discussion}}

In this work, we have reviewed some of our previous works on probing
the interior structure and singularity resolution of the Schwarzschild
black hole in LQG and GUP frameworks using expansion scalar and Raychaudhuri
equation. We have also presented new results regarding the Kretschmann
scalar in both frameworks in addition to new numerical results about
GUP models, which leads to eliminating some of them with respect to
their ability to resolve the singularity of Schwarzschild interior.

We first present the general form of timelike and null geodesics in
the Kantowski-Sachs spacetime which is isometric to the interior of
the Schwarzschild black hole. We then derive the timelike and null
expansion scalars and their rate of change, i.e., the Raychaudhuri
equation for congruences in this metric. All of these quantities depend
on the time derivative of metric components or canonical variables,
and hence on the solutions to the equations of motion. If the classical
Hamiltonian gets modified due to quantum effects, the resulting effective
Hamiltonian would yield different equations of motion and hence we
will obtain modified expansion scalar and Raychaudhuri equation. This
is the route we follow.

We first write the Hamiltonian in terms of the Ashtekar-Barbero connection
adapted to this model. Using the equations of motion derived from
this Hamiltonian, we compute the expansion scalar $\theta$ and its
rate of change $\frac{d\theta}{d\tau}$ ($\frac{d\theta}{d\lambda}$
in the null case, with $\lambda$ the curve parameter) for this classical
system. Not surprisingly, both $\theta$ and $\frac{d\theta}{d\tau}$
diverse at the center of the black hole where $t\to0$, where $t$
is the Schwarzschild time in the interior. We then turn into LQG,
or more precisely polymer quantization of the Hamiltonian. We use
three main schemes of this method, which are different based on the
form of the minimum scale introduced in the theory. Using the resulting
effective Hamiltonian, we find the new modified solutions to the equations
of motion, and find timelike and null $\theta$ and $\frac{d\theta}{d\tau}$
($\frac{d\theta}{d\lambda}$). We show that these effective quantities
remain finite everywhere in the interior and never diverge even at
$t=0$. This is shown to be true for all three schemes we studied.
We also compute the Kretschmann scalar $K$ analytically and numerically
and show that in all three schemes, $K$ remain finite everywhere
in the interior. This shows that LQG definitely resolves the singularity.

We then turn to the GUP approach. By modifying the algebra of the
canonical variables using GUP parameters $\alpha,\,\beta$ we study
four most commonly considered model in GUP. Using modified equations
of motion, once again we derive timelike and null $\theta$ and $\frac{d\theta}{d\tau}$
($\frac{d\theta}{d\lambda}$) for all the four models. Interestingly,
we find that only two of these models, for which the canonical algebra
is modified by introducing additional configuration-related terms,
have the ability to resolve the singularity. Furthermore not all of
the cases of GUP parameters $\alpha,\,\beta$ in these two models
lead to the singularity resolution. In one of the models (model 1)
where the canonical algebra is modified by introducing quadratic terms
in configuration variables, the GUP parameters should be both negative
to achieve singularity resolution. In the other model (model 3) where
the canonical algebra is modified by linear terms in configuration
variables, one of the parameters should be negative while the other
should be positive, i.e., $\alpha_{b}<0$ and $\alpha_{c}>0$. Even
the inverse of this case, where $\alpha_{b}>0$ and $\alpha_{c}<0$,
does not work. In both of these models with these specific choice
of the GUP parameters, not only timelike and null $\theta$ and $\frac{d\theta}{d\tau}$
($\frac{d\theta}{d\lambda}$) remain finite everywhere in the interior,
but also the Kretschmann scalar does so.

Hence, we have reaffirmed previous works that claim that all of the
LQG models based on a certain polymer quantization resolve the Schwarzschild
singularity. Furthermore, we have presented new results excluding
certain GUP models by means of their abilities to resolve such a singularity
in this Hamiltonian first order framework.

It is interesting to apply this method to works within LQG that use
approaches other than the polymer quantization or models which deal
with the full spacetime of the Schwrazschild balck hole. One can also
apply this method to the full spacetime of the Schwarzschild model
but using the GUP framework. These are all interesting works that
can be followed in the future.
	
\begin{acknowledgments}
	S. D. and S. R. acknowledge the support of the Natural Sciences and Engineering
	Research Council of Canada (NSERC), [S. D.: funding reference number RGPIN-2019-
	05404, S. R. funding reference numbers RGPIN-2021-03644 and DGECR-2021-00302]
\end{acknowledgments}

\appendix
\section{Raychaudhuri equation\label{sec:Raychaudhuri-equation}}

One of the ways to probe the classical and effective structure of
spacetime is by investigating the behavior of geodesics. More precisely,
how a congruence of timelike or null geodesic evolves over time. This
analysis is intimately related to the geodesic deviation and to the
so-called expansion scalar and its rate of change, the Raychaudhuri
equation, as we will see below.

Consider a family of curves in a region of spacetime such that through
each point in that region, one and only one geodesic passes. This
is called a congruence. If every curve in the congruence is timelike/null,
then conguence is called timelike/null. In what follows we briefly
review both type of these congruences and how they evolve over time.

\subsection{Timelike congruence}

Suppose we have a congruence of timelike geodesics with unit timelike
tangent vectors\footnote{We use lower case Latin letter for abstract indices and Greek indices
	for components.} $\left\{ U^{a}\right\} $, where
\begin{align}
	g_{ab}U^{a}U^{b}= & -1, & U^{a}\nabla_{a}U^{b}= & 0.\label{eq:U-props}
\end{align}
Using these curves, we can decompose the spacetime metric $g_{ab}$
as
\begin{equation}
	g_{ab}=h_{ab}-U_{a}U_{b},\label{eq:g-h-uu}
\end{equation}
where $h_{ab}$ is called the transverse metric. The metric $h_{ab}$
is spatial in the sense that it is orthogonal to the timelike $U^{a}$
\begin{equation}
	h_{ab}U^{b}=0=h_{ab}U^{a}
\end{equation}
which is simple to check from \eqref{eq:g-h-uu}. It is also essentially
a three dimensional metric on the hypersurfaces transverse to $U^{a}$,
and this can be seen by taking the trace of \eqref{eq:g-h-uu} which
leads to $h^{a}{}_{a}=3$. The $(1,1)$ tensor $h^{a}{}_{b}$ is a
projection operator onto the transverse hypersurfaces to $U^{a}$
since $B^{a}{}_{c}B^{c}{}_{b}=B^{a}{}_{b}$, which can also be seen
from \eqref{eq:g-h-uu}.

In order to study the evolution of the congruence, we consider the
tensor
\begin{equation}
	B_{ab}=\nabla_{b}U_{a},
\end{equation}
which is sometimes called the expansion tensor, and its $B^{a}{}_{b}$
version measures the amount of failure of the deviation vector between
the geodesics in the congruence from being parallel transported. This
tensor is also spatial since it is orthogonal to $U^{a}$
\begin{equation}
	B_{ab}U^{b}=0=B_{ab}U^{a}.
\end{equation}
This is the result of the curves being geodesics, i.e., \eqref{eq:U-props}.

We algebraically decompose $B_{ab}$ into its trace part, symmetric
traceless part and antisymmetric part as
\begin{equation}
	B_{ab}=\frac{1}{3}\theta h_{ab}+\sigma_{ab}+\omega_{ab}.
\end{equation}
Here 
\begin{equation}
	\theta=B^{a}{}_{a}
\end{equation}
is the trace of $B_{ab}$ and is called the expansion scalar, and
describes the fractional change of the area of the cross-section area
of the congruence per unit time. The symmetric traceless part
\begin{equation}
	\sigma_{ab}=B_{(ab)}-\frac{1}{3}\theta h_{ab}
\end{equation}
is called the shear tensor. It measures how the cross-section is deformed
from a circle. The antisymmetric part
\begin{equation}
	\omega_{ab}=B_{[ab]}
\end{equation}
is called the vorticity (or rotation) tensor, and encodes the overall
rotation of the cross-section while the area remains unchanged. The
factor $\frac{1}{3}$ is the result of the transverse hypersurfaces
being three dimensional. Both $\sigma_{ab}$ and $\omega_{ab}$ are
spatial tensors, i.e., and their contraction with $U^{a}$ vanishes.

These quantities and their rates of change along the geodesic in proper
time incorporate important information about the structure of spacetime
particularly, the geodesics incompleteness and singularities. The
most important of these rates of change is the rate of change of the
expansion scalar along the geodesics, $\frac{d\theta}{d\tau}$, where
$\tau$ is the proper time along the geodesic. It can be computed
to yield
\begin{equation}
	U^{a}\nabla_{a}\theta\coloneqq\frac{d\theta}{d\tau}=-\frac{1}{3}\theta^{2}-\sigma_{ab}\sigma^{ab}+\omega_{ab}\omega^{ab}-R_{ab}U^{a}U^{b}.\label{eq:RE-timelike}
\end{equation}
Here, $\sigma^{2}=\sigma_{ab}\sigma^{ab}$ is called the shear parameter
and $\omega^{2}=\omega_{ab}\omega^{ab}$ is the vorticity parameter.
In the presence of matter that obeys strong energy condition the last
term $R_{ab}U^{a}U^{b}$ is always nonnegative, and it vanishes for
the vacuum case.

The above equation is called the Raychaudhuri equation, and is a purely
geometrical identity. Since $\sigma_{ab}$ and $\omega_{ab}$ are
spatial tensors, $\sigma^{2}>0,\,\omega^{2}>0$. So, in cases where
the strong energy condition holds, the first second and the fourth
terms on the right hand side of \eqref{eq:RE-timelike} (including
the signs behind them) are all negative and contribute to the convergence
of geodesics as we move along them. The only term with a positive
sign is the third term $\omega^{2}$, which contributes to divergence
of geodesics. Hence, if it was not for the vorticity parameter, the
rate of change of the expansion scalar would always have been negative,
which leads to the geodesics increasingly converge as we move along
them. In fact this is the case where $U^{a}$ are hypersurface-orthogonal.
In that case we have
\begin{equation}
	\frac{d\theta}{d\tau}\leq-\frac{1}{3}\theta^{2}
\end{equation}
which can be solved to yield
\begin{equation}
	\frac{1}{\theta(\tau)}\geq\frac{1}{\theta\left(\tau_{0}\right)}+\frac{1}{3}\tau.
\end{equation}
Then if we have an initially-converging congruence, i.e., $\theta\left(\tau_{0}\right)<0$,
we arrive at a caustic point where $\theta\to-\infty$ in a finite
proper time 
\begin{equation}
	\tau\leq-3\frac{1}{\theta\left(\tau_{0}\right)}.
\end{equation}
This caustic point could be the result of bad coordinates or a real
physical singularity in spacetime. The Raychaudhuri equation is the
backbone of the theorems of Hawking and Penrose about geodesic incompleteness
and singularities in general spacetimes.

\subsection{Null congruence}

For the null congruences with geodesics parametrized by $\lambda$,
we consider the subspace normal to the null vector field $k^{a}$
tangent to the geodesics. To do that, introduce a auxiliary null vector
field $l^{a}$ such that 
\begin{equation}
	k_{a}l^{a}=-1.\label{eq:k-l-inner}
\end{equation}
Using these two null vectors, we can decompose the metric as
\begin{equation}
	g_{ab}=h_{ab}-2k_{(a}l_{b)},\label{eq:transverse-h-null}
\end{equation}
where once again $h_{ab}$ is the transverse metric on the hypersurface
transverse to $k^{a}$ and $l^{a}$. This hypersurface and its metric
$h_{ab}$ is two dimensional which can be confirmed by noticing $h^{a}{}_{a}=2$.
The tensor $h^{a}{}_{b}$ is again a projection operator onto these
transverse hypersurfaces.

We can once again define 
\begin{equation}
	B_{ab}=\nabla_{b}k_{a},\label{eq:B-k}
\end{equation}
but while this tensor is orthogonal to $k^{a}$, it is not orthogonal
to $l^{a}$. It turns out the purely transverse part $B_{ab}$,
\begin{equation}
	\tilde{B}_{ab}=h^{c}{}_{a}h^{d}{}_{b}B_{cd}\label{eq:B-tilde-gen}
\end{equation}
is the tensor that is orthogonal to both $k^{a}$ and $l^{a}$. This
tensor can explicitly written as
\begin{equation}
	\tilde{B}_{ab}=B_{ab}+k_{a}l^{c}B_{cb}+k_{b}B_{ac}l^{c}+k_{a}k_{b}B_{cd}l^{c}l^{d}.\label{eq:B-tilde-gen-2}
\end{equation}
Once again we can decompose $\tilde{B}_{ab}$ as
\begin{equation}
	\tilde{B}_{ab}=\frac{1}{2}\tilde{\theta}h_{ab}+\tilde{\sigma}_{ab}+\tilde{\omega}_{ab},
\end{equation}
where
\begin{align}
	\tilde{\theta}= & \tilde{B}^{a}{}_{a}\\
	\tilde{\sigma}_{ab}= & \tilde{B}_{(ab)}-\frac{1}{2}\theta h_{ab}\\
	\tilde{\omega}_{ab}= & \tilde{B}_{[ab]}
\end{align}
are the expansion scalar, and shear and vorticity tenors respectively.
The factor $\frac{1}{2}$ is the result of the transverse hypersurfaces
being two dimensional. Again, both $\tilde{\sigma}_{ab}$ and $\tilde{\omega}_{ab}$
are spatial tensors.

The Raychaudhuri equation is now derived by considering the evolution
of $\tilde{\theta}$ along the geodesics
\begin{equation}
	k^{a}\nabla_{a}\tilde{\theta}\coloneqq\frac{d\tilde{\theta}}{d\lambda}=-\frac{1}{2}\tilde{\theta}^{2}-\tilde{\sigma}_{ab}\tilde{\sigma}^{ab}+\tilde{\omega}_{ab}\tilde{\omega}^{ab}-R_{ab}k^{a}k^{b}.\label{eq:RE-null}
\end{equation}
Notice that $\tilde{\theta}$ is unique and independent of $l^{a}$
since $\tilde{\theta}=\tilde{B}^{a}{}_{a}=B^{a}{}_{a}=\nabla_{a}k^{a}$,
and so is the Raychaudhuri equation above. The interpretation of \eqref{eq:RE-null}
is similar to the to the timelike case and leads to the existence
of caustic points in many situations.

In what follows we will use both the expansion and the Raychaudhuri
equation for both the null and timelike cases to study the effective
behavior of the interior of the Schwarzschild black hole.

	\bibliographystyle{apsrev4-2}
	\bibliography{mainbib}
	
\end{document}